\documentclass[
 reprint,
superscriptaddress,
 amsmath,amssymb,
 aps,
 pre,tightenlines,longbibliography
]{revtex4-2}
\usepackage[table,xcdraw]{xcolor}
\usepackage{lipsum}
\usepackage{graphicx}
\usepackage{dcolumn}
\usepackage{bm}
\usepackage{natbib}
\usepackage{graphicx}
\usepackage{textcomp}
\usepackage{amsmath}
\usepackage{amsfonts}
\usepackage{mathtools}
\usepackage{mathrsfs}
\usepackage{empheq}
\usepackage{tcolorbox}
\usepackage{fancyhdr}
\usepackage{datetime}
\usepackage{lipsum}
\usepackage{graphicx}
\usepackage{dcolumn}
\usepackage{bm}
\usepackage{subcaption}
\usepackage{xspace}
\usepackage{amsmath}
\usepackage{nccmath}
\usepackage{booktabs}
\usepackage[export]{adjustbox}
\usepackage[margin=0.68in]{geometry}
\usepackage[hidelinks]{hyperref}
\usepackage{booktabs}

\hypersetup{
    colorlinks=false,
    linkcolor=black,
    urlcolor=black}

\begin{document}

\title{Exact coherent structures and phase space geometry of pre-turbulent 2D active nematic channel flow}

\author{Caleb G. Wagner}
\affiliation{
Mechanical and Materials Engineering, University of Nebraska - Lincoln, Lincoln, NE 68588}

\author{Michael M. Norton}
\affiliation{
School of Physics and Astronomy, Rochester Institute of Technology, Rochester, NY 14623
}
\author{Jae Sung Park}
\affiliation{
Mechanical and Materials Engineering, University of Nebraska - Lincoln, Lincoln, NE 68588}

\author{Piyush Grover}
\affiliation{
Mechanical and Materials Engineering, University of Nebraska - Lincoln, Lincoln, NE 68588}
\begin{abstract}
Confined active nematics exhibit rich dynamical behavior, including spontaneous flows, periodic defect dynamics, and chaotic `active turbulence'. Here, we study these phenomena using the framework of Exact Coherent Structures, which has been successful in characterizing the routes to high Reynolds number turbulence of passive fluids. Exact Coherent Structures are stationary, periodic, quasiperiodic, or traveling wave solutions of the hydrodynamic equations that, together with their invariant manifolds, serve as an organizing template of the dynamics. We compute the dominant Exact Coherent Structures and connecting orbits in a pre-turbulent active nematic channel flow, which enables a fully nonlinear but highly reduced order description in terms of a directed graph. Using this reduced representation, we compute instantaneous perturbations that switch the system between disparate spatiotemporal states occupying distant regions of the infinite dimensional phase space. Our results lay the groundwork for a systematic means of understanding and controlling active nematic flows in the moderate to high activity regime.
\end{abstract}
\maketitle
Active matter is a class of materials composed of interacting and energy-consuming constituents. The past two decades have seen active matter grow into a new paradigm of nonequilibrium matter, with applications to both synthetic and biological systems \cite{Gompper2020}. Under the influence of particle-level driving forces, the emergent spatiotemporal structures of active matter are free to explore a much larger state space than available to passive equilibrium materials. Behaviors with no known equilibrium analogue include flocking and swarming \cite{Toner1995,Toner2005,Vicsek2012,Attanasi2014b,Attanasi2014c,Cavagna2017,vanderVaart2019}, athermal clustering of spheres \cite{Fily2012,Redner2013,Stenhammar2013,Buttinoni2013,Cates2015}, spontaneous flows \cite{Wan2008,Tailleur2009,Angelani2011,Ghosh2013,giomi2012banding,Marenduzzo2007rheology,Giomi2008concentration,Bartolo2013,Baek2018}, and low Reynolds number `active' turbulence \cite{Dombrowski2004,Wensink2012,doostmohammadi2018active,BlanchMercader2018,alert2021active}.

There is an extensive theoretical framework for understanding and manipulating emergent structures in materials at or near equilibrium. However, there is not yet an equivalent framework for active matter. In this paper, we make progress towards this goal in the context of active nematics (AN), which are suspensions of active, rod-like, and apolar components \cite{marchetti2013hydrodynamics,doostmohammadi2018active}; examples include bacterial films and cell colonies \cite{Yaman2019,DellArciprete2018}. Some of the most distinct phenomenology of AN occurs under confinement, in which case diverse spatiotemporal flow patterns are observed, including states of active turbulence \citep{giomi2012banding,Keber2014,doostmohammadi2018active,opathalage2019self,Wu2017transition,shendruk2017dancing,duclos2020topological}.
 There is much interest in learning to navigate this large space of spatiotemporal structures, for example steering a system toward a desired end state or switching between states \cite{ross2019controlling, zhang2019structuring,norton2020optimal}. In addition, there are fundamental unanswered questions related to active turbulence: how active fluids become turbulent, how to characterize them, and how to promote or inhibit transition to turbulence \cite{bowick2021symmetry,alert2021active}.

\begin{figure*}
\includegraphics[width=\textwidth,height=1.8in]{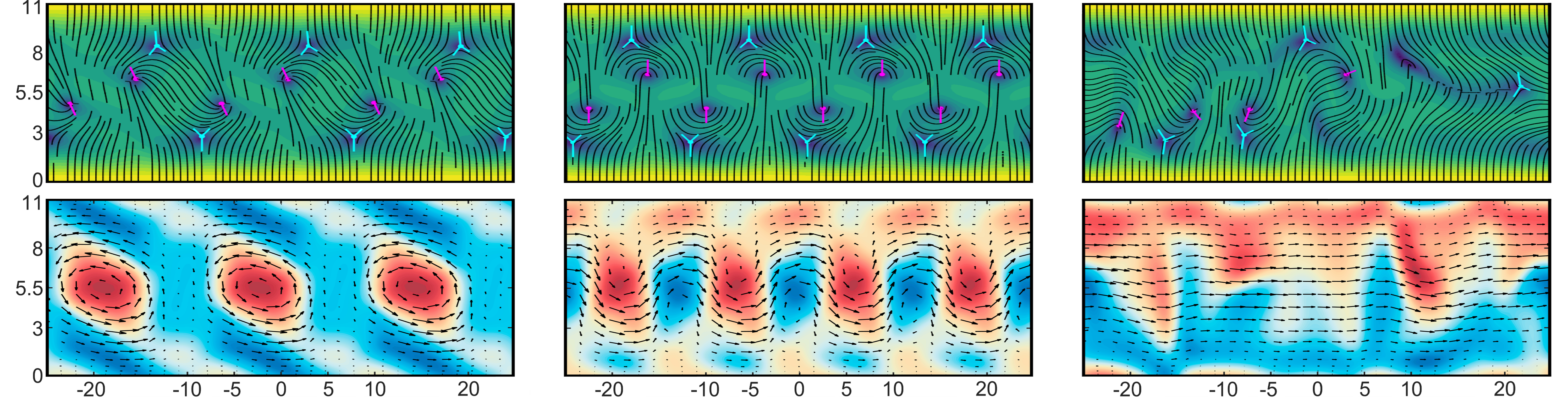}
\caption{\footnotesize{Snapshots of three Exact Coherent Structures (ECS). The top plots show the nematic director field overlayed on the nematic order parameter (color gradient), and the bottom plots show the velocity field overlayed on the vorticity. \textbf{Left column:} $\text{PO}_\text{3a}$, a periodic orbit with a 3-fold translational symmetry. \textbf{Middle column:} $\text{RPO}_\text{u4}$, a relative periodic orbit that cycles between a vortex lattice (as shown) and a nearly unidirectional, defect-less flow. \textbf{Right column:} $\text{RPO}_\text{1c}$, a relative periodic orbit without obvious spatial structure. See \cite{Note1} for videos of each ECS.}} \label{fig:ECS_snapshots}
\end{figure*}

Here we take a deterministic dynamical systems approach to these questions, beginning with the hydrodynamic equations governing AN. The dynamical systems approach 
has provided fresh insight into the long-standing problem of transition to turbulence in \textit{passive, high Reynolds number} fluid flows \citep{graham2020exact}: the core premise, going back to \citep{hopf1948mathematical,ruelle1971nature}, considers the fluid to be a deterministic dynamical system evolving in an infinite dimensional phase space \citep{cvitanovic2013recurrent}. The dominant flow structures are understood in terms of \emph{Exact Coherent Structures} (ECS) and the dynamical pathways connecting them. An ECS is a (generically unstable) stationary, periodic, quasiperiodic, or traveling wave solution of the hydrodynamic equations. Each ECS possesses invariant manifolds that are dynamical pathways connecting regions of phase space. A finite set of ECS, together with their invariant manifolds, constitutes a reduced-order but exact characterization of the global phase space. Though each ECS is non-turbulent, this representation is fully adequate for describing turbulent flows, which appear as chaotic trajectories meandering through the phase space and visiting the neighborhoods of different ECS in a recurring fashion \citep{park2015exact,budanur2019geometry,suri2020capturing}. Therefore, the ECS and their invariant manifolds act as an organizing template for the complicated spatiotemporal motion of the fluid. In inertial fluids, control strategies using this framework \cite{davis2020dynamics,linkmann2020linear,lucas2020stabilisation} are being explored for suppressing or delaying the transition to turbulence and reducing viscous dissipation. 
Recently, the approach has also been extended to elasto-inertial \cite{dubief2020first} and viscoelastic \cite{page2020exact} turbulence.

However, similar insight is missing in \textit{active, low Reynolds number fluids}. Previous work on pre-turbulent flows has focused on discovering stable solutions and tracking equilibria through primary bifurcations \cite{giomi2012banding,shendruk2017dancing,norton2018insensitivity,walton2020pressure}, while fully developed turbulence has been studied using coarse-grained statistical descriptions \citep{wensink2012meso,linkmann2019phase,linkmann2020condensate,mukherjee2021anomalous,alert2021active} that do not deal primarily with deterministic dynamics.

In this work, we take a first step toward developing a dynamical systems picture of AN turbulence. Specifically, we undertake a detailed study of ECS and heteroclinic connections in a 2D channel in the pre-turbulent regime. We find three coexisting attractors---two periodic orbits and a low-dimensional chaotic set---and over 40 unstable ECS. Away from the attractors, the phase space has complex global structure shaped by the unstable ECS and their invariant manifolds. In particular, the ECS dictate which of the three attractors a given flow configuration will evolve toward.

Our results go beyond previous work on AN in that they generate a reduced-order picture of the exact nonlinear dynamics: because the ECS framework is based on global relationships among exact time-dependent structures, it does not involve phenomenological approximations or restrictions to locally linear analysis. Moreover, our computation of \textit{unstable} structures generates new insight into the origin of stable structures and the dynamical pathways leading to them. Finally, we show how this understanding allows control of AN flows using minimal external input.

\begin{figure*}[!htbp]
\includegraphics[width=7.032727272in,height=3in]{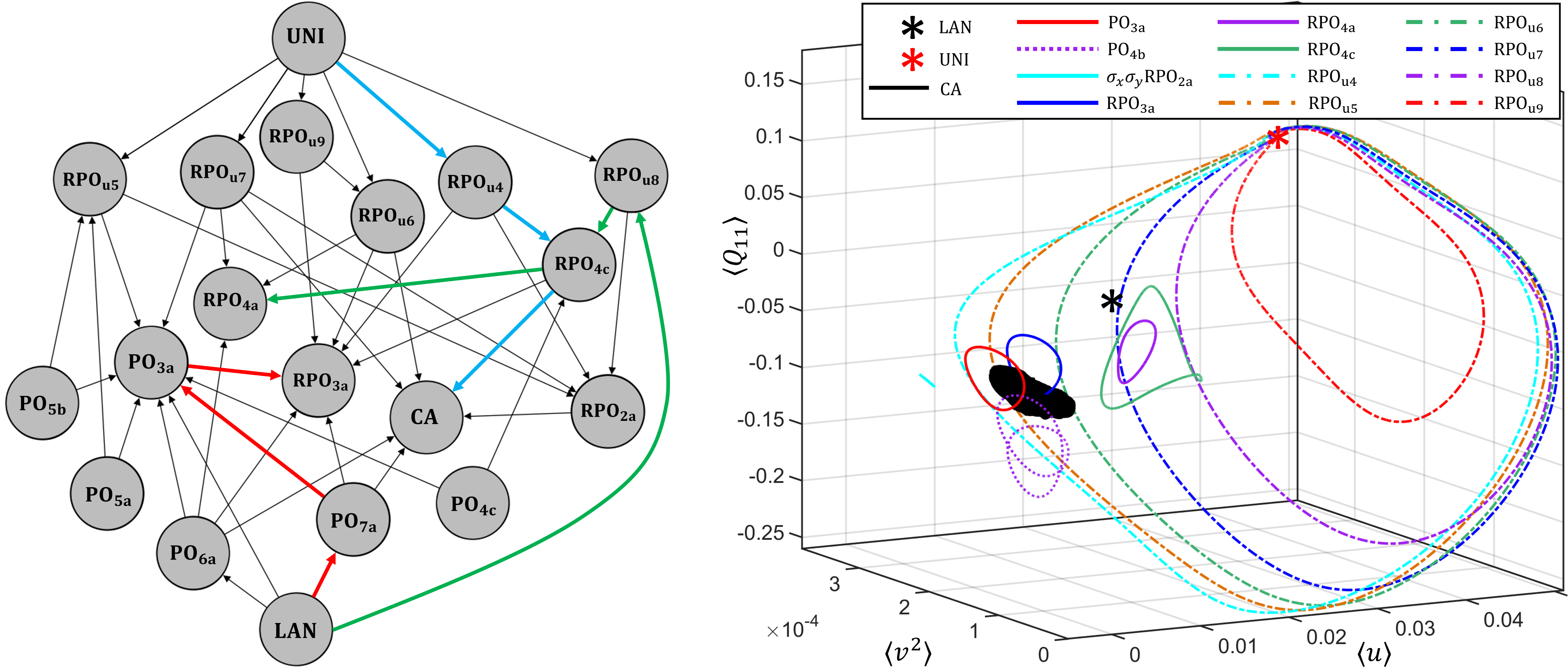}
\caption{\footnotesize{\textbf{Left:} Directed graph representation of the phase space, with ECS as nodes and heteroclinic connections as edges. \textbf{Right:} ECS in a reduced 3D phase space $(\langle u\rangle,\langle v^2\rangle,\langle Q_{11}\rangle)$, where $\langle . \rangle$ denotes the instantaneous channel average. The channel average removes the continuous translation symmetry \cite{cvitanovic2005chaos}, such that RPOs are closed orbits rather than tori in this representation. The laning equilibrium (LAN) and POs have no net streamwise flow and therefore lie on the $\langle u \rangle =0 $ plane. The right drifting ($\langle u \rangle >0$) unidirectional equilibrium, RPOs, and chaotic attractor have left drifting counterparts ($\langle u \rangle <0$) (not shown).}}\label{fig:layer}
\end{figure*}

\textbf{Nematohydrodynamic Model.}---We model the AN in terms of the velocity $\mathbf{u}(\mathbf{r},t)$ and nematic alignment tensor $\mathbf{Q}(\mathbf{r},t)$. The latter is symmetric and traceless and can be parameterized as $\mathbf{Q} = q \left(\hat{\mathbf{n}}\otimes \hat{\mathbf{n}} - \mathbf{I}/2 \right)$, where the scalar $q$ and unit vector $\hat{\mathbf{n}}$ describe the degree and direction of nematic ordering, respectively. The domain is a periodic 2D channel, parameterized as $(x,y) \in \left[-L/2, L/2 \right] \times \left[0, h \right]$, with $x$ the periodic coordinate. The channel walls impose a no-slip boundary condition on $\mathbf{u}$ and strong perpendicular anchoring on $\mathbf{Q}$.
Following earlier work, we describe the dynamics using the hydrodynamic equations
\begin{align}\begin{split}
&\rho \left(\partial_t + \mathbf{u} \cdot \boldsymbol{\nabla}\right)\mathbf{u} = -\boldsymbol{\nabla} p + \boldsymbol{\nabla} \cdot \left( 2 \eta \mathbf{E} - \alpha \mathbf{Q} \right), \\
&\left(\partial_t + \mathbf{u} \cdot \boldsymbol{\nabla}\right)\mathbf{Q} + \mathbf{Q} \cdot \boldsymbol{\Omega} - \boldsymbol{\Omega} \cdot \mathbf{Q} =  \Gamma \, \mathbf{H},\\
 &\qquad \boldsymbol{\nabla} \cdot \mathbf{u} = 0.\end{split}\label{eq:main}
\end{align}
The first and last lines are the incompressible Navier--Stokes equations, with $p$ the pressure, $\mathbf{E}$ and $\boldsymbol{\Omega}$ the strain rate and vorticity tensors, and $\eta$ the viscosity.The term $\boldsymbol{\nabla} \! \cdot \! \left(\alpha \mathbf{Q} \right)$ is the active dipolar density that drives the system. Recent work has shown that the resulting energy fluxes are dominated by viscous dissipation and inertial energy transfer \cite{koch2021role}; hence, we omit terms associated with passive elastic stresses.
The dynamics of $\mathbf{Q}$ consists of: (1) advective and rotational coupling to the velocity and the vorticity, and (2) relaxation via the molecular field $\mathbf{H} \mathop{=} \mathscr{A} \mathbf{Q} \mathbin{-} \mathscr{B} \mathbf{Q} \mathrm{Tr}\left(\mathbf{Q}^2\right) \mathbin{+} K \nabla^2 \mathbf{Q}$ toward configurations that minimize an effective free energy functional. Here $\mathscr{A}$, $\mathscr{B}$, and $\Gamma$ are material constants describing bulk properties of the nematic, and $K$ is an elastic constant characterizing the energy cost of spatial variations in $\mathbf{Q}$. We focus on a single parameter set, working in units such that $\rho \mathop{=} \eta \mathop{=} 1$, $\mathscr{A} \mathop{=} 0.1$, $\mathscr{B} \mathop{=} 0.5$, $\Gamma \mathop{=} 0.34$, $K \mathop{=} 0.04$, and $\alpha \mathop{=} K(31h/2)^2$, and choose channel dimensions $L \mathop{=} 50$ and $h \mathop{=} 11$ in these units. For comparison, the nematic has an intrinsic length $L_n \mathop{=} \sqrt{K / \mathscr{A}} \simeq 0.63$, which is roughly the radius of a defect core, and activity induces the length scale $L_a \mathop{=} \sqrt{K/\alpha}\simeq 0.71$, which measures the balance between active and elastic stresses. We also observe that the velocity magnitude is roughly $0.01$--$0.1$, which corresponds to Reynolds number $\text{Re} \sim 0.1$--$1$. Finally, we note that Ref.~\cite{shendruk2017dancing} and others incorporate additional terms in Eqs.~\ref{eq:main} that account for \textit{flow alignment}, which is the coupling between $\mathbf{Q}$ and the symmetric part of the flow gradients. Here, we neglect these terms to focus on the essential aspects of the problem \cite{shankar2018defect,blow2017motility}.

To emphasize the phase space approach, we rewrite Eqs.~\ref{eq:main} as $\dot{X}\mathop{=}F(X)$, where $X\mathop{=}[\mathbf{u},\mathbf{Q}]$ denotes the state of the system. The associated flow map is $f^t(X_0)\mathop{=}X_0+\int_{0}^tF(X(\tau))d\tau$, where $X_0$ is the initial condition. Since ECS are generically unstable, they cannot be computed from direct time-dependent simulations; rather, one searches for solutions to certain fixed point equations (FPEs). The FPE for an equilibrium solution $X_{\text{eq}}$ is just $F(X_{\text{eq}})\mathop{=}0$, while any point $X_{\text{P}}$ on a periodic orbit (PO) satisfies $f^{T}(X_{\text{P}})\mathop{=}X_{\text{P}}$, where $T$ is the time period. Similarly, a point $X_{\text{RP}}$ on a relative periodic orbit (RPO) satisfies $f^{T}(X_{\text{RP}})\mathop{=}\tau_{x}(\ell) X_{\text{RP}}$, where $\tau_x(\ell)$ is a streamwise translation by $\ell$. Hence, an RPO is a field profile that recurs at a streamwise-shifted location after time $T$. In phase space, an RPO densely covers the surface of a two-torus. We also compute heteroclinic connections between pairs of ECS \cite{suri2019heteroclinic}, which are trajectories that depart the `source' ECS along its unstable manifold and converge to the `destination' ECS along its stable manifold.

\textbf{Symmetries.}---Eqs. \eqref{eq:main} are equivariant under the one-parameter group of $x$ translations, $\tau_x(\ell)$, as well as the following $x$ and $y$ reflections: 
\begin{align} \sigma_x[u,v,Q_{11},Q_{12}](x,y)=[u,-v,Q_{11},-Q_{12}](x,h-y),\nonumber \\
\sigma_y[u,v,Q_{11},Q_{12}](x,y)=[-u,v,Q_{11},-Q_{12}](L-x,y).\nonumber
\end{align}
If an initial condition is invariant under the action of a subgroup of the group generated by $(\sigma_x,\sigma_y,\tau_x(\ell))$, then its future iterates will also respect the subgroup symmetries. Some of the ECS and heteroclinic connections fall into such invariant subspaces, while others possess no symmetries (see Fig.~\ref{fig:ECS_snapshots}). As our results below illustrate, such symmetries are powerful tools for analyzing the phase space geometry.

\textbf{Methods.}---Our computations use the open-source pseudospectral code Dedalus \citep{burns2020dedalus}. 
For channel geometries, Dedalus implements a Fourier basis for the periodic directions and a Chebyshev polynomial basis for the wall-normal direction. All ECS and connections reported here were computed using $256$ Fourier modes and $64$ Chebyshev modes, corresponding to a phase space dimension of $4 \mathbin{\times} 256 \mathbin{\times} 64 \mathop{=} 65536$.
To solve the FPEs, we use modified Newton-Raphson algorithms \citep{viswanath2007recurrent}. Two key ingredients are adaptive `hookstep' step-size selection to improve global convergence \citep{Dennis1996}, and a matrix-free GMRES \citep{saad1986gmres,chandler2013invariant} algorithm for solving the linear BVP at each iteration. The matrix-free methods are essential because they scale efficiently to the large problem dimensions encountered in hydrodynamic simulations.
Finally, finding a new ECS requires a good initial guess for the FPE solver. Here, we devise initial guesses using a combination of (1) the global search method of \cite{chandler2013invariant} that samples arbitrary time-dependent trajectories for approximate solutions to the FPEs, (2) symmetry reduction \cite{willis2013revealing}, and (3) branch continuation in channel width; see supplemental material for details.

\begin{figure*}[!t]
\includegraphics[width=6.9in,height=4in]{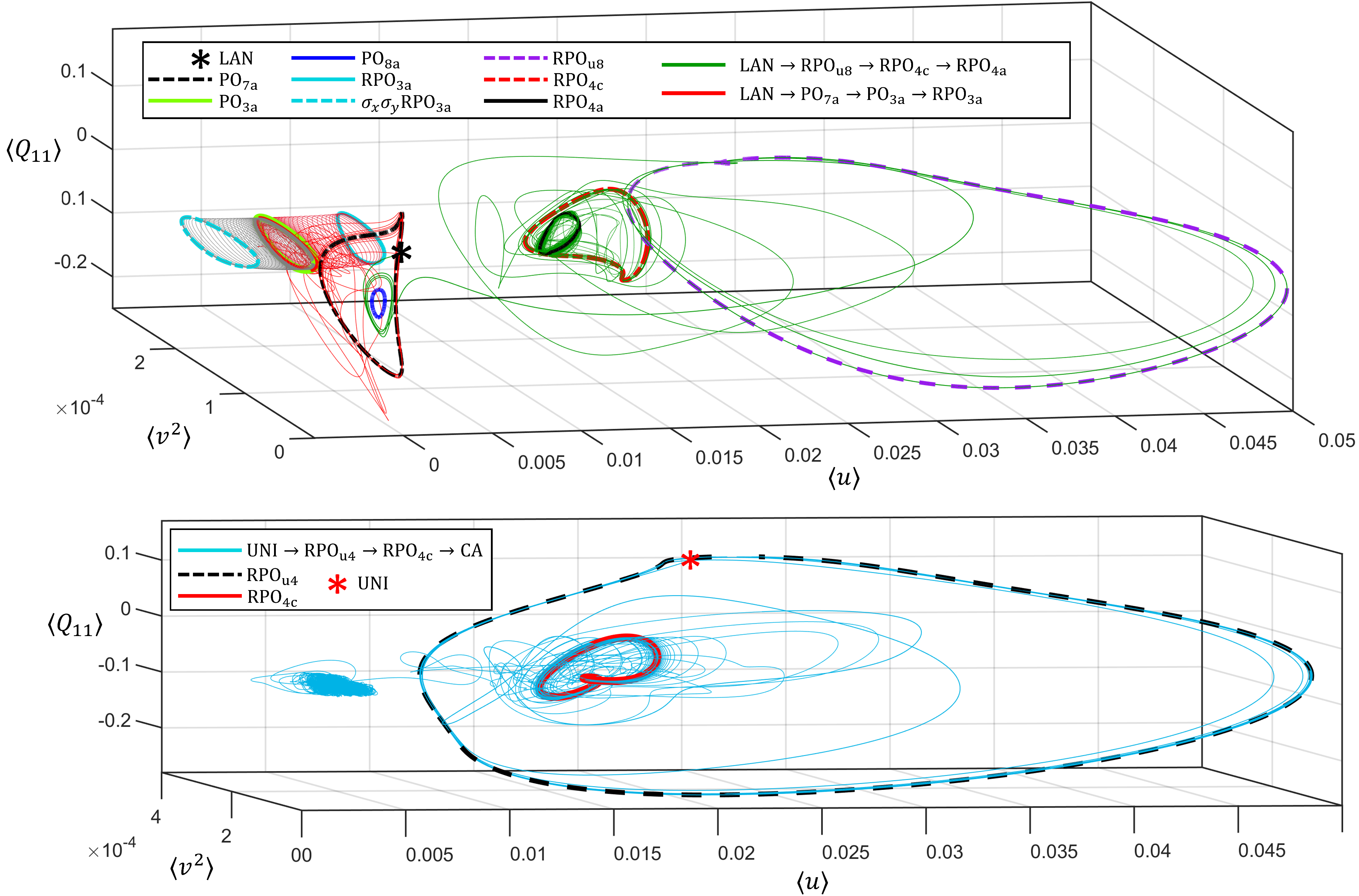}
\caption{\footnotesize{Four connecting orbits in the reduced phase space obtained from the graph representation in Fig.~\ref{fig:layer}. Such orbits can be constructed for a pair of ECS if a directed path exists between them in the graph. Each orbit is formed by patching together heteroclinic connections between successive ECS along the path using small perturbations. \textbf{Top:} Green shows the connection $\text{LAN} \rightarrow \text{RPO}_{\text{u8}}\rightarrow \text{RPO}_{\text{4c}} \rightarrow \text{RPO}_{\text{4a}}$. The spiral segment between $\text{LAN}$ and $\text{RPO}_{\text{u8}}$ appears to be a near pass to $\text{PO}_{\text{8a}}$. Red shows two connections: both starting with $\text{LAN} \rightarrow \text{PO}_{\text{7a}} \rightarrow \text{PO}_{\text{3a}}$ and then perturbed either toward $\text{RPO}_{\text{3a}}$ (red) or $\sigma_x \sigma_y \text{RPO}_{\text{3a}}$ (grey). These last segments highlight the two-dimensional unstable manifold of $\text{PO}_{\text{3a}}$. \textbf{Bottom:} Blue shows the connection $\text{UNI} \rightarrow \text{RPO}_{\text{u4}}\rightarrow \text{RPO}_{\text{4c}} \rightarrow \text{CA (chaotic attractor)}$. See \cite{Note1} for movies of each connection.}}\label{fig:transfer}
\end{figure*}
 
\textbf{Results.}---In time-dependent simulations, the dominant attracting state roughly passes through the following sequence of transitions as activity is increased: (1) zero-flow state; (2) defect-less, unidirectional flow; (3) vortex lattice with motile defects (`dancing disclinations'); (4) spatiotemporal chaos (turbulence). Our results generally agree with \cite{shendruk2017dancing}, which considers a similar AN model in channel confinement. The main difference is that the stable vortex lattice is an RPO in our case and a PO in \cite{shendruk2017dancing}. This difference appears to arise from the effects of flow alignment, as we recover the results of \cite{shendruk2017dancing} at sufficiently large values of flow alignment.

In this article, we discuss the phase space structure at an intermediate non-dimensional activity, $A \mathop{\equiv} \sqrt{\frac{\alpha h^2}{K}}\mathop{=}15.5$, where the system has several co-existing attractors and saddle-type ECS. While this system is pre-turbulent, the phase space is quite rich, and there are numerous heteroclinic connections between ECS. Fig.~\ref{fig:layer} shows the dominant ECS in a reduced 3D phase space, alongside several connections as a directed graph. In addition to the unidirectional equilibria (UNI), we also found a pair of laning equilibria (LAN), in which the upper and lower halves of the channel flow in opposite directions, $\pm u(y) = \mp u(h-y)$ and $v \mathop{=} 0$. Both UNI and LAN are independent of $x$.

\textit{Periodic Orbits.}---We found 11 unstable POs with $k$-vortex lattice structure for $3 \mathop{\leq} k \mathop{\leq} 8$, which we label $\text{PO}_{k\zeta}$ for $\zeta \mathbin{=} a,b, \ldots$. Each $\text{PO}_{k\zeta}$ has $2k$ defect pairs and is invariant under the action of $\mathbb{T}_k \equiv \tau_x(L/k)$ and $\sigma_x \sigma_y$. They are unstable versions of the previously reported stable `dancing disclinations' solutions \cite{shendruk2017dancing,tan2019topological}.

\textit{Relative Periodic Orbits.}---We found over 30 RPOs, which we grouped based on their symmetries and relation to each other in phase space. One family, labeled $\text{RPO}_{\text{uk}}$ for $4 \mathop{\leq} k \mathop{\leq} 9$, cycles between a $k$-fold vortex-like structure and a defect-free, nearly unidirectional flow. The time period of these RPOs diverges as activity is decreased from $A \mathop{=} 15.5$, which leads us to conjecture that they are born as homoclinic orbits to the unidirectional equilibrium (UNI). The remaining RPOs are grouped based on their (exact or approximate) discrete translational symmetry $\mathbb{T}_k$, and labeled as $\text{RPO}_{k\zeta}$ for $\zeta \mathbin{=} a,b, \ldots$. Some are left and right drifting versions of the $\text{PO}_{k\zeta}$ family; others appear more closely related to the $\text{RPO}_{\text{uk}}$ family or lack distinct structure altogether. Under the action of $\sigma_x \sigma_y$, an RPO is transformed into its `opposite drifting' counterpart, changing the sign on the shift $\ell$.

\textit{Attractors.}---We find three attractors, not counting copies related by symmetry transformations. Two are RPOs labeled $\text{RPO}_{\text{3a}}$ and $\text{RPO}_{\text{4a}}$, and one is a chaotic attractor labeled `CA'. $\text{RPO}_{\text{3a}}$ and $\text{RPO}_{\text{4a}}$ consist of 3 and 4-fold `rolling vortices', each with left and right flowing versions related by the transformation $\sigma_x \sigma_y$. CA is a higher-dimensional set localized to a cigar-shaped region of the 3D phase space projection. Trajectories within CA appear to be broken or frustrated versions of a $3$-fold `dancing disclinations' configuration, and the power spectrum contains broadband components, which suggests CA is chaotic. We have confirmed this using the 0-1 test, which takes a time series as input and outputs a binary indicator for the presence of chaos; see the supplemental material for details.

\textit{Heteroclinic connections.}---Individual ECS lend structure to localized regions of phase space. To understand the \textit{global} structure, we compute heteroclinic connections, which are special dynamical pathways connecting ECS. These reveal, for instance, the relationship between the $\text{RPO}_{\text{uk}}$ family and the UNI equilibrium: by choosing a perturbation with $k$-fold translational symmetry, a trajectory starting on UNI passes directly onto $\text{RPO}_{\text{uk}}$. There are myriad other connections both inside and outside the ECS families. Some involve relatively little change in structure; for instance, $\text{PO}_{\text{3a}}$ and $\text{PO}_{\text{4b}}$ connect to their left and right flowing RPO counterparts. Others display striking changes in structure along nontrivial paths in phase space that, at first glance, seem unlikely to be found by our search strategy. For example, some trajectories starting from a $k$-fold PO have their $k$-fold symmetry destroyed before eventually landing on the \textit{unstable} $\text{PO}_\text{3a}$ and acquiring a $3$-fold symmetry. In reality, these and similar connections are not accidental: in most cases they occur because the target ECS is stable in an invariant subspace. $\text{PO}_\text{3a}$, for example, is stable in the $\sigma_x \sigma_y$ subspace. Nonetheless, there may be nontrivial connections that have little to do with invariant subspaces and require more systematic search tools, such as the nonlinear adjoint method \cite{farano2019computing}. See Tables S1-S10 for a list of ECS and heteroclinic connections.

\textbf{Directed graph representation.}---In experiments, one might wish to direct the system toward a specific attractor. In fact, our framework allows for more complex control objectives involving unstable ECS, which is a necessary prelude to engineering turbulent AN flows, where \textit{all} ECS are unstable. The centerpiece of this control capability is a reduced-order representation of the phase space in terms of a \textit{directed graph}, in which ECS are nodes and heteroclinic connections are edges (Fig.~\ref{fig:layer}). This representation uncovers nontrivial relationships in phase space, which can be exploited to induce desired transitions using minimal external control input. For instance, such methods have previously been used to discover low-energy dynamical channels forming an `interplanetary superhighway' in the solar system \cite{koon2000heteroclinic}. Fig.~\ref{fig:transfer} provides four examples of connecting orbits between distant ECS that were found by patching together heteroclinic connections using small perturbations.

\textbf{Conclusion.}---Successful characterization and control of far-from-equilibrium dynamics is a key step in realizing the promise of active matter. We have employed the Exact Coherent Structure approach to obtain a tractable, reduced-order representation of a model AN system. At higher activities, this approach can lead to a better understanding of transitional turbulence in active fluids. In experiments, the reduced order representation can be exploited by applying external vorticity \citep{rivas2020driventransition}, light \cite{ross2019controlling, zhang2019structuring}, or pressure \cite{walton2020pressure} modulation to reach and maintain otherwise inaccessible spatiotemporal states. 

\begin{acknowledgments}
\textbf{Acknowledgements.}---We thank Aparna Baskaran, Predrag Cvitanovi\'{c}, Seth Fraden, Mike Hagan and Ashley Willis for helpful discussions.
\end{acknowledgments}

\nocite{apsrev41Control}
\bibliographystyle{apsrev4-1}

%


\clearpage
\newpage

\onecolumngrid

\newcommand{\beginsupplement}{%
        \setcounter{table}{0}
        \renewcommand{\thetable}{S\arabic{table}}%
        \setcounter{figure}{0}
        \renewcommand{\thefigure}{S\arabic{figure}}%
     }
\def\theequation{S\arabic{equation}}

\makeatletter
\renewcommand\@bibitem[1]{\item\if@filesw \immediate\write\@auxout
    {\string\bibcite{#1}{A\the\value{\@listctr}}}\fi\ignorespaces}
\def\@biblabel#1{[A#1]}
\makeatother

\section{Supplementary Material}
\beginsupplement

\subsection{Dedalus implementation}
Dedalus implements a Fourier basis in the periodic direction ($x$ coordinate) and a Chebyshev basis in the wall normal direction ($y$ coordinate). For computing ECS, we use $N_x = 256$ Fourier and $N_y = 64$ Chebyshev modes, together with a dealiasing factor of $3/2$. $N_x$ was chosen to accommodate the characteristic length scale set by the observed ECS -- an ECS with 9-fold translation symmetry, such as $\text{RPO}_\text{u9}$, would be allotted about $N_x/9$ Fourier modes per each of the 9 unit cells. Assuming $32$ Fourier modes are required to resolve the characteristic scales within such a unit cell, we would need $288$ Fourier modes. To keep simulations tractable, we use instead $N_x = 256$ and check empirically that using larger $N_x$ does not significantly change our results. Similarly, for $N_y$ we observe that $32$ modes are adequate in many cases, but fail to resolve all unstable directions in the laning equilibrium (labeled `LAN'; see table \ref{table:ECS_data_EQ}). Therefore, we use $N_y = 64$ for computing ECS.

For the initial sweeps of the Newton-Rhapson searches and for other exploratory runs, we use $N_x = 128$ and $N_y = 32$. We assume this lower resolution is sufficient to uncover the majority of relevant structures, which can then be verified in higher resolution simulations.

For timestepping, we used the `RK222' timestepper provided by Dedalus, which is a second-order, two-step, implicit/explicit Runga-Kutta method. The nominal timestep was $dt = 0.05$, adjusted slightly for each ECS to allow for an integer number of timesteps. This value of $dt$ is significantly smaller than the limit imposed by the Courant–Friedrichs–Lewy (CFL) condition for stability, which typically varies approximately between $1$ and $10$, depending on the ECS. 

\subsection{Computation of ECS}
\subsubsection{Modified Newton-Raphson algorithm}
Computation of an ECS amounts to solving the $N$-dimensional, nonlinear, algebraic system defined by the corresponding fixed point equation. Here $N$ is the number of degrees of freedom in the spectral representation, and is usually large -- in our case $N = 65536$, and in 3d turbulent flows $N$ may exceed $10^6$. 

At the time of writing, Dedalus can natively solve only 1d nonlinear BVPs, corresponding to the fixed point equation for 1d equilibria. To compute POs and RPOs as well as any 2d ECS, we implement a modified Newton-Raphson iteration. This method has been successful in ECS calculations of inertial fluids [S5]. 
The core element of the standard Newton-Raphson method is the successive solution of linear approximations of the nonlinear manifold defined by the fixed point equation, until convergence is achieved. In the best-case scenario, convergence sufficiently near the fixed point is quadratic; however, this feature may hold only very close to the fixed point, and in general little can be said about global convergence. Moreover, direct solution of the linear approximating equation quickly becomes impractical for high-dimensional systems, as doing so requires linear algebraic operations on $N \times N$ matrices. 

Following earlier work on inertial fluids [S5], we address these limitations with two modifications to the standard Newton-Raphson iteration. First, we approximate the solution of the linear system using GMRES, which solves the system in a low-dimensional Krylov subspace. The minimum subspace dimension required to obtain satisfactory results is independent of $N$, and usually quite small. In our case, we find satisfactory results using a subspace dimension as small as $20$. Second, we constrain the step size of each Newton iteration by imposing a fixed norm $\rho$ on the solution of the linear approximating equation. This improves global convergence by avoiding the situation where the step would otherwise extend past the region in which the linearization is valid. The linear system is then replaced by a least-squares problem, which can also be solved in a Krylov subspace. Note that this is distinct from first solving the linear system and then fixing the norm after the fact by scalar multiplication. The latter is sometimes called a damped Newton method, and performs worse than the constrained optimization because the \emph{direction} of the step is chosen based on the unconstrained system.

We use an adaptive approach to step size selection, testing various step sizes until the error on the \emph{nonlinear} problem decreases from the previous iterate. Using a $l^2$ norm on the spectral coefficients, this amounts to a step size typically $\mathcal{O}(10^{-2}\text{--}10^{-1})$, but occasionally as small as $\mathcal{O}(10^{-12})$.

With the same $l^2$ norm, we carry out the Newton iterations until the error of the fixed point equation is $\leq 10^{-12}$. The exception is $\text{PO}_{\text{8a}}$, for which the residual could not be decreased past $10^{-7}$. Possible explanations are that $\text{PO}_{\text{8a}}$ is simply too unstable for a single-point shooting method, or that it is not strictly a PO---for example, it could be a very thin invariant torus. Finally, an ECS is considered to be invariant under a group operation $g$ if the $l^2$ norm of the difference between the ECS and its transformation under the group operation is less than $10^{-10}$. 

It is worth noting that we do not observe quadratic convergence even very close to a fixed point---rather, the error after each iteration decreases by a constant
factor, typically around $10$. We conjecture that this happens due to the continuous translational symmetry in $x$, which renders the fixed point equation degenerate.

\subsubsection{Finding initial guesses} 

While the adaptive step size of the Newton-Raphson iteration improves global convergence, good initial guesses are still required to catalogue all the dynamically relevant ECS. Besides being close to an ECS, initial guesses must not exclude any dynamically important regions of the phase space, including attractors as well as repelling regions in which a trajectory may linger long enough to be physically relevant; examples of the latter are provided below. Here we employ a multifaceted search strategy to establish with reasonable confidence that all dynamically relevant ECS have been catalogued.

Most ECS were discovered using initial guesses from time-dependent trajectories, which naturally favor the most dynamically relevant ECS. The initial guesses were determined by 1) choosing a random state $X(t_0)$ from an initial set of trajectories, and 2) scanning for period $T$ (up to a given threshold) and shift $\ell$ that minimize the residual $||\tau_{\ell} X(t_0 + T) - X(t_0)||^2$. The resulting $(X(t_0)$, $T$, $\ell)$ ideally constitute a `near pass' to an ECS.

The initial set of trajectories themselves were generated using several methods:
\begin{itemize} 
\item `Quenching' from states that are physically relevant at other parameter values--for example, stable equilibria, POs, and RPOs; or snapshots from long-time turbulent trajectories. As an example, initial conditions derived from turbulent snapshots at large activity manage to uncover all three attractors at the value $\alpha = 15.5$ considered here.
\item Trajectories falling within dynamical equivariance classes. Some ECS are unstable in the full space, but stable or nearly stable in a symmetry subspace, e.g., $\mathbb{T}_3$, the subspace of solutions possessing three-fold translational symmetry. Taking initial guesses from within this subspace will increase the probability of finding such ECS. Because the equivariance classes are intrinsic to the dynamics (and hence, the subspaces are invariant, as explained in the main text), ECS with the corresponding symmetries are important for mapping out the global phase space geometry--many heteroclinic connections, for instance, fall entirely within a symmetry subspace.
\item Trajectories following unstable manifolds of known ECS. Since heteroclinic orbits are contained in the intersections of unstable and stable manifolds of different ECS, initializing trajectories on the unstable manifold can uncover globally distant, but dynamically connected regions of phase space.
\end{itemize}

Hundreds of instances of the Newton solver were run along such trajectories. In some cases a significant percentage failed to converge within a time limit of about 3-5 days. However, usually at least 10\% converged, sometimes much more. We continued searching in this way until discovering a new ECS became infrequent ($\lesssim 1$ new ECS per 100 solver instances).

A smaller number of ECS were computed by parameter continuation, i.e., initializing the solver using a known ECS from a nearby set of parameters. We found it particularly effective to use the domain width $L$ as the continuation parameter. Given a $k$-fold symmetric ECS, this continuation method can be used to look for $(k\pm1)$-fold ECS with similar structure. The idea is the following: because the dynamics is equivariant under translations in $x$, a $k$-fold ECS can be decomposed into $k$ unit cells with identical time evolution, such that the dynamics of a single unit cell on a domain of width $w/k$ completely characterizes the full, $k$-fold ECS. Then, if one can use parameter continuation to shrink or grow the domain and obtain corresponding ECS with width $w/(k+1)$ or $w/(k-1)$, these can be stitched back together to obtain $(k\pm1)$-fold ECS in the full space. In this way, for example, we were able to discover $\text{PO}_{\text{5b}}$ starting from $\text{PO}_{\text{4b}}$.

\subsubsection{Linear stability}

Linear stability is defined in terms of the linear operator that governs the dynamics of a small perturbation $\delta X$ about an ECS. For equilibria, the linear operator is the Jacobian matrix, and its eigenvalues $\lambda_n$ describe the evolution of small perturbations $\delta Y_n$ along the corresponding eigenvector $Y_n$ via $\delta Y_n(t) \sim e^{\mu_n t}\delta Y_n(0)$. An unstable equilibrium therefore has at least one positive eigenvalue. For a PO, the linear operator is the monodromy matrix [S4] $\mathbf{M}$ that, for a reference point $X_0$ on the ECS, maps a small perturbation $\delta X$ onto the corresponding perturbation $\delta X'$ after evolving for a single period $T$. For an RPO, there is the additional step of shifting back to the frame of the initial state, i.e., translating the final state by $-\ell$. In terms of the flow map $\phi(X,T)$, we have
\begin{align}
\delta X' &= \tau_{-\ell} \phi(X + \delta X, T) - X \\
&\equiv \mathbf{M} \delta X
\label{eq:linear-operator-definition}
\end{align}
It can be shown that the eigenvalues of $\mathbf{M}$, referred to as Floquet multipliers, do not depend on the reference point. In our system, POs and RPOs have at least two multipliers equal to $1$, one for a shift along the orbit and one for a pure translation in $x$. The remaining multipliers determine the asymptotic behavior of a given perturbation, at least in the region where the linearization is valid. If all the multipliers are $\leq 1$, we say the ECS is asymptotically stable. If at least one multiplier is strictly greater than $1$, we say the ECS is unstable because \emph{generic} perturbations (as might arise due to experimental noise or finite-precision arithmetic) grow exponentially. 

Computationally, we obtain the stability of equilibria using the eigenvalue solver provided by Dedalus, which can handle 1d equilibria. For everything else, we again use Krylov subspace methods--this time calculating the leading eigenvalues of $\mathbf{M}$ within a $60$-dimensional Krylov subspace. The action of $\mathbf{M}$ on a vector $Y$ is calculated by fixing the $l^2$ norm of $Y$ to $10^{-7}$ and applying Eq. \ref{eq:linear-operator-definition}. We use the value $10^{-7}$ because it is sufficiently small to ensure the action of $\phi$ on $\delta X$ is linear, but large enough to avoid round-off error.

 Tables \ref{table:ECS_data_POs}--\ref{table:ECS_data_RPOs} list the real and imaginary parts $\text{Re}(\lambda)$ and $\text{Im}(\lambda)$ of the largest multiplier for each POs and RPOs. One can equivalently describe the behavior of perturbations by $\delta X_n(t) \sim e^{\mu_n t}X_n(0)$, where $\mu_n$ is called the Floquet exponent. The largest exponent is also listed in same tables. Table \ref{table:ECS_data_EQ} lists the real and imaginary parts of the largest eigenvalue $\mu$ of the Jacobian evaluated at the various equilibria.

\subsection{Computation of heteroclinic connections}

Finding heteroclinic connections proceeds naturally from the global search strategy mentioned above. In the process of following time-dependent trajectories---especially those along the unstable manifold of an ECS---candidate connections can be identified as those passing nearby an ECS in the 3D phase space projection.

To quantitatively check convergence towards the candidate `target ECS', we scan for the closest approach of the connecting orbit to any one point on the target ECS. Because of the continuous translational symmetry, we must recognize a situation where the connection approaches an $x$-translated version of the original ECS. To do so, we imbue the phase space with a metric that `reduces' the symmetry by replacing the distance between two states $X_1$ and $X_2$ with the smallest distance between $X_1$ and any of the continuously translated copies of $X_2$:
\begin{equation}
\text{Distance between } X_1 \text{ and } X_2 = \text{min}_s ||\tau_s X_2 - X_1|| < \epsilon
\label{eq:connection-verification-criterion}
\end{equation}
This number is given in the rightmost column of Table \ref{tab:SI-connections} for each connection. We have used $\epsilon = 10^{-3}$ as the threshold for convergence in \eqref{eq:connection-verification-criterion}. (See Ref. [S3] for application of a similar convergence criterion for computing heteroclinic connections.) 

 This search strategy allows for easy discovery and verification of connections that end on an attractor. However, some candidate connections appear in time-dependent simulations only as inconclusive near-passes to an unstable ECS, e.g., local minima in the distance metric \eqref{eq:connection-verification-criterion} that do not quite fall within the threshold $\epsilon$. We have found that such a near pass often occurs because the trajectory approximately evolves in an invariant subspace (corresponding to a discrete symmetry), and the candidate target ECS is stable in that subspace. Then, the lack of conclusive numerical convergence occurs because the subspace is unstable under symmetry-breaking perturbations, which could arise in numerical simulations due to round-off error. To circumvent this problem, we configure the time-dependent solver to remain within the symmetry subspace for all time, by projecting out any symmetry-breaking components that may accumulate from round-off error. To enforce discrete translational symmetries $\mathbb{T}_k$, we shrink the domain size by a factor of $k$. The convergence criterion \eqref{eq:connection-verification-criterion} can then be verified within the target precision $\epsilon = 10^{-3}$. In this way, we have discovered several heteroclinic connections terminating in ECS that are unstable in the full space, but stable in an invariant subspace.
 
 Connections to the chaotic attractor (CA) are difficult to verify using a direct distance metric, as the manifold defining CA is not known in its entirety. One possibility is to compute the smallest distance between a candidate heteroclinic orbit and a reference trajectory within CA. However, we have not had success with this approach, possibly because the reference trajectory would have to be extremely long to adequately sample the set. For example, the closest approach we could find for the connection $\text{RPO}_{\text{4c}} \rightarrow \text{CA}$ was about $0.015$.
 
 Instead, we verify that the asymptotic behavior of a candidate connection matches a few key properties of a reference trajectory defining the attractor---specifically, that it contains an attracting set, is chaotic, has the same average number of defects, and occupies roughly the same region of phase space (closest approach $< 0.05$); see the following section for details on how these properties are established for a set of reference trajectories.

\begin{figure*}
\includegraphics[width=5.488087631in,height=2.3in]{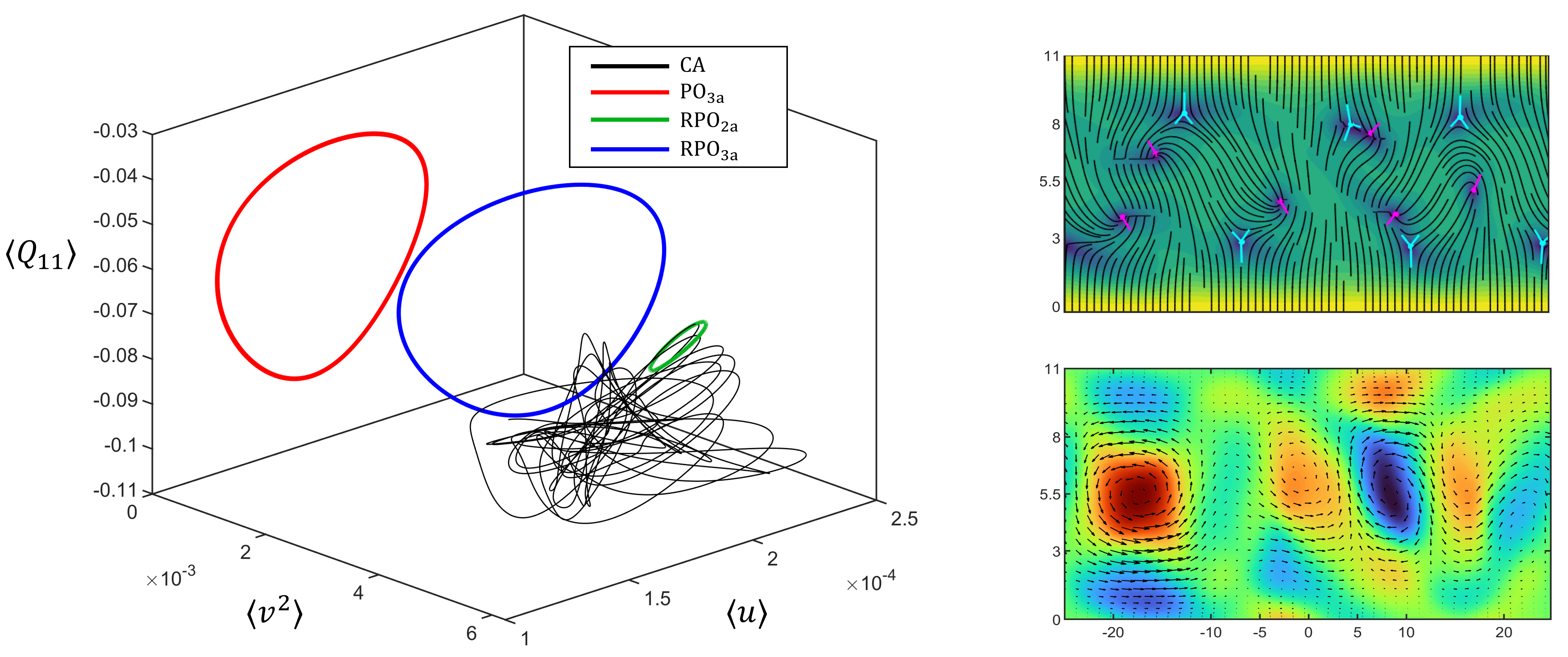}
\caption\footnotesize{Left: A typical trajectory within the chaotic attractor (CA). Right: An illustrative snapshot along the trajectory. The top right shows the nematic director field (black lines) overlayed on the nematic order parameter field (color gradient), and the bottom right shows the velocity field (black arrows) overlayed on the vorticity (color gradient).} \label{fig:CA_plots}
\end{figure*}

\subsection{Test for chaos}
One of the three attractors is a chaotic set--labeled `CA' for `chaotic attractor'--that occupies an oblong region of the 3D phase space projection (Fig. \ref{fig:CA_plots}). It is structurally stable in the sense that it is robust against variation in activity, grid resolution, and timestep. Our evidence that the set is an attractor comes from long-time trajectories that remain localized to the same region of phase space and exhibit similar properties, such as average number of defects. Depending on the grid resolution and timestepping, the duration of these reference trajectories range from $10^6$ time units ($N_x = 256$, $N_y = 64$, $dt = 0.05$) to $10^8$ time units ($N_x = 64$, $N_y = 32$, $dt = 1.0$).

To establish that the set is chaotic, we apply the 0-1 test [S1,S2]. The test takes a time series $\phi(n)$ as input and uses it to drive the 2-dimensional system
\begin{align}
    p(n+1) &= p(n) + \phi(n) \cos c n \\
    q(n+1) &= q(n) + \phi(n) \sin c n
\end{align}
where $c \in (0, 2 \phi)$ is fixed. If $\phi(n)$ is sufficiently long to sample the entire attractor, then the behavior of $p(n)$ and $q(n)$ qualitatively distinguish between regular and chaotic dynamics: for regular dynamics, $p(n)$ and $q(n)$ are typically bounded, whereas for chaotic dynamics, they typically behave asymptotically as a 2D Brownian motion. In the latter case, $p(n)$ and $q(n)$ are unbounded, and their mean-squared-displacement (MSD) scales as $n$. Except at certain isolated values of $c$, which correspond to resonances in $\phi(n)$, the two cases can be distinguished by computing the correlation $\kappa_c$ between linear growth and the MSD: regular and chaotic dynamics lead to $\kappa_c = 0$ and $\kappa_c = 1$, respectively. Therefore, one can test for chaos in the original dynamics using a single number $\kappa_c$.

Here we present the results for $N_x = 128$ Fourier modes, $N_y = 32$ Chebyshev modes, and timestep $dt = 0.25$. Similar results were obtained for both higher and lower resolution simulations. We choose $\phi(n)$ to be the channel-averaged x-velocity $\langle U \rangle$, sampled at fixed intervals $\Delta t = 10^3$ from a trajectory of duration $T \approx 7 \times 10^6$. Following Ref. [S2], we use a modified MSD that regularizes the linear scaling with $n$ by subtracting out an oscillatory component. To avoid the bias that would result from inadvertently choosing $c$ near a resonance, we compute the median of $\kappa_c$ for 100 randomly selected values of $c$ in the interval $(\pi/5, 4\pi/5)$. Other details of our implementation follow Ref. [S2]. In the end, we find $\kappa = \text{median} \left( \{\kappa_c \} \right) = 0.9985$, indicating chaotic dynamics. To test that our implementation is correct, we repeated the same procedure for $\phi(n)$ sampled from a quasiperiodic attractor that appears at larger activity. Here we find $\kappa = 0.01723$, which is close to the expected $\kappa \approx 0$.

\begin{figure*}
\begin{subfigure}[b]{0.5\textwidth}
\includegraphics[width=\textwidth]{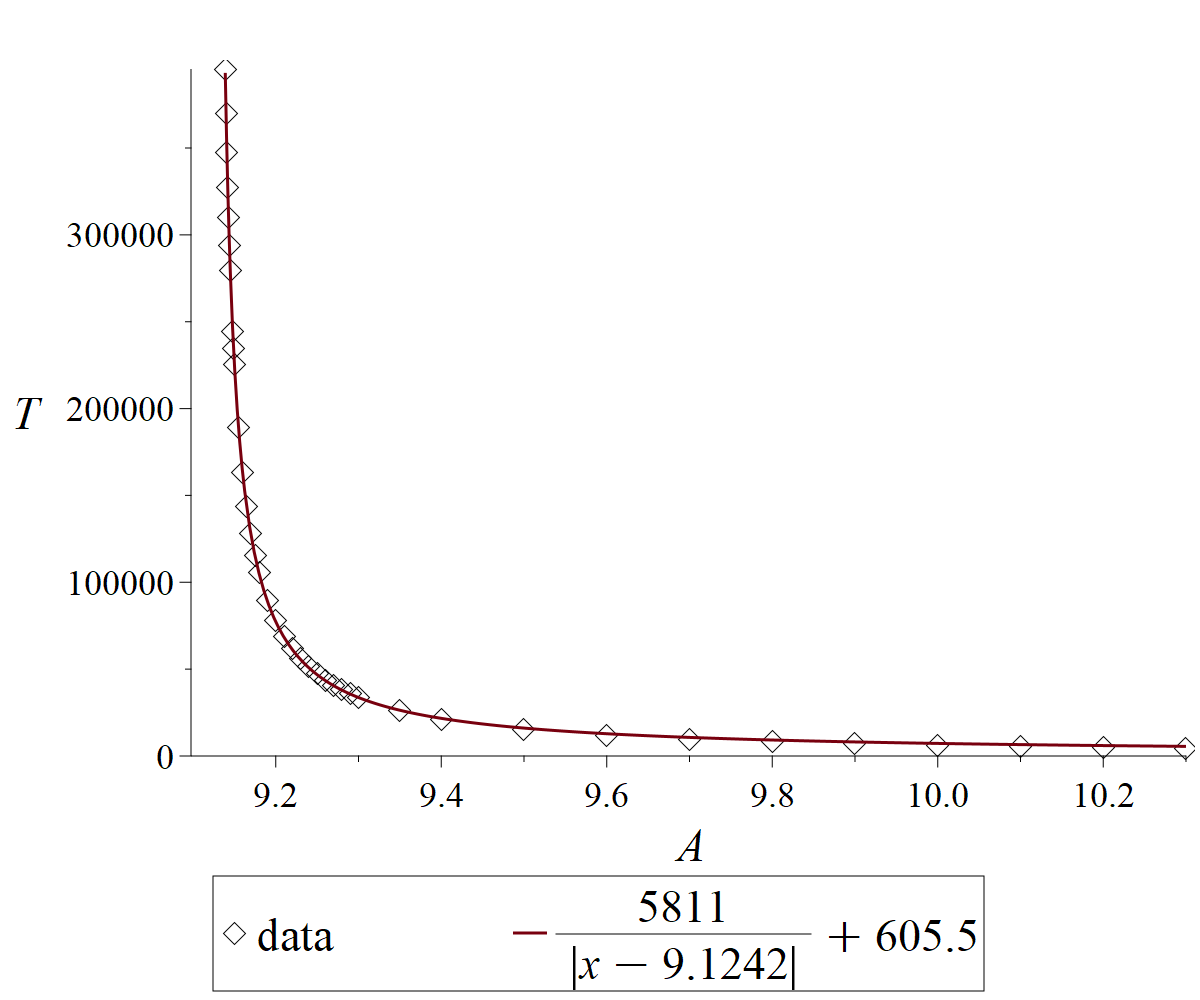}
 \end{subfigure}
     \hfill
\begin{subfigure}[b]{.49\textwidth}
\includegraphics[width=\textwidth]{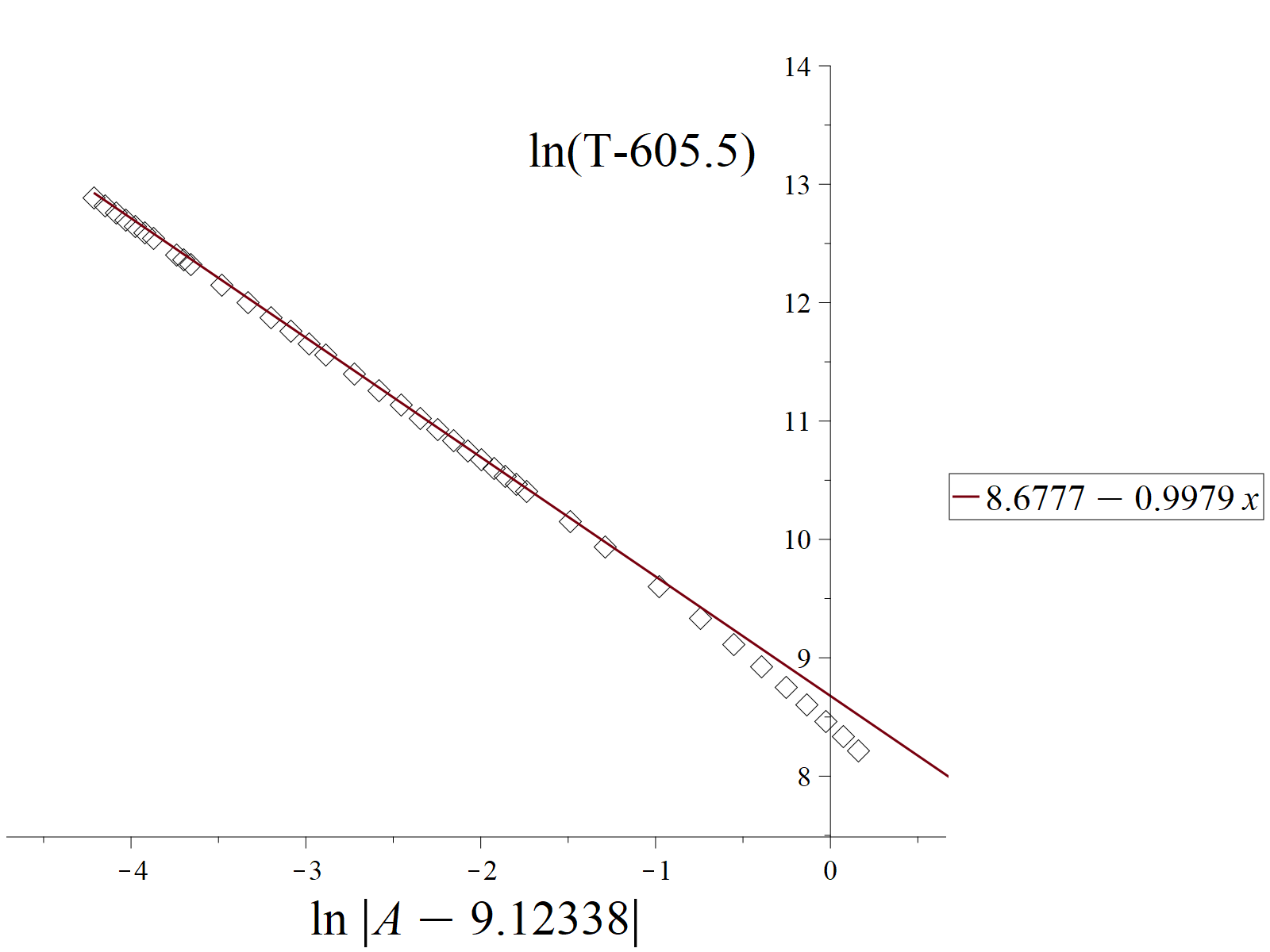}
 \end{subfigure}
\caption{$\text{RPO}_{\text{u4}}$ period scaling with non-dimensional activity parameter $A$}
\label{fig:RPOu4_period_scaling}
\end{figure*}

\subsection{Homoclinic bifurcation}

Branch continuation of $\text{RPO}_\text{u4}$ towards lower activity suggests that the ECS in this family are created by successive infinite-period bifurcations that are homoclinic to the unidirectional equilibrium (labeled `UNI'). Visually, this setup is suggested by the fact that $\text{RPO}_\text{u4}$ passes very close to UNI along its orbit. Near such a bifurcation, the period diverges as $T \sim z^{-1/2}$, where $z$ is the bifurcation parameter. In our case, the activity number $A$ is the square root of the bifurcation parameter $\alpha$, so we expect a scaling $T \sim A^{-1}$. Because $\text{RPO}_\text{u4}$ is stable just above the conjectured bifurcation, we have been able to check this scaling over 2 decades, from $T \sim 10^3$ to $T \sim 10^5$. The evidence is indeed consistent with the $T \sim A^{-1}$ scaling (Fig. \ref{fig:RPOu4_period_scaling}). To complete the picture, one should identify all the solutions involved in the bifurcation -- here, at least one other solution would be required in addition to UNI and $\text{RPO}_\text{u4}$, either an equilibrium or a traveling wave. We leave such an investigation to future work. 
\clearpage
\section{Data tables}

Tables \ref{table:ECS_data_POs}--\ref{tab:ECS_data_T5_subspace} summarize all the ECS and their properties, including their linear stability in the full phase space and relevant symmetry subspaces. The family of RPOs that are nearly homoclinic to the unidirectional flow are notated as $\text{RPO}_\text{uk}$ for each $\mathbb{T}_k$ symmetry represented. All the other RPOs and the POs are named and grouped according to their exact or approximate translational symmetry $\mathbb{T}_k$. This categorization includes certain ECS that break a translational symmetry to a small degree, e.g., by deforming one of the otherwise $k$-fold vortices. The POs and RPOs within a group are separately ordered by lowercase letters a, b, c, ... For example, $\text{RPO}_\text{4a}$ and $\text{RPO}_\text{4b}$ are the first and second RPOs in the $\mathbb{T}_4$ grouping. The grouping $\text{RPO}_\text{1x}$ corresponds to RPOs without any discrete translational symmetry, except for the trivial case where the shift equals the channel width.

Each RPO also has an opposite-shifted counterpart, obtained by the transformation $\sigma_x \sigma_y$ along with flipping the sign of the shift $\ell$. $\text{RPO}_\text{4c}$ also has $x$-reflected counterpart, i.e., $\sigma_x \text{RPO}_\text{4c}$ is also an ECS with the same period and shift.

\begin{table}[h!]
\centering
\begin{tabular}{p{15mm}>{\centering\arraybackslash}p{11mm}>{\centering\arraybackslash}p{30mm}>{\centering\arraybackslash}p{10mm}>{\centering\arraybackslash}p{20mm}>{\centering\arraybackslash}p{20mm}>{\centering\arraybackslash}p{20mm}>{\centering\arraybackslash}p{20mm}}  \hline \hline \vspace{.1mm}
ECS  & \vspace{.1mm} $T$ & \vspace{.1mm} Symmetries & \vspace{.1mm} $N_u$ & \vspace{.1mm} $\mathrm{Re}(\mu) \times 10^2$ & \vspace{.1mm} $\mathrm{Im}(\mu) \times 10^2$ & \vspace{.1mm} $\mathrm{Re}(\lambda)$ & \vspace{.1mm} $\mathrm{Im}(\lambda)$ \\
\hline \addlinespace[1.2mm]
$\text{PO}_{\text{3a}}$  & 343.09  & $\mathbb{T}_3$, $\sigma_x \sigma_y$               & 1  & 0.0824 & 0 & 1.3268    & 0      \\
\addlinespace[3mm]
$\text{PO}_{\text{4a}}$ & 974.68  & $\mathbb{T}_2$, $\mathbb{T}_4$, $\sigma_x \sigma_y$        & 5  & 0.1114 & 0 & 2.9611    & 0      \\
$\text{PO}_{\text{4b}}$ & 422.85  & $\mathbb{T}_2$, $\mathbb{T}_4$, $\sigma_x \sigma_y$        & 8  & 0.5020 & 0 & 8.3544    & 0      \\
$\text{PO}_{\text{4c}}$ & 587.23  & $\mathbb{T}_2$, $\mathbb{T}_4$, $\sigma_x \sigma_y$        & 13 & 0.6703 & 0 & 51.2255   & 0      \\
$\text{PO}_{\text{4d}}$ & 758.53  & $\mathbb{T}_2$, $\sigma_x \mathbb{T}_4$, $\sigma_x \sigma_y$ & 14 & 1.1904 & 0 & 8345.4 & 0 \\
\addlinespace[3mm]
$\text{PO}_{\text{5a}}$ & 1137.55 & $\mathbb{T}_5$, $\sigma_x \sigma_y$               & 14 & 0.5155 & 0 & 352.2035  & 0      \\
$\text{PO}_{\text{5b}}$ & 538.75  & $\mathbb{T}_5$, $\sigma_x \sigma_y$               & 15 & 1.0473 & 0 & 282.1622  & 0      \\
$\text{PO}_{\text{5c}}$ & 621.56  & $\mathbb{T}_5$, $\sigma_x \sigma_y$               & 25 & 1.0437 & 0 & 656.75    & 0 \\
\addlinespace[3mm]
$\text{PO}_{\text{6a}}$  & 1274.17 & $\mathbb{T}_2$, $\mathbb{T}_3$, $\mathbb{T}_6$, $\sigma_x \sigma_y$ & 6  & 0.5170 & 0 & 725.8385  & 0      \\
\addlinespace[3mm]
$\text{PO}_{\text{7a}}$  & 1234.85 & $\mathbb{T}_7$, $\sigma_x \sigma_y$               & 9  & 0.5921 & 0 & 1497.7805 & 0.0005 \\
\addlinespace[3mm]
$\text{PO}_{\text{8a}}$*  & 970.46 & $\mathbb{T}_2$, $\mathbb{T}_4$, $\mathbb{T}_8$, $\sigma_x \sigma_y$ &   &  &  &  & 
\\
\addlinespace[2mm]
\multicolumn{2}{l}{\hspace{2mm}*residual = $3.4 \times 10^{-7}$} &               &  &   &   &   \\
 \addlinespace[1mm]
 \hline \hline
\end{tabular}
\caption{List of Periodic orbits (POs). $N_u$ is the number of unstable directions in the full phase space. $\lambda$ is the largest Floquet multiplier, and $\mu$ is the corresponding Floquet exponent.}
\label{table:ECS_data_POs}
\end{table}

\begin{table}[]
\centering
\begin{tabular}{p{13mm}>{\centering\arraybackslash}p{11mm}>{\centering\arraybackslash}p{15mm}>{\centering\arraybackslash}p{30mm}>{\centering\arraybackslash}p{10mm}>{\centering\arraybackslash}p{17mm}>{\centering\arraybackslash}p{17mm}>{\centering\arraybackslash}p{17mm}>{\centering\arraybackslash}p{17mm}}  \hline \hline \addlinespace[2mm]
ECS  & $T$ & $\ell$ & Symmetries & $N_u$ & $\mathrm{Re}(\mu) \times 10^2$ & $\mathrm{Im}(\mu) \times 10^2$ & $\mathrm{Re}(\lambda)$ & $\mathrm{Im}(\lambda)$ \\
\hline \addlinespace[1.2mm]
$\text{RPO}_{\text{u4}}$ & 738.19  & 12.936  & $\mathbb{T}_2$, $\mathbb{T}_4$, $\sigma_x \mathbb{T}_8$          & 13 & 0.3045  & 0.4088 & -9.3933 & 1.1703 \\
$\text{RPO}_{\text{u5}}$ & 607.38  & 20.321  & $\mathbb{T}_5$, $\sigma_x \mathbb{T}_2$                 & 6  & 0.0579  & 0.1550 & 0.8369  & 1.1487 \\
$\text{RPO}_{\text{u6}}$ & 666.18  & -3.642  & $\mathbb{T}_2$, $\mathbb{T}_3$, $\mathbb{T}_6$, $\sigma_x \mathbb{T}_4$   & 6  & 0.0243  & 0.3703 & -0.9184 & 0.7347 \\
$\text{RPO}_{\text{u7}}$ & 812.02  & 31.589  & $\mathbb{T}_7$, $\sigma_x \mathbb{T}_2$                 & 8  & 0.0816  & 0.3221 & -1.6780 & 0.9746 \\
$\text{RPO}_{\text{u8}}$ & 986.91  & 22.221  & $\mathbb{T}_2$, $\mathbb{T}_4$, $\mathbb{T}_8$                   & 10 & 0.4848  & 0.1976 & -44.329 & 111.10 \\
$\text{RPO}_{\text{u9}}$ & 1137.00 & 27.373  & $\mathbb{T}_3$, $\sigma_x \mathbb{T}_2$, $\sigma_x \mathbb{T}_6$ & 15 & 0.7235  & 0.2313 & -3257.0 & 1831.4 \\
\addlinespace[3mm]
$\text{RPO}_{\text{1a}}$ & 84.93   & -10.329 & -                                     & 6  & 0.1259  & 1.8204 & 0.0276  & 1.1125 \\
$\text{RPO}_{\text{1b}}$ & 399.03  & 6.564   & -                                     & 4  & 0.6265  & 0      & 12.182  & 0      \\
$\text{RPO}_{\text{1c}}$ & 518.89  & 21.262  & -                                     & 7  & 0.3445  & 0.3268 & -0.7449 & 5.9276 \\
$\text{RPO}_{\text{1d}}$ & 455.57  & 37.527  & -                                     & 5  & 0.2468  & 0.4503 & -1.4228 & 2.7292 \\
$\text{RPO}_{\text{1e}}$ & 276.37  & -5.080  & -                                     & 2  & 0.0175  & 0.4852 & 0.2391  & 1.0219 \\
$\text{RPO}_{\text{1f}}$ & 678.75  & 3.558   & -                                     & 3  & 0.0999  & 0      & 1.9707  & 0      \\
$\text{RPO}_{\text{1g}}$ & 383.31  & 9.822   & -                                     & 2  & 0.0696  & 0.0842 & 1.2382  & 0.4142 \\
\addlinespace[3mm]
$\text{RPO}_{\text{2a}}$ & 108.35  & 17.870  & $\sigma_x \mathbb{T}_2$                        & 2  & 0.0281  & 1.6527 & -0.2249 & 1.0061 \\
$\text{RPO}_{\text{2b}}$ & 173.91  & 6.426   & $\mathbb{T}_2$                                 & 4  & 0.1446  & 0.7852 & 0.2621  & 1.2589 \\
$\text{RPO}_{\text{2c}}$ & 178.34  & 12.282  & $\mathbb{T}_2$, $\sigma_x \mathbb{T}_4$                 & 6  & 0.2932  & 1.3412 & -1.2346 & 1.1496 \\
$\text{RPO}_{\text{2d}}$ & 641.01  & 25.114  & $\mathbb{T}_2$                                 & 14 & 0.5296  & 0      & 29.818  & 0      \\
$\text{RPO}_{\text{2e}}$ & 494.88  & -3.449  & $\mathbb{T}_2$                                 & 14 & 0.8775  & 0      & 76.890  & 0      \\
$\text{RPO}_{\text{2f}}$ & 576.71  & 21.566  & $\mathbb{T}_2$                                 & 19 & 0.7026  & 0      & 57.511  & 0      \\
\addlinespace[3mm]
$\text{RPO}_{\text{3a}}$ & 357.45  & 2.296   & $\mathbb{T}_3$                                 & 0  & -0.0546 & 0.1859 & 0.6478  & 0.5074 \\
$\text{RPO}_{\text{3b}}$ & 636.45  & -26.361 & $\mathbb{T}_3$, $\sigma_x \mathbb{T}_2$, $\sigma_x \mathbb{T}_6$ & 12 & 0.6290  & 0.4936 & -54.787 & 0      \\
$\text{RPO}_{\text{3c}}$ & 262.57  & -9.542  & $\mathbb{T}_3$, $\sigma_x \mathbb{T}_2$, $\sigma_x \mathbb{T}_6$ & 8  & 0.2676  & 1.1544 & -2.0065 & 0.2225 \\
$\text{RPO}_{\text{3d}}$ & 403.72  & 43.284  & -                                     & 7  & 0.4339  & 0      & 5.7639  & 0      \\
\addlinespace[3mm]
$\text{RPO}_{\text{4a}}$ & 372.33  & 8.423   & $\mathbb{T}_2$, $\mathbb{T}_4$                          & 0  & -0.0314 & 0      & 0.8898  & 0      \\
$\text{RPO}_{\text{4b}}$ & 382.89  & 8.858   & $\mathbb{T}_2$, $\mathbb{T}_4$                          & 3  & 0.0550  & 0      & 1.2345  & 0      \\
$\text{RPO}_{\text{4c}}$ & 418.40  & -1.240  & $\mathbb{T}_2$, $\mathbb{T}_4$                          & 7  & 0.1395  & 0.7401 & -1.7904 & 0.0808 \\
$\text{RPO}_{\text{4d}}$ & 389.25  & 9.262   & $\mathbb{T}_2$                                 & 3  & 0.1137  & 0      & 1.5570  & 0      \\
$\text{RPO}_{\text{4e}}$ & 387.90  & 9.166   & -                                     & 4  & 0.0831  & 0      & 1.3802  & 0      \\
\addlinespace[3mm]
$\text{RPO}_{\text{5a}}$ & 931.67  & 22.975 & $\mathbb{T}_5$ & 7  & 0.2607 & 0.0167 & 11.206  & 1.7531 \\
$\text{RPO}_{\text{5b}}$ & 757.66  & 44.652 & $\mathbb{T}_5$ & 14 & 0.1948 & 0.0663 & 3.8327  & 2.1069 \\
$\text{RPO}_{\text{5c}}$ & 743.36  & 24.607 & $\mathbb{T}_5$ & 9  & 0.1830 & 0.2157 & -0.1262 & 3.8955 \\
$\text{RPO}_{\text{5d}}$ & 1653.64 & -2.411 & $\mathbb{T}_5$ & 13 & 0.2203 & 0.1900 & -38.225 & 0      \\
$\text{RPO}_{\text{5e}}$ & 1613.55 & 45.759 & $\mathbb{T}_5$ & 9  & 0.2043 & 0.0338 & 23.094  & 14.009  
\\
\addlinespace[1mm]
\hline \hline            
\end{tabular}
\caption{List of Relative periodic orbits (RPOs). $N_u$ is the number of unstable directions in the full phase space. $\lambda$ is the largest Floquet multiplier, and $\mu$ is the corresponding Floquet exponent.}
\label{table:ECS_data_RPOs}
\end{table}

\begin{table}[]
\centering
\begin{tabular}{p{15mm}>{\centering\arraybackslash}p{25mm}>{\centering\arraybackslash}p{20mm}>{\centering\arraybackslash}p{10mm}>{\centering\arraybackslash}p{20mm}>{\centering\arraybackslash}p{20mm}}  \hline \hline \vspace{.1mm}
ECS ID & \vspace{.1mm}  Symmetries & \vspace{.1mm}  $N_u$ & \vspace{.1mm}  $\mathrm{Re}(\mu)$ & \vspace{.1mm}  $\mathrm{Im}(\mu)$ \\
\hline \addlinespace[1.2mm]
UNI & $\mathbb{T}_{\infty}$, $\sigma_x$ & 8 & 0.0111 & -0.0193 \\
LAN & $\mathbb{T}_{\infty}$, $\sigma_x \sigma_y$ & 7 & 0.0108 & 0  \\
\addlinespace[1mm]
\hline \hline  
\end{tabular}
\caption{List of Equilibria (EQ). $\mathbb{T}_{\infty}$ means the solution is invariant under arbitrary translations in $x$. $N_u$ is the number of unstable directions in the full phase space. $\mu$ is the largest eigenvalue of the Jacobian evaluated at the equilibria.}
\label{table:ECS_data_EQ}
\end{table}

\newpage

\begin{table}[]
\centering
\begin{tabular}{p{11mm}>{\centering\arraybackslash}p{11mm}>{\centering\arraybackslash}p{10mm}>{\centering\arraybackslash}p{20mm}>{\centering\arraybackslash}p{20mm}>{\centering\arraybackslash}p{20mm}>{\centering\arraybackslash}p{20mm}>{\centering\arraybackslash}p{20mm}}  \hline \hline \vspace{.1mm}
ECS  & \vspace{.1mm} $T$ & \vspace{.1mm} $N_u$ & \vspace{.1mm} $\mathrm{Re}(\mu) \times 10^2$ & \vspace{.1mm} $\mathrm{Im}(\mu) \times 10^2$ & \vspace{.1mm} $\mathrm{Re}(\lambda)$ & \vspace{.1mm} $\mathrm{Im}(\lambda)$ \\
\hline \addlinespace[1.2mm]
$\text{PO}_{\text{3a}}$  & 343.09  & 0 & -0.0136 & 0      & 0.9545  & 0 \\
$\text{PO}_{\text{4a}}$ & 974.68  & 3 & 0.1114  & 0      & 2.9611  & 0 \\
$\text{PO}_{\text{4b}}$ & 422.85  & 5 & 0.1946  & 0      & 2.2766  & 0 \\
$\text{PO}_{\text{4c}}$ & 587.23  & 6 & 0.5577  & 0.5350 & -26.436 & 0 \\
$\text{PO}_{\text{5a}}$ & 1137.08 & 6 & 0.5143  & 0      & 346.57  & 0 \\
$\text{PO}_{\text{5b}}$ & 538.75  & 7 & 0.4177  & 0      & 9.4928  & 0 \\
$\text{PO}_{\text{5c}}$ & 621.56  & 12 & 0.4437 & 0.0819 & 13.764 & 7.6809 \\
$\text{PO}_{\text{6a}}$  & 1274.17 & 4 & 0.5170  & 0      & 725.86  & 0 \\
$\text{PO}_{\text{7a}}$  & 1234.85 & 5 & 0.5921  & 0      & 1497.8  & 0   
\\
\addlinespace[1mm]
\hline \hline                 
\end{tabular}
\caption{List of ECS invariant under $\sigma_x \sigma_y$. $N_u$ is the number of unstable directions in the corresponding invariant subspace. Here, $\lambda$ and $\mu$ are computed by restricting perturbations within that subspace.}
\label{table:ECS_data_sigma_x_sigma_y_subspace}
\end{table}

\begin{table}[]
\centering
\begin{tabular}{p{14mm}>{\centering\arraybackslash}p{12mm}>{\centering\arraybackslash}p{15mm}>{\centering\arraybackslash}p{10mm}>{\centering\arraybackslash}p{20mm}>{\centering\arraybackslash}p{20mm}>{\centering\arraybackslash}p{20mm}>{\centering\arraybackslash}p{20mm}>{\centering\arraybackslash}p{20mm}}  \hline \hline \vspace{.1mm}
ECS  & \vspace{.1mm} $T$& \vspace{.1mm} $\ell$ & \vspace{.1mm} $N_u$ & \vspace{.1mm} $\mathrm{Re}(\mu) \times 10^2$ & \vspace{.1mm} $\mathrm{Im}(\mu) \times 10^2$ & \vspace{.1mm} $\mathrm{Re}(\lambda)$ & \vspace{.1mm} $\mathrm{Im}(\lambda)$ \\
\hline \addlinespace[1.2mm]
$\text{RPO}_{\text{u5}}$ & 607.38  & 20.321  & 2 & 0.0415  & 0.1367 & 0.8684  & 0.9497 \\
$\text{RPO}_{\text{u7}}$ & 812.02  & 31.589  & 4 & 0.0291  & 0.1228 & 0.6873  & 1.0634 \\
$\text{RPO}_{\text{u9}}$ & 1137.00 & 27.373  & 6 & 0.7065  & 0.2517 & -2962.1 & 849.30 \\
$\text{RPO}_{\text{2a}}$ & 108.35  & 17.870  & 0 & -0.1018 & 2.1731 & -0.6322 & 0.6343 \\
$\text{RPO}_{\text{3b}}$ & 636.45  & -26.361 & 6 & 0.5839  & 0.2417 & 1.3397  & 41.092 \\
$\text{RPO}_{\text{3c}}$ & 262.57  & -9.542  & 4 & 0.2676  & 1.1544 & -2.0065 & 0.2225    
\\
\addlinespace[1mm]
\hline \hline                 
\end{tabular}
\caption{List of ECS invariant under $\sigma_x \mathbb{T}_2$. $N_u$ is the number of unstable directions in the corresponding invariant subspace. Here, $\lambda$ and $\mu$ are computed by restricting perturbations within that subspace.}
\label{table:ECS_data_sigma_x_T2_subspace}
\end{table}

\begin{table}[]
\centering
\begin{tabular}{p{15mm}>{\centering\arraybackslash}p{15mm}>{\centering\arraybackslash}p{15mm}>{\centering\arraybackslash}p{10mm}>{\centering\arraybackslash}p{20mm}>{\centering\arraybackslash}p{20mm}>{\centering\arraybackslash}p{20mm}>{\centering\arraybackslash}p{20mm}}  \hline \hline \vspace{.1mm}
ECS  & \vspace{.1mm} $T$ & \vspace{.1mm} $\ell$ &\vspace{.1mm} $N_u$ & \vspace{.1mm} $\mathrm{Re}(\mu) \times 10^2$ & \vspace{.1mm} $\mathrm{Im}(\mu) \times 10^2$ & \vspace{.1mm} $\mathrm{Re}(\lambda)$ & \vspace{.1mm} $\mathrm{Im}(\lambda)$ \\
\hline \addlinespace[1.2mm]
\rowcolor[HTML]{FCE4D6} 
UNI   & \multicolumn{1}{c}{\cellcolor[HTML]{FCE4D6}-} & \multicolumn{1}{c}{\cellcolor[HTML]{FCE4D6}-} & 4 & 1.11  & -1.93  & \multicolumn{1}{c}{\cellcolor[HTML]{FCE4D6}-} & \multicolumn{1}{c}{\cellcolor[HTML]{FCE4D6}-}  \\
\rowcolor[HTML]{FCE4D6} 
LAN   & \multicolumn{1}{c}{\cellcolor[HTML]{FCE4D6}-} & \multicolumn{1}{c}{\cellcolor[HTML]{FCE4D6}-} & 3 & 1.01  & 0      & \multicolumn{1}{c}{\cellcolor[HTML]{FCE4D6}-} & \multicolumn{1}{c}{\cellcolor[HTML]{FCE4D6}-}   \\
\addlinespace[1.2mm]
\rowcolor[HTML]{E2EFDA} 
$\text{PO}_{\text{4a}}$  & 974.68                                        & \multicolumn{1}{c}{\cellcolor[HTML]{E2EFDA}-} & 3 & 0.0956  & 0.3223  & -2.5387                                       & 0                                             \\
\rowcolor[HTML]{E2EFDA} 
$\text{PO}_{\text{4b}}$  & 422.85                                        & \multicolumn{1}{c}{\cellcolor[HTML]{E2EFDA}-} & 6 & 0.5020  & 0       & 8.3544                                        & 0                                             \\
\rowcolor[HTML]{E2EFDA} 
$\text{PO}_{\text{4c}}$  & 587.23                                        & \multicolumn{1}{c}{\cellcolor[HTML]{E2EFDA}-} & 7 & 0.6703  & 0       & 51.225                                        & 0                                             \\
\rowcolor[HTML]{E2EFDA} 
$\text{PO}_{\text{6a}}$   & 1274.17                                       & \multicolumn{1}{c}{\cellcolor[HTML]{E2EFDA}-} & 3 & 0.5170  & 0       & 725.84                                        & 0                                             \\
\addlinespace[1.2mm]
\rowcolor[HTML]{DDEBF7} 
$\text{RPO}_{\text{u4}}$ & 738.19                                        & 12.936                                        & 5 & 0.3045  & 0.4088  & -9.3933                                       & 1.1703                                        \\
\rowcolor[HTML]{DDEBF7} 
$\text{RPO}_{\text{u6}}$ & 666.18                                        & -3.642                                        & 2 & 0.0196  & 0.0513  & 1.0733                                        & 0.3817                                        \\
\rowcolor[HTML]{DDEBF7} 
$\text{RPO}_{\text{u8}}$ & 986.91                                        & 22.221                                        & 4 & 0.4632  & 0.2364  & -66.706                                       & 69.925                                        \\
\rowcolor[HTML]{DDEBF7} 
$\text{RPO}_{\text{2b}}$ & 173.91                                        & 6.426                                         & 2 & 0.1296  & 1.4777  & -1.0534                                       & 0.6779                                        \\
\rowcolor[HTML]{DDEBF7} 
$\text{RPO}_{\text{2c}}$ & 178.34                                        & 12.282                                        & 2 & 0.2932  & 1.3412  & -1.2346                                       & 1.1496                                        \\
\rowcolor[HTML]{DDEBF7} 
$\text{RPO}_{\text{2d}}$ & 641.01                                        & 25.114                                        & 8 & 0.5255  & 0.0749  & 25.758                                        & 13.418                                        \\
\rowcolor[HTML]{DDEBF7} 
$\text{RPO}_{\text{4a}}$ & 372.33                                        & 8.423                                         & 0 & -0.0314 & 0       & 0.8898                                        & 0                                             \\
\rowcolor[HTML]{DDEBF7} 
$\text{RPO}_{\text{4b}}$ & 382.89                                        & 8.858                                         & 1 & 0.0550  & 0       & 1.2345                                        & 0                                             \\
\rowcolor[HTML]{DDEBF7} 
$\text{RPO}_{\text{4c}}$ & 418.40                                        & -1.240                                        & 3 & 0.1288  & 0.7509  & -1.7138                                       & 0                                             \\
\rowcolor[HTML]{DDEBF7} 
$\text{RPO}_{\text{4d}}$ & 389.25                                        & 9.262                                         & 2 & 0.0541  & 0.0901  & 1.1591                                        & 0.4243  
\\
\addlinespace[1mm]
\hline \hline
\end{tabular}
\caption{List of ECS invariant under $\mathbb{T}_2$. $N_u$ is the number of unstable directions in the corresponding invariant subspace. Here, $\lambda$ and $\mu$ are computed by restricting perturbations within that subspace.}
\label{tab:ECS_data_T2_subspace}
\end{table}

\begin{table}[]
\centering
\begin{tabular}{p{15mm}>{\centering\arraybackslash}p{15mm}>{\centering\arraybackslash}p{15mm}>{\centering\arraybackslash}p{10mm}>{\centering\arraybackslash}p{20mm}>{\centering\arraybackslash}p{20mm}>{\centering\arraybackslash}p{20mm}>{\centering\arraybackslash}p{20mm}}  \hline \hline \vspace{.1mm}
ECS  & \vspace{.1mm} $T$ & \vspace{.1mm} $\ell$ &\vspace{.1mm} $N_u$ & \vspace{.1mm} $\mathrm{Re}(\mu) \times 10^2$ & \vspace{.1mm} $\mathrm{Im}(\mu) \times 10^2$ & \vspace{.1mm} $\mathrm{Re}(\lambda)$ & \vspace{.1mm} $\mathrm{Im}(\lambda)$ \\
\hline \addlinespace[1.2mm]
\rowcolor[HTML]{FCE4D6} 
UNI   & \multicolumn{1}{c}{\cellcolor[HTML]{FCE4D6}-} & \multicolumn{1}{c}{\cellcolor[HTML]{FCE4D6}-} & 3 & 1.111  & -1.93 & \multicolumn{1}{c}{\cellcolor[HTML]{FCE4D6}-} & \multicolumn{1}{c}{\cellcolor[HTML]{FCE4D6}-} \\
\rowcolor[HTML]{FCE4D6} 
LAN   & \multicolumn{1}{c}{\cellcolor[HTML]{FCE4D6}-} & \multicolumn{1}{c}{\cellcolor[HTML]{FCE4D6}-} & 3 & 0.97  & 0        & \multicolumn{1}{c}{\cellcolor[HTML]{FCE4D6}-} & \multicolumn{1}{c}{\cellcolor[HTML]{FCE4D6}-} \\
\addlinespace[1.2mm]
\rowcolor[HTML]{E2EFDA} 
$\text{PO}_{\text{3a}}$   & 343.09                                        & \multicolumn{1}{c}{\cellcolor[HTML]{E2EFDA}-} & 1 & 0.0824  & 0       & 1.3268                                        & 0                                             \\
\rowcolor[HTML]{E2EFDA} 
$\text{PO}_{\text{6a}}$   & 1274.17                                       & \multicolumn{1}{c}{\cellcolor[HTML]{E2EFDA}-} & 2 & 0.3178  & 0       & 57.330                                        & 0                                             \\
\addlinespace[1.2mm]
\rowcolor[HTML]{DDEBF7} 
$\text{RPO}_{\text{u6}}$ & 666.18                                        & -3.642                                        & 0 & -0.1217 & 0.4337  & -0.4303                                       & 0.1111                                        \\
\rowcolor[HTML]{DDEBF7} 
$\text{RPO}_{\text{u9}}$ & 1137.00                                       & 27.373                                        & 4 & 0.7235  & 0.2313  & -3257.0                                       & 1831.4                                        \\
\rowcolor[HTML]{DDEBF7} 
$\text{RPO}_{\text{3a}}$ & 357.45                                        & 2.296                                         & 0 & -0.1819 & 0       & 0.5220                                        & 0                                             \\
\rowcolor[HTML]{DDEBF7} 
$\text{RPO}_{\text{3b}}$ & 636.45                                        & -26.361                                       & 4 & 0.6290  & 0.4936  & -54.787                                       & 0                                             \\
\rowcolor[HTML]{DDEBF7} 
$\text{RPO}_{\text{3c}}$ & 262.6                                         & -9.542                                        & 2 & 0.1451  & 0.5784  & 0.0762                                        & 1.4619                                        \\
\addlinespace[1mm]
\hline \hline
\end{tabular}
\caption{List of ECS invariant under $\mathbb{T}_3$. $N_u$ is the number of unstable directions in the corresponding invariant subspace. Here, $\lambda$ and $\mu$ are computed by restricting perturbations within that subspace.}
\label{tab:ECS_data_T3_subspace}
\end{table}

\begin{table}[]
\centering
\begin{tabular}{p{15mm}>{\centering\arraybackslash}p{15mm}>{\centering\arraybackslash}p{15mm}>{\centering\arraybackslash}p{10mm}>{\centering\arraybackslash}p{20mm}>{\centering\arraybackslash}p{20mm}>{\centering\arraybackslash}p{20mm}>{\centering\arraybackslash}p{20mm}}  \hline \hline \vspace{.1mm}
ECS  & \vspace{.1mm} $T$ & \vspace{.1mm} $\ell$ &\vspace{.1mm} $N_u$ & \vspace{.1mm} $\mathrm{Re}(\mu) \times 10^2$ & \vspace{.1mm} $\mathrm{Im}(\mu) \times 10^2$ & \vspace{.1mm} $\mathrm{Re}(\lambda)$ & \vspace{.1mm} $\mathrm{Im}(\lambda)$ \\
\hline \addlinespace[1.2mm]
\rowcolor[HTML]{FCE4D6} 
UNI   & \multicolumn{1}{c}{\cellcolor[HTML]{FCE4D6}-} & \multicolumn{1}{c}{\cellcolor[HTML]{FCE4D6}-} & 2 & 0.87  & -1.36                    & \multicolumn{1}{c}{\cellcolor[HTML]{FCE4D6}-} & \multicolumn{1}{c}{\cellcolor[HTML]{FCE4D6}-} \\
\rowcolor[HTML]{FCE4D6} 
LAN   & \multicolumn{1}{c}{\cellcolor[HTML]{FCE4D6}-} & \multicolumn{1}{c}{\cellcolor[HTML]{FCE4D6}-} & 2 & 1.01  & 0                          & \multicolumn{1}{c}{\cellcolor[HTML]{FCE4D6}-} & \multicolumn{1}{c}{\cellcolor[HTML]{FCE4D6}-} \\
\addlinespace[1.2mm]
\rowcolor[HTML]{E2EFDA} 
$\text{PO}_{\text{4a}}$  & 974.68                                        & \multicolumn{1}{c}{\cellcolor[HTML]{E2EFDA}-} & 1                    & 0.0812  & 0       & 2.2060                                        & 0                                             \\
\rowcolor[HTML]{E2EFDA} 
$\text{PO}_{\text{4b}}$  & 422.85                                        & \multicolumn{1}{c}{\cellcolor[HTML]{E2EFDA}-} & 3                    & 0.5020  & 0       & 8.3544                                        & 0                                             \\
\rowcolor[HTML]{E2EFDA} 
$\text{PO}_{\text{4c}}$  & 587.23                                        & \multicolumn{1}{c}{\cellcolor[HTML]{E2EFDA}-} & 3                    & 0.6703  & 0       & 51.226                                        & 0                                             \\
\addlinespace[1.2mm]
\rowcolor[HTML]{DDEBF7} 
$\text{RPO}_{\text{u4}}$ & 738.19                                        & 12.936                                        & 1                    & 0.0492  & 0.4256  & -1.4377                                       & 0                                             \\
\rowcolor[HTML]{DDEBF7} 
$\text{RPO}_{\text{u8}}$ & 986.91                                        & 22.221                                        & 2                    & 0.2987  & 0.2411  & -13.792                                       & 13.166                                        \\
\rowcolor[HTML]{DDEBF7} 
$\text{RPO}_{\text{4a}}$ & 372.33                                        & 8.423                                         & 0                    & -0.0314 & 0       & 0.8898                                        & 0                                             \\
\rowcolor[HTML]{DDEBF7} 
$\text{RPO}_{\text{4b}}$ & 382.89                                        & 8.858                                         & 1                    & 0.0550  & 0       & 1.2345                                        & 0                                             \\
\rowcolor[HTML]{DDEBF7} 
$\text{RPO}_{\text{4c}}$ & 418.40                                        & -1.240                                        & 0                    & -0.0227 & 0.7509  & -0.9092                                       & 0                                             \\
\addlinespace[1mm]
\hline \hline
\end{tabular}
\caption{List of ECS invariant under $\mathbb{T}_4$. $N_u$ is the number of unstable directions in the corresponding invariant subspace. Here, $\lambda$ and $\mu$ are computed by restricting perturbations within that subspace.}
\label{tab:ECS_data_T4_subspace}
\end{table}

\begin{table}[]
\centering
\begin{tabular}{p{15mm}>{\centering\arraybackslash}p{15mm}>{\centering\arraybackslash}p{15mm}>{\centering\arraybackslash}p{10mm}>{\centering\arraybackslash}p{20mm}>{\centering\arraybackslash}p{20mm}>{\centering\arraybackslash}p{20mm}>{\centering\arraybackslash}p{20mm}}  \hline \hline \vspace{.1mm}
ECS  & \vspace{.1mm} $T$ & \vspace{.1mm} $\ell$ &\vspace{.1mm} $N_u$ & \vspace{.1mm} $\mathrm{Re}(\mu) \times 10^2$ & \vspace{.1mm} $\mathrm{Im}(\mu) \times 10^2$ & \vspace{.1mm} $\mathrm{Re}(\lambda)$ & \vspace{.1mm} $\mathrm{Im}(\lambda)$ \\
\hline \addlinespace[1.2mm]
\rowcolor[HTML]{FCE4D6} 
UNI   & \multicolumn{1}{c}{\cellcolor[HTML]{FCE4D6}-} & \multicolumn{1}{c}{\cellcolor[HTML]{FCE4D6}-} & 2 & 1.09  & -1.64                    & \multicolumn{1}{c}{\cellcolor[HTML]{FCE4D6}-} & \multicolumn{1}{c}{\cellcolor[HTML]{FCE4D6}-} \\
\rowcolor[HTML]{FCE4D6} 
LAN   & \multicolumn{1}{c}{\cellcolor[HTML]{FCE4D6}-} & \multicolumn{1}{c}{\cellcolor[HTML]{FCE4D6}-} & 1 & 1.08  & 0                         & \multicolumn{1}{c}{\cellcolor[HTML]{FCE4D6}-} & \multicolumn{1}{c}{\cellcolor[HTML]{FCE4D6}-}  \\
\addlinespace[1.2mm]
\rowcolor[HTML]{E2EFDA} 
$\text{PO}_{\text{5a}}$  & 1137.08                                       & \multicolumn{1}{c}{\cellcolor[HTML]{E2EFDA}-} & 1                    & 0.3929  & 0       & 87.1113                                       & 0                                             \\
\rowcolor[HTML]{E2EFDA} 
$\text{PO}_{\text{5b}}$  & 538.75                                        & \multicolumn{1}{c}{\cellcolor[HTML]{E2EFDA}-} & 3                    & 1.0473  & 0       & 282.162                                       & 0                                             \\
\rowcolor[HTML]{E2EFDA} 
$\text{PO}_{\text{5c}}$  & 622.08                                        & \multicolumn{1}{c}{\cellcolor[HTML]{E2EFDA}-} & 3                    & 1.0436  & 0       & 659.985                                       & 0                                             \\
\addlinespace[1.2mm]
\rowcolor[HTML]{DDEBF7} 
$\text{RPO}_{\text{u5}}$ & 607.38  & 20.321                                        & 0 & -0.0134 & 0.5172 & -0.9220 & 0      \\
\rowcolor[HTML]{DDEBF7} 
$\text{RPO}_{\text{5a}}$ & 931.67  & 22.975                                        & 1 & 0.2322  & 0      & 8.7028  & 0      \\
\rowcolor[HTML]{DDEBF7} 
$\text{RPO}_{\text{5b}}$ & 757.66  & 44.652                                        & 2 & 0.0471  & 0.3680 & -1.3410 & 0.4947 \\
\rowcolor[HTML]{DDEBF7} 
$\text{RPO}_{\text{5c}}$ & 743.36  & 24.607                                        & 1 & 0.1169  & 0      & 2.3839  & 0      \\
\rowcolor[HTML]{DDEBF7} 
$\text{RPO}_{\text{5d}}$ & 1653.64 & -2.411                                        & 1 & 0.2203  & 0.1900 & -38.225 & 0      \\
\rowcolor[HTML]{DDEBF7} 
$\text{RPO}_{\text{5e}}$ & 1613.55 & 45.759                                        & 1 & 0.1769  & 0      & 17.377  & 0      \\
\addlinespace[1mm]
\hline \hline
\end{tabular}
\caption{List of ECS invariant under $\mathbb{T}_5$. $N_u$ is the number of unstable directions in the corresponding invariant subspace. Here, $\lambda$ and $\mu$ are computed by restricting perturbations within that subspace.}
\label{tab:ECS_data_T5_subspace}
\end{table}

\newpage

 \begin{table}[]
 \centering
 \begin{tabular}{p{18mm}>{\centering\arraybackslash}p{25mm}>{\centering\arraybackslash}p{35mm}>{\centering\arraybackslash}p{10mm}>{\centering\arraybackslash}p{10mm}>{\centering\arraybackslash}p{15mm}>{\centering\arraybackslash}p{30mm}} \hline \hline \addlinespace[2mm]
Source ECS & Target ECS & Verified in subspace? (Y/N) & $d_{\text{u1}}$ & $N_{\text{u2}}$ & $d_{\text{intersection}}$ & Residual \\  \hline \addlinespace[1.2mm]
UNI   & $\text{RPO}_{\text{u4}}$                     & N             & 16 & 13 & 3  & 2.92E-05 \\
UNI   & $\text{RPO}_{\text{u5}}$                     & Y ($\mathbb{T}_5$)               & 16 & 6  & 10 & 7.36E-07 \\
UNI   & $\text{RPO}_{\text{u6}}$                     & Y ($\mathbb{T}_6$)               & 16 & 6  & 10 & 4.64E-06 \\
UNI   & $\text{RPO}_{\text{u7}}$                     & Y ($\mathbb{T}_7$)               & 16 & 8  & 8  & 4.22E-05 \\
UNI   & $\text{RPO}_{\text{u8}}$                     & Y ($\mathbb{T}_8$)               & 16 & 10 & 6  & 6.58E-06 \\
UNI   & $\text{RPO}_{\text{u9}}$                     & Y ($\mathbb{T}_9$)               & 16 & 15 & 1  & 1.78E-06 \\
UNI   & $\text{RPO}_{\text{2a}}$                     & Y ($\sigma_x \mathbb{T}_2$)      & 16 & 2  & 14 & 1.22E-04 \\
UNI   & $\sigma_x \sigma_y$ $\text{RPO}_{\text{2a}}$ & Y ($\sigma_x \mathbb{T}_2$)      & 16 & 2  & 14 & 1.69E-04 \\
      &                           &                     &    &    &    &          \\
LAN   & $\text{PO}_{\text{3a}}$                       & Y ($\sigma_x \sigma_y$) & 14 & 1  & 13 & 4.15E-04 \\
LAN   & $\text{PO}_{\text{6a}}$                       & Y ($\mathbb{T}_6$)               & 14 & 6  & 8  & 3.01E-05 \\
LAN   & $\text{PO}_{\text{7a}}$                       & Y ($\mathbb{T}_7$)               & 14 & 9  & 5  & 8.96E-05 \\
LAN   & $\text{RPO}_{\text{u5}}$                     & Y ($\mathbb{T}_5$)               & 14 & 6  & 8  & 3.92E-06 \\
LAN   & $\text{RPO}_{\text{u8}}$                     & Y ($\mathbb{T}_8$)               & 14 & 10 & 4  & 5.89E-06 \\
      &                           &                     &    &    &    &          \\
$\text{RPO}_{\text{u4}}$ & $\text{RPO}_{\text{3a}}$                     & N                   & 14 & 0  & 14 & 1.31E-04 \\
$\text{RPO}_{\text{u4}}$ & $\text{RPO}_{\text{4a}}$   & Y ($\mathbb{T}_2$)               & 14 & 0  & 14 & 1.40E-04 \\
$\text{RPO}_{\text{u4}}$ & $\sigma_x$ $\text{RPO}_{\text{4c}}$  & Y ($\mathbb{T}_4$)               & 14 & 7  & 7  & 6.93E-05 \\
$\text{RPO}_{\text{u5}}$ & $\text{RPO}_{\text{3a}}$                     & N                  & 7  & 0  & 7  & 1.61E-04 \\
$\text{RPO}_{\text{u6}}$ & $\text{RPO}_{\text{3a}}$                     & N                   & 7  & 0  & 7  & 4.28E-05 \\
$\text{RPO}_{\text{u6}}$ & $\text{RPO}_{\text{4a}}$                     & Y ($\mathbb{T}_2$)               & 7  & 0  & 7  & 1.38E-04 \\
$\text{RPO}_{\text{u7}}$ & $\text{RPO}_{\text{3a}}$                     & N                  & 9  & 0  & 9  & 8.45E-05 \\
$\text{RPO}_{\text{u7}}$ & $\text{RPO}_{\text{2a}}$                     & Y ($\sigma_x \mathbb{T}_2$)      & 9  & 2  & 7  & 6.64E-04 \\
$\text{RPO}_{\text{u8}}$ & $\text{RPO}_{\text{3a}}$                     & N                   & 11 & 0  & 11 & 6.35E-05 \\
$\text{RPO}_{\text{u8}}$ & $\text{RPO}_{\text{4a}}$                     & Y ($\mathbb{T}_2$)               & 11 & 0  & 11 & 1.13E-04 \\
$\text{RPO}_{\text{u8}}$ & $\sigma_x$ $\text{RPO}_{\text{4c}}$          & Y ($\mathbb{T}_4$)               & 11 & 7  & 4  & 8.19E-05 \\
$\text{RPO}_{\text{u9}}$ & $\text{RPO}_{\text{3a}}$                     & Y ($\mathbb{T}_3$)               & 16 & 0  & 16 & 3.94E-05 \\
$\text{RPO}_{\text{u9}}$ & $\text{RPO}_{\text{u6}}$                     & Y ($\mathbb{T}_3$)               & 16 & 6  & 10 & 6.34E-05 \\
      &                           &                     &    &    &    &          \\
$\text{PO}_{\text{3a}}$   & $\text{RPO}_{\text{3a}}$                     & Y ($\mathbb{T}_3$)               & 2  & 0  & 2  & 5.46E-05 \\
$\text{RPO}_{\text{3c}}$ & $\text{RPO}_{\text{2a}}$                     & Y ($\sigma_x \mathbb{T}_2$)      & 9  & 2  & 7  & 1.14E-04 \\
$\text{RPO}_{\text{3c}}$ & $\sigma_x \sigma_y$ $\text{RPO}_{\text{2a}}$ & Y ($\sigma_x \mathbb{T}_2$)      & 9  & 2  & 7  & 1.19E-04 \\
      &                           &                     &    &    &    &          \\
$\text{PO}_{\text{4a}}$  & $\text{RPO}_{\text{3a}}$                     & N                   & 6  & 0  & 6  & 4.10E-05 \\
$\text{PO}_{\text{4a}}$  & $\text{RPO}_{\text{4a}}$                     & N                  & 6  & 0  & 6  & 1.43E-04 \\
$\text{PO}_{\text{4a}}$  & $\sigma_x \sigma_y$ $\text{RPO}_{\text{4c}}$ & Y ($\mathbb{T}_4$)               & 2* & 0* & 2* & 8.45E-05 \\
$\text{PO}_{\text{4b}}$  & $\text{RPO}_{\text{4a}}$                     & Y ($\mathbb{T}_4$)            & 9  & 0  & 9  & 1.28E-04 \\
$\text{PO}_{\text{4c}}$  & $\text{PO}_{\text{3a}}$                       & Y ($\sigma_x \sigma_y$) & 14 & 1  & 13 & 1.85E-03 \\
$\text{PO}_{\text{4c}}$  & $\text{RPO}_{\text{3a}}$                     & N                    & 14 & 0  & 14 & 8.83E-05         \\
$\text{PO}_{\text{4c}}$  & $\text{RPO}_{\text{4a}}$                     & Y ($\mathbb{T}_4$)                   & 14 & 0  & 14 & 1.30E-04         \\
$\text{PO}_{\text{4c}}$  & $\text{RPO}_{\text{4c}}$                     & Y ($\mathbb{T}_4$)               & 14 & 7  & 7  & 3.05E-05 \\
$\text{RPO}_{\text{4c}}$ & $\text{RPO}_{\text{3a}}$                     & N                   & 8  & 0  & 8  & 1.38E-04 \\
$\text{RPO}_{\text{4c}}$ & $\text{RPO}_{\text{4a}}$                     & N                   & 8  & 0  & 8  & 1.47E-04 \\
      &                           &                     &    &    &    &          \\
$\text{PO}_{\text{5a}}$  & $\text{PO}_{\text{3a}}$                       & Y ($\sigma_x \sigma_y$) & 15 & 1  & 14 & 5.94E-04 \\
$\text{PO}_{\text{5b}}$  & $\text{RPO}_{\text{3a}}$                     & N                   & 16 & 0  & 16 & 3.09E-04 \\
$\text{PO}_{\text{5b}}$  & $\text{RPO}_{\text{u5}}$                     & Y ($\mathbb{T}_5$)               & 16 & 6  & 10 & 1.08E-04 \\
$\text{PO}_{\text{5c}}$  & $\text{RPO}_{\text{u5}}$                     & Y ($\mathbb{T}_5$)               & 26 & 6  & 20 & 9.54E-05         \\
$\text{RPO}_{\text{5b}}$ & $\text{RPO}_{\text{u5}}$                     & Y ($\mathbb{T}_5$)               & 15 & 6  & 9  & 9.19E-05         \\
$\text{RPO}_{\text{5c}}$ & $\text{RPO}_{\text{u5}}$                     & Y ($\mathbb{T}_5$)            & 10 & 6  & 4  & 8.87E-05  \\
\addlinespace[1.5mm]
\multicolumn{2}{l}{\hspace{2mm}*In \char'164he $\mathbb{T}_4$ subspace} &               &  &   &   &   \\
 \addlinespace[1mm]
 \hline \hline
 \end{tabular}
 \caption{Heteroclinic connections. Except for $\text{UNI} \rightarrow \text{RPO}_{\text{u4}}$, the target ECS of all connections is stable either in the full space or a symmetry subspace. In the latter case, the symmetry subspace is given in column 3. In columns 4 and 5, $d_{\text{u1}}$ is the dimension of unstable manifold of the source ECS, and $N_{\text{u2}}$ is the number of unstable direction of the target ECS, both computed in the full phase space. $d_{\text{intersection}}=d_{\text{u1}}-N_{\text{u2}}$ is the expected dimension of the heteroclinic connection.}
 \label{tab:SI-connections}
 \end{table}

\clearpage

\bibliographystyle{plain}


\begin{thebibliography}{73}%
\makeatletter
\providecommand \@ifxundefined [1]{%
 \@ifx{#1\undefined}
}%
\providecommand \@ifnum [1]{%
 \ifnum #1\expandafter \@firstoftwo
 \else \expandafter \@secondoftwo
 \fi
}%
\providecommand \@ifx [1]{%
 \ifx #1\expandafter \@firstoftwo
 \else \expandafter \@secondoftwo
 \fi
}%
\providecommand \natexlab [1]{#1}%
\providecommand \enquote  [1]{``#1''}%
\providecommand \bibnamefont  [1]{#1}%
\providecommand \bibfnamefont [1]{#1}%
\providecommand \citenamefont [1]{#1}%
\providecommand \href@noop [0]{\@secondoftwo}%
\providecommand \href [0]{\begingroup \@sanitize@url \@href}%
\providecommand \@href[1]{\@@startlink{#1}\@@href}%
\providecommand \@@href[1]{\endgroup#1\@@endlink}%
\providecommand \@sanitize@url [0]{\catcode `\\12\catcode `\$12\catcode
  `\&12\catcode `\#12\catcode `\^12\catcode `\_12\catcode `\%12\relax}%
\providecommand \@@startlink[1]{}%
\providecommand \@@endlink[0]{}%
\providecommand \url  [0]{\begingroup\@sanitize@url \@url }%
\providecommand \@url [1]{\endgroup\@href {#1}{\urlprefix }}%
\providecommand \urlprefix  [0]{URL }%
\providecommand \Eprint [0]{\href }%
\providecommand \doibase [0]{http://dx.doi.org/}%
\providecommand \selectlanguage [0]{\@gobble}%
\providecommand \bibinfo  [0]{\@secondoftwo}%
\providecommand \bibfield  [0]{\@secondoftwo}%
\providecommand \translation [1]{[#1]}%
\providecommand \BibitemOpen [0]{}%
\providecommand \bibitemStop [0]{}%
\providecommand \bibitemNoStop [0]{.\EOS\space}%
\providecommand \EOS [0]{\spacefactor3000\relax}%
\providecommand \BibitemShut  [1]{\csname bibitem#1\endcsname}%
\let\auto@bib@innerbib\@empty

\bibitem [{\citenamefont {Gompper}\ \emph {et~al.}(2020)\citenamefont
  {Gompper}, \citenamefont {Winkler}, \citenamefont {Speck}, \citenamefont
  {Solon}, \citenamefont {Nardini}, \citenamefont {Peruani}, \citenamefont
  {Löwen}, \citenamefont {Golestanian}, \citenamefont {Kaupp}, \citenamefont
  {Alvarez}, \citenamefont {Ki{\o}rboe}, \citenamefont {Lauga}, \citenamefont
  {Poon}, \citenamefont {DeSimone}, \citenamefont {Mui{\~{n}}os-Landin},
  \citenamefont {Fischer}, \citenamefont {Söker}, \citenamefont {Cichos},
  \citenamefont {Kapral}, \citenamefont {Gaspard}, \citenamefont {Ripoll},
  \citenamefont {Sagues}, \citenamefont {Doostmohammadi}, \citenamefont
  {Yeomans}, \citenamefont {Aranson}, \citenamefont {Bechinger}, \citenamefont
  {Stark}, \citenamefont {Hemelrijk}, \citenamefont {Nedelec}, \citenamefont
  {Sarkar}, \citenamefont {Aryaksama}, \citenamefont {Lacroix}, \citenamefont
  {Duclos}, \citenamefont {Yashunsky}, \citenamefont {Silberzan}, \citenamefont
  {Arroyo},\ and\ \citenamefont {Kale}}]{Gompper2020}%
  \BibitemOpen
  \bibfield  {author} {\bibinfo {author} {\bibfnamefont {Gerhard}\ \bibnamefont
  {Gompper}}, \bibinfo {author} {\bibfnamefont {Roland~G}\ \bibnamefont
  {Winkler}}, \bibinfo {author} {\bibfnamefont {Thomas}\ \bibnamefont {Speck}},
  \bibinfo {author} {\bibfnamefont {Alexandre}\ \bibnamefont {Solon}}, \bibinfo
  {author} {\bibfnamefont {Cesare}\ \bibnamefont {Nardini}}, \bibinfo {author}
  {\bibfnamefont {Fernando}\ \bibnamefont {Peruani}}, \bibinfo {author}
  {\bibfnamefont {Hartmut}\ \bibnamefont {Löwen}}, \bibinfo {author}
  {\bibfnamefont {Ramin}\ \bibnamefont {Golestanian}}, \bibinfo {author}
  {\bibfnamefont {U~Benjamin}\ \bibnamefont {Kaupp}}, \bibinfo {author}
  {\bibfnamefont {Luis}\ \bibnamefont {Alvarez}}, \bibinfo {author}
  {\bibfnamefont {Thomas}\ \bibnamefont {Ki{\o}rboe}}, \bibinfo {author}
  {\bibfnamefont {Eric}\ \bibnamefont {Lauga}}, \bibinfo {author}
  {\bibfnamefont {Wilson C~K}\ \bibnamefont {Poon}}, \bibinfo {author}
  {\bibfnamefont {Antonio}\ \bibnamefont {DeSimone}}, \bibinfo {author}
  {\bibfnamefont {Santiago}\ \bibnamefont {Mui{\~{n}}os-Landin}}, \bibinfo
  {author} {\bibfnamefont {Alexander}\ \bibnamefont {Fischer}}, \bibinfo
  {author} {\bibfnamefont {Nicola~A}\ \bibnamefont {Söker}}, \bibinfo {author}
  {\bibfnamefont {Frank}\ \bibnamefont {Cichos}}, \bibinfo {author}
  {\bibfnamefont {Raymond}\ \bibnamefont {Kapral}}, \bibinfo {author}
  {\bibfnamefont {Pierre}\ \bibnamefont {Gaspard}}, \bibinfo {author}
  {\bibfnamefont {Marisol}\ \bibnamefont {Ripoll}}, \bibinfo {author}
  {\bibfnamefont {Francesc}\ \bibnamefont {Sagues}}, \bibinfo {author}
  {\bibfnamefont {Amin}\ \bibnamefont {Doostmohammadi}}, \bibinfo {author}
  {\bibfnamefont {Julia~M}\ \bibnamefont {Yeomans}}, \bibinfo {author}
  {\bibfnamefont {Igor~S}\ \bibnamefont {Aranson}}, \bibinfo {author}
  {\bibfnamefont {Clemens}\ \bibnamefont {Bechinger}}, \bibinfo {author}
  {\bibfnamefont {Holger}\ \bibnamefont {Stark}}, \bibinfo {author}
  {\bibfnamefont {Charlotte~K}\ \bibnamefont {Hemelrijk}}, \bibinfo {author}
  {\bibfnamefont {Fran{\c{c}}ois~J}\ \bibnamefont {Nedelec}}, \bibinfo {author}
  {\bibfnamefont {Trinish}\ \bibnamefont {Sarkar}}, \bibinfo {author}
  {\bibfnamefont {Thibault}\ \bibnamefont {Aryaksama}}, \bibinfo {author}
  {\bibfnamefont {Mathilde}\ \bibnamefont {Lacroix}}, \bibinfo {author}
  {\bibfnamefont {Guillaume}\ \bibnamefont {Duclos}}, \bibinfo {author}
  {\bibfnamefont {Victor}\ \bibnamefont {Yashunsky}}, \bibinfo {author}
  {\bibfnamefont {Pascal}\ \bibnamefont {Silberzan}}, \bibinfo {author}
  {\bibfnamefont {Marino}\ \bibnamefont {Arroyo}}, \ and\ \bibinfo {author}
  {\bibfnamefont {Sohan}\ \bibnamefont {Kale}},\ }\bibfield  {title} {\enquote
  {\bibinfo {title} {The 2020 motile active matter roadmap},}\ }\href {\doibase
  10.1088/1361-648x/ab6348} {\bibfield  {journal} {\bibinfo  {journal} {J.
  Phys. Condens. Matter}\ }\textbf {\bibinfo {volume} {32}},\ \bibinfo {pages}
  {193001} (\bibinfo {year} {2020})}\BibitemShut {NoStop}%
\bibitem [{\citenamefont {Toner}\ and\ \citenamefont {Tu}(1995)}]{Toner1995}%
  \BibitemOpen
  \bibfield  {author} {\bibinfo {author} {\bibfnamefont {John}\ \bibnamefont
  {Toner}}\ and\ \bibinfo {author} {\bibfnamefont {Yuhai}\ \bibnamefont {Tu}},\
  }\bibfield  {title} {\enquote {\bibinfo {title} {Long-range order in a
  two-dimensional {DynamicalXYModel}: How birds fly together},}\ }\href
  {\doibase 10.1103/physrevlett.75.4326} {\bibfield  {journal} {\bibinfo
  {journal} {Phys. Rev. Lett.}\ }\textbf {\bibinfo {volume} {75}},\ \bibinfo
  {pages} {4326--4329} (\bibinfo {year} {1995})}\BibitemShut {NoStop}%
\bibitem [{\citenamefont {Toner}\ \emph {et~al.}(2005)\citenamefont {Toner},
  \citenamefont {Tu},\ and\ \citenamefont {Ramaswamy}}]{Toner2005}%
  \BibitemOpen
  \bibfield  {author} {\bibinfo {author} {\bibfnamefont {John}\ \bibnamefont
  {Toner}}, \bibinfo {author} {\bibfnamefont {Yuhai}\ \bibnamefont {Tu}}, \
  and\ \bibinfo {author} {\bibfnamefont {Sriram}\ \bibnamefont {Ramaswamy}},\
  }\bibfield  {title} {\enquote {\bibinfo {title} {Hydrodynamics and phases of
  flocks},}\ }\href {\doibase 10.1016/j.aop.2005.04.011} {\bibfield  {journal}
  {\bibinfo  {journal} {Annals of Physics}\ }\textbf {\bibinfo {volume}
  {318}},\ \bibinfo {pages} {170--244} (\bibinfo {year} {2005})}\BibitemShut
  {NoStop}%
\bibitem [{\citenamefont {Vicsek}\ and\ \citenamefont
  {Zafeiris}(2012)}]{Vicsek2012}%
  \BibitemOpen
  \bibfield  {author} {\bibinfo {author} {\bibfnamefont {Tam{\'{a}}s}\
  \bibnamefont {Vicsek}}\ and\ \bibinfo {author} {\bibfnamefont {Anna}\
  \bibnamefont {Zafeiris}},\ }\bibfield  {title} {\enquote {\bibinfo {title}
  {Collective motion},}\ }\href {\doibase 10.1016/j.physrep.2012.03.004}
  {\bibfield  {journal} {\bibinfo  {journal} {Phys. Rep.}\ }\textbf {\bibinfo
  {volume} {517}},\ \bibinfo {pages} {71--140} (\bibinfo {year}
  {2012})}\BibitemShut {NoStop}%
\bibitem [{\citenamefont {Attanasi}\ \emph
  {et~al.}(2014{\natexlab{a}})\citenamefont {Attanasi}, \citenamefont
  {Cavagna}, \citenamefont {Castello}, \citenamefont {Giardina}, \citenamefont
  {Grigera}, \citenamefont {Jeli\'{c}}, \citenamefont {Melillo}, \citenamefont
  {Parisi}, \citenamefont {Pohl}, \citenamefont {Shen},\ and\ \citenamefont
  {Viale}}]{Attanasi2014b}%
  \BibitemOpen
  \bibfield  {author} {\bibinfo {author} {\bibfnamefont {Alessandro}\
  \bibnamefont {Attanasi}}, \bibinfo {author} {\bibfnamefont {Andrea}\
  \bibnamefont {Cavagna}}, \bibinfo {author} {\bibfnamefont {Lorenzo~Del}\
  \bibnamefont {Castello}}, \bibinfo {author} {\bibfnamefont {Irene}\
  \bibnamefont {Giardina}}, \bibinfo {author} {\bibfnamefont {Tomas~S}\
  \bibnamefont {Grigera}}, \bibinfo {author} {\bibfnamefont {Asja}\
  \bibnamefont {Jeli\'{c}}}, \bibinfo {author} {\bibfnamefont {Stefania}\
  \bibnamefont {Melillo}}, \bibinfo {author} {\bibfnamefont {Leonardo}\
  \bibnamefont {Parisi}}, \bibinfo {author} {\bibfnamefont {Oliver}\
  \bibnamefont {Pohl}}, \bibinfo {author} {\bibfnamefont {Edward}\ \bibnamefont
  {Shen}}, \ and\ \bibinfo {author} {\bibfnamefont {Massimiliano}\ \bibnamefont
  {Viale}},\ }\bibfield  {title} {\enquote {\bibinfo {title} {{Information
  transfer and behavioural inertia in starling flocks}},}\ }\href {\doibase
  10.1038/NPHYS3035} {\bibfield  {journal} {\bibinfo  {journal} {Nat. Phys.}\
  }\textbf {\bibinfo {volume} {10}},\ \bibinfo {pages} {691} (\bibinfo {year}
  {2014}{\natexlab{a}})}\BibitemShut {NoStop}%
\bibitem [{\citenamefont {Attanasi}\ \emph
  {et~al.}(2014{\natexlab{b}})\citenamefont {Attanasi}, \citenamefont
  {Cavagna}, \citenamefont {{Del Castello}}, \citenamefont {Giardina},
  \citenamefont {Melillo}, \citenamefont {Parisi}, \citenamefont {Pohl},
  \citenamefont {Rossaro}, \citenamefont {Shen}, \citenamefont {Silvestri},\
  and\ \citenamefont {Viale}}]{Attanasi2014c}%
  \BibitemOpen
  \bibfield  {author} {\bibinfo {author} {\bibfnamefont {Alessandro}\
  \bibnamefont {Attanasi}}, \bibinfo {author} {\bibfnamefont {Andrea}\
  \bibnamefont {Cavagna}}, \bibinfo {author} {\bibfnamefont {Lorenzo}\
  \bibnamefont {{Del Castello}}}, \bibinfo {author} {\bibfnamefont {Irene}\
  \bibnamefont {Giardina}}, \bibinfo {author} {\bibfnamefont {Stefania}\
  \bibnamefont {Melillo}}, \bibinfo {author} {\bibfnamefont {Leonardo}\
  \bibnamefont {Parisi}}, \bibinfo {author} {\bibfnamefont {Oliver}\
  \bibnamefont {Pohl}}, \bibinfo {author} {\bibfnamefont {Bruno}\ \bibnamefont
  {Rossaro}}, \bibinfo {author} {\bibfnamefont {Edward}\ \bibnamefont {Shen}},
  \bibinfo {author} {\bibfnamefont {Edmondo}\ \bibnamefont {Silvestri}}, \ and\
  \bibinfo {author} {\bibfnamefont {Massimiliano}\ \bibnamefont {Viale}},\
  }\bibfield  {title} {\enquote {\bibinfo {title} {{Collective Behaviour
  without Collective Order in Wild Swarms of Midges}},}\ }\href {\doibase
  10.1371/journal.pcbi.1003697} {\bibfield  {journal} {\bibinfo  {journal}
  {PLoS Comput. Biol.}\ }\textbf {\bibinfo {volume} {10}},\ \bibinfo {pages}
  {e1003697} (\bibinfo {year} {2014}{\natexlab{b}})}\BibitemShut {NoStop}%
\bibitem [{\citenamefont {Cavagna}\ \emph {et~al.}(2017)\citenamefont
  {Cavagna}, \citenamefont {Conti}, \citenamefont {Creato}, \citenamefont
  {Castello}, \citenamefont {Giardina}, \citenamefont {Grigera}, \citenamefont
  {Melillo}, \citenamefont {Parisi},\ and\ \citenamefont
  {Viale}}]{Cavagna2017}%
  \BibitemOpen
  \bibfield  {author} {\bibinfo {author} {\bibfnamefont {Andrea}\ \bibnamefont
  {Cavagna}}, \bibinfo {author} {\bibfnamefont {Daniele}\ \bibnamefont
  {Conti}}, \bibinfo {author} {\bibfnamefont {Chiara}\ \bibnamefont {Creato}},
  \bibinfo {author} {\bibfnamefont {Lorenzo~Del}\ \bibnamefont {Castello}},
  \bibinfo {author} {\bibfnamefont {Irene}\ \bibnamefont {Giardina}}, \bibinfo
  {author} {\bibfnamefont {Tomas~S.}\ \bibnamefont {Grigera}}, \bibinfo
  {author} {\bibfnamefont {Stefania}\ \bibnamefont {Melillo}}, \bibinfo
  {author} {\bibfnamefont {Leonardo}\ \bibnamefont {Parisi}}, \ and\ \bibinfo
  {author} {\bibfnamefont {Massimiliano}\ \bibnamefont {Viale}},\ }\bibfield
  {title} {\enquote {\bibinfo {title} {Dynamic scaling in natural swarms},}\
  }\href {\doibase 10.1038/nphys4153} {\bibfield  {journal} {\bibinfo
  {journal} {Nature Physics}\ }\textbf {\bibinfo {volume} {13}},\ \bibinfo
  {pages} {914--918} (\bibinfo {year} {2017})}\BibitemShut {NoStop}%
\bibitem [{\citenamefont {van~der Vaart}\ \emph {et~al.}(2019)\citenamefont
  {van~der Vaart}, \citenamefont {Sinhuber}, \citenamefont {Reynolds},\ and\
  \citenamefont {Ouellette}}]{vanderVaart2019}%
  \BibitemOpen
  \bibfield  {author} {\bibinfo {author} {\bibfnamefont {Kasper}\ \bibnamefont
  {van~der Vaart}}, \bibinfo {author} {\bibfnamefont {Michael}\ \bibnamefont
  {Sinhuber}}, \bibinfo {author} {\bibfnamefont {Andrew~M.}\ \bibnamefont
  {Reynolds}}, \ and\ \bibinfo {author} {\bibfnamefont {Nicholas~T.}\
  \bibnamefont {Ouellette}},\ }\bibfield  {title} {\enquote {\bibinfo {title}
  {Mechanical spectroscopy of insect swarms},}\ }\href {\doibase
  10.1126/sciadv.aaw9305} {\bibfield  {journal} {\bibinfo  {journal} {Sci.
  Adv.}\ }\textbf {\bibinfo {volume} {5}},\ \bibinfo {pages} {eaaw9305}
  (\bibinfo {year} {2019})}\BibitemShut {NoStop}%
\bibitem [{\citenamefont {Fily}\ and\ \citenamefont
  {Marchetti}(2012)}]{Fily2012}%
  \BibitemOpen
  \bibfield  {author} {\bibinfo {author} {\bibfnamefont {Yaouen}\ \bibnamefont
  {Fily}}\ and\ \bibinfo {author} {\bibfnamefont {M.~Cristina}\ \bibnamefont
  {Marchetti}},\ }\bibfield  {title} {\enquote {\bibinfo {title} {{Athermal
  Phase Separation of Self-Propelled Particles with No Alignment}},}\ }\href
  {\doibase 10.1103/PhysRevLett.108.235702} {\bibfield  {journal} {\bibinfo
  {journal} {Phys. Rev. Lett.}\ }\textbf {\bibinfo {volume} {108}},\ \bibinfo
  {pages} {235702} (\bibinfo {year} {2012})}\BibitemShut {NoStop}%
\bibitem [{\citenamefont {Redner}\ \emph {et~al.}(2013)\citenamefont {Redner},
  \citenamefont {Hagan},\ and\ \citenamefont {Baskaran}}]{Redner2013}%
  \BibitemOpen
  \bibfield  {author} {\bibinfo {author} {\bibfnamefont {Gabriel~S.}\
  \bibnamefont {Redner}}, \bibinfo {author} {\bibfnamefont {Michael~F.}\
  \bibnamefont {Hagan}}, \ and\ \bibinfo {author} {\bibfnamefont {Aparna}\
  \bibnamefont {Baskaran}},\ }\bibfield  {title} {\enquote {\bibinfo {title}
  {Structure and dynamics of a phase-separating active colloidal fluid},}\
  }\href {\doibase 10.1103/physrevlett.110.055701} {\bibfield  {journal}
  {\bibinfo  {journal} {Phys. Rev. Lett.}\ }\textbf {\bibinfo {volume} {110}}
  (\bibinfo {year} {2013}),\ 10.1103/physrevlett.110.055701}\BibitemShut
  {NoStop}%
\bibitem [{\citenamefont {Stenhammar}\ \emph {et~al.}(2013)\citenamefont
  {Stenhammar}, \citenamefont {Tiribocchi}, \citenamefont {Allen},
  \citenamefont {Marenduzzo},\ and\ \citenamefont {Cates}}]{Stenhammar2013}%
  \BibitemOpen
  \bibfield  {author} {\bibinfo {author} {\bibfnamefont {Joakim}\ \bibnamefont
  {Stenhammar}}, \bibinfo {author} {\bibfnamefont {Adriano}\ \bibnamefont
  {Tiribocchi}}, \bibinfo {author} {\bibfnamefont {Rosalind~J.}\ \bibnamefont
  {Allen}}, \bibinfo {author} {\bibfnamefont {Davide}\ \bibnamefont
  {Marenduzzo}}, \ and\ \bibinfo {author} {\bibfnamefont {Michael~E.}\
  \bibnamefont {Cates}},\ }\bibfield  {title} {\enquote {\bibinfo {title}
  {Continuum theory of phase separation kinetics for active brownian
  particles},}\ }\href {\doibase 10.1103/PhysRevLett.111.145702} {\bibfield
  {journal} {\bibinfo  {journal} {Phys. Rev. Lett.}\ }\textbf {\bibinfo
  {volume} {111}},\ \bibinfo {pages} {145702} (\bibinfo {year}
  {2013})}\BibitemShut {NoStop}%
\bibitem [{\citenamefont {Buttinoni}\ \emph {et~al.}(2013)\citenamefont
  {Buttinoni}, \citenamefont {Bialk\'e}, \citenamefont {K\"ummel},
  \citenamefont {L\"owen}, \citenamefont {Bechinger},\ and\ \citenamefont
  {Speck}}]{Buttinoni2013}%
  \BibitemOpen
  \bibfield  {author} {\bibinfo {author} {\bibfnamefont {Ivo}\ \bibnamefont
  {Buttinoni}}, \bibinfo {author} {\bibfnamefont {Julian}\ \bibnamefont
  {Bialk\'e}}, \bibinfo {author} {\bibfnamefont {Felix}\ \bibnamefont
  {K\"ummel}}, \bibinfo {author} {\bibfnamefont {Hartmut}\ \bibnamefont
  {L\"owen}}, \bibinfo {author} {\bibfnamefont {Clemens}\ \bibnamefont
  {Bechinger}}, \ and\ \bibinfo {author} {\bibfnamefont {Thomas}\ \bibnamefont
  {Speck}},\ }\bibfield  {title} {\enquote {\bibinfo {title} {Dynamical
  clustering and phase separation in suspensions of self-propelled colloidal
  particles},}\ }\href {\doibase 10.1103/PhysRevLett.110.238301} {\bibfield
  {journal} {\bibinfo  {journal} {Phys. Rev. Lett.}\ }\textbf {\bibinfo
  {volume} {110}},\ \bibinfo {pages} {238301} (\bibinfo {year}
  {2013})}\BibitemShut {NoStop}%
\bibitem [{\citenamefont {Cates}\ and\ \citenamefont
  {Tailleur}(2015)}]{Cates2015}%
  \BibitemOpen
  \bibfield  {author} {\bibinfo {author} {\bibfnamefont {Michael~E.}\
  \bibnamefont {Cates}}\ and\ \bibinfo {author} {\bibfnamefont {Julien}\
  \bibnamefont {Tailleur}},\ }\bibfield  {title} {\enquote {\bibinfo {title}
  {Motility-induced phase separation},}\ }\href {\doibase
  10.1146/annurev-conmatphys-031214-014710} {\bibfield  {journal} {\bibinfo
  {journal} {Annu. Rev. Condens. Matter Phys.}\ }\textbf {\bibinfo {volume}
  {6}},\ \bibinfo {pages} {219--244} (\bibinfo {year} {2015})}\BibitemShut
  {NoStop}%
\bibitem [{\citenamefont {Wan}\ \emph {et~al.}(2008)\citenamefont {Wan},
  \citenamefont {Olson~Reichhardt}, \citenamefont {Nussinov},\ and\
  \citenamefont {Reichhardt}}]{Wan2008}%
  \BibitemOpen
  \bibfield  {author} {\bibinfo {author} {\bibfnamefont {M.~B.}\ \bibnamefont
  {Wan}}, \bibinfo {author} {\bibfnamefont {C.~J.}\ \bibnamefont
  {Olson~Reichhardt}}, \bibinfo {author} {\bibfnamefont {Z.}~\bibnamefont
  {Nussinov}}, \ and\ \bibinfo {author} {\bibfnamefont {C.}~\bibnamefont
  {Reichhardt}},\ }\bibfield  {title} {\enquote {\bibinfo {title}
  {Rectification of swimming bacteria and self-driven particle systems by
  arrays of asymmetric barriers},}\ }\href {\doibase
  10.1103/PhysRevLett.101.018102} {\bibfield  {journal} {\bibinfo  {journal}
  {Phys. Rev. Lett.}\ }\textbf {\bibinfo {volume} {101}},\ \bibinfo {pages}
  {018102} (\bibinfo {year} {2008})}\BibitemShut {NoStop}%
\bibitem [{\citenamefont {Tailleur}\ and\ \citenamefont
  {Cates}(2009)}]{Tailleur2009}%
  \BibitemOpen
  \bibfield  {author} {\bibinfo {author} {\bibfnamefont {J.}~\bibnamefont
  {Tailleur}}\ and\ \bibinfo {author} {\bibfnamefont {M.~E.}\ \bibnamefont
  {Cates}},\ }\bibfield  {title} {\enquote {\bibinfo {title} {Sedimentation,
  trapping, and rectification of dilute bacteria},}\ }\href
  {http://stacks.iop.org/0295-5075/86/i=6/a=60002} {\bibfield  {journal}
  {\bibinfo  {journal} {EPL (Europhysics Letters)}\ }\textbf {\bibinfo {volume}
  {86}},\ \bibinfo {pages} {60002} (\bibinfo {year} {2009})}\BibitemShut
  {NoStop}%
\bibitem [{\citenamefont {Angelani}\ \emph {et~al.}(2011)\citenamefont
  {Angelani}, \citenamefont {Maggi}, \citenamefont {Bernardini}, \citenamefont
  {Rizzo},\ and\ \citenamefont {Di~Leonardo}}]{Angelani2011}%
  \BibitemOpen
  \bibfield  {author} {\bibinfo {author} {\bibfnamefont {L.}~\bibnamefont
  {Angelani}}, \bibinfo {author} {\bibfnamefont {C.}~\bibnamefont {Maggi}},
  \bibinfo {author} {\bibfnamefont {M.~L.}\ \bibnamefont {Bernardini}},
  \bibinfo {author} {\bibfnamefont {A.}~\bibnamefont {Rizzo}}, \ and\ \bibinfo
  {author} {\bibfnamefont {R.}~\bibnamefont {Di~Leonardo}},\ }\bibfield
  {title} {\enquote {\bibinfo {title} {Effective interactions between colloidal
  particles suspended in a bath of swimming cells},}\ }\href {\doibase
  10.1103/PhysRevLett.107.138302} {\bibfield  {journal} {\bibinfo  {journal}
  {Phys. Rev. Lett.}\ }\textbf {\bibinfo {volume} {107}},\ \bibinfo {pages}
  {138302} (\bibinfo {year} {2011})}\BibitemShut {NoStop}%
\bibitem [{\citenamefont {Ghosh}\ \emph {et~al.}(2013)\citenamefont {Ghosh},
  \citenamefont {Misko}, \citenamefont {Marchesoni},\ and\ \citenamefont
  {Nori}}]{Ghosh2013}%
  \BibitemOpen
  \bibfield  {author} {\bibinfo {author} {\bibfnamefont {Pulak~K.}\
  \bibnamefont {Ghosh}}, \bibinfo {author} {\bibfnamefont {Vyacheslav~R.}\
  \bibnamefont {Misko}}, \bibinfo {author} {\bibfnamefont {Fabio}\ \bibnamefont
  {Marchesoni}}, \ and\ \bibinfo {author} {\bibfnamefont {Franco}\ \bibnamefont
  {Nori}},\ }\bibfield  {title} {\enquote {\bibinfo {title} {Self-propelled
  janus particles in a ratchet: Numerical simulations},}\ }\href {\doibase
  10.1103/PhysRevLett.110.268301} {\bibfield  {journal} {\bibinfo  {journal}
  {Phys. Rev. Lett.}\ }\textbf {\bibinfo {volume} {110}},\ \bibinfo {pages}
  {268301} (\bibinfo {year} {2013})}\BibitemShut {NoStop}%
\bibitem [{\citenamefont {Giomi}\ \emph {et~al.}(2012)\citenamefont {Giomi},
  \citenamefont {Mahadevan}, \citenamefont {Chakraborty},\ and\ \citenamefont
  {Hagan}}]{giomi2012banding}%
  \BibitemOpen
  \bibfield  {author} {\bibinfo {author} {\bibfnamefont {L}~\bibnamefont
  {Giomi}}, \bibinfo {author} {\bibfnamefont {L}~\bibnamefont {Mahadevan}},
  \bibinfo {author} {\bibfnamefont {B}~\bibnamefont {Chakraborty}}, \ and\
  \bibinfo {author} {\bibfnamefont {MF}~\bibnamefont {Hagan}},\ }\bibfield
  {title} {\enquote {\bibinfo {title} {Banding, excitability and chaos in
  active nematic suspensions},}\ }\href@noop {} {\bibfield  {journal} {\bibinfo
   {journal} {Nonlinearity}\ }\textbf {\bibinfo {volume} {25}},\ \bibinfo
  {pages} {2245} (\bibinfo {year} {2012})}\BibitemShut {NoStop}%
\bibitem [{\citenamefont {Marenduzzo}\ \emph {et~al.}(2007)\citenamefont
  {Marenduzzo}, \citenamefont {Orlandini},\ and\ \citenamefont
  {Yeomans}}]{Marenduzzo2007rheology}%
  \BibitemOpen
  \bibfield  {author} {\bibinfo {author} {\bibfnamefont {D.}~\bibnamefont
  {Marenduzzo}}, \bibinfo {author} {\bibfnamefont {E.}~\bibnamefont
  {Orlandini}}, \ and\ \bibinfo {author} {\bibfnamefont {J.~M.}\ \bibnamefont
  {Yeomans}},\ }\bibfield  {title} {\enquote {\bibinfo {title} {Hydrodynamics
  and rheology of active liquid crystals: A numerical investigation},}\ }\href
  {\doibase 10.1103/physrevlett.98.118102} {\bibfield  {journal} {\bibinfo
  {journal} {Phys. Rev. Lett.}\ }\textbf {\bibinfo {volume} {98}} (\bibinfo
  {year} {2007}),\ 10.1103/physrevlett.98.118102}\BibitemShut {NoStop}%
\bibitem [{\citenamefont {Giomi}\ \emph {et~al.}(2008)\citenamefont {Giomi},
  \citenamefont {Marchetti},\ and\ \citenamefont
  {Liverpool}}]{Giomi2008concentration}%
  \BibitemOpen
  \bibfield  {author} {\bibinfo {author} {\bibfnamefont {Luca}\ \bibnamefont
  {Giomi}}, \bibinfo {author} {\bibfnamefont {M.~Cristina}\ \bibnamefont
  {Marchetti}}, \ and\ \bibinfo {author} {\bibfnamefont {Tanniemola~B.}\
  \bibnamefont {Liverpool}},\ }\bibfield  {title} {\enquote {\bibinfo {title}
  {Complex spontaneous flows and concentration banding in active polar
  films},}\ }\href {\doibase 10.1103/physrevlett.101.198101} {\bibfield
  {journal} {\bibinfo  {journal} {Phys. Rev. Lett.}\ }\textbf {\bibinfo
  {volume} {101}} (\bibinfo {year} {2008}),\
  10.1103/physrevlett.101.198101}\BibitemShut {NoStop}%
\bibitem [{\citenamefont {Bricard}\ \emph {et~al.}(2013)\citenamefont
  {Bricard}, \citenamefont {Caussin}, \citenamefont {Desreumaux}, \citenamefont
  {Dauchot},\ and\ \citenamefont {Bartolo}}]{Bartolo2013}%
  \BibitemOpen
  \bibfield  {author} {\bibinfo {author} {\bibfnamefont {Antoine}\ \bibnamefont
  {Bricard}}, \bibinfo {author} {\bibfnamefont {Jean-Baptiste}\ \bibnamefont
  {Caussin}}, \bibinfo {author} {\bibfnamefont {Nicolas}\ \bibnamefont
  {Desreumaux}}, \bibinfo {author} {\bibfnamefont {Olivier}\ \bibnamefont
  {Dauchot}}, \ and\ \bibinfo {author} {\bibfnamefont {Denis}\ \bibnamefont
  {Bartolo}},\ }\bibfield  {title} {\enquote {\bibinfo {title} {Emergence of
  macroscopic directed motion in populations of motile colloids},}\ }\href
  {\doibase 10.1038/nature12673} {\bibfield  {journal} {\bibinfo  {journal}
  {Nature}\ }\textbf {\bibinfo {volume} {503}},\ \bibinfo {pages} {95--98}
  (\bibinfo {year} {2013})}\BibitemShut {NoStop}%
\bibitem [{\citenamefont {Baek}\ \emph {et~al.}(2018)\citenamefont {Baek},
  \citenamefont {Solon}, \citenamefont {Xu}, \citenamefont {Nikola},\ and\
  \citenamefont {Kafri}}]{Baek2018}%
  \BibitemOpen
  \bibfield  {author} {\bibinfo {author} {\bibfnamefont {Yongjoo}\ \bibnamefont
  {Baek}}, \bibinfo {author} {\bibfnamefont {Alexandre~P.}\ \bibnamefont
  {Solon}}, \bibinfo {author} {\bibfnamefont {Xinpeng}\ \bibnamefont {Xu}},
  \bibinfo {author} {\bibfnamefont {Nikolai}\ \bibnamefont {Nikola}}, \ and\
  \bibinfo {author} {\bibfnamefont {Yariv}\ \bibnamefont {Kafri}},\ }\bibfield
  {title} {\enquote {\bibinfo {title} {Generic long-range interactions between
  passive bodies in an active fluid},}\ }\href {\doibase
  10.1103/PhysRevLett.120.058002} {\bibfield  {journal} {\bibinfo  {journal}
  {Phys. Rev. Lett.}\ }\textbf {\bibinfo {volume} {120}},\ \bibinfo {pages}
  {058002} (\bibinfo {year} {2018})}\BibitemShut {NoStop}%
\bibitem [{\citenamefont {Dombrowski}\ \emph {et~al.}(2004)\citenamefont
  {Dombrowski}, \citenamefont {Cisneros}, \citenamefont {Chatkaew},
  \citenamefont {Goldstein},\ and\ \citenamefont {Kessler}}]{Dombrowski2004}%
  \BibitemOpen
  \bibfield  {author} {\bibinfo {author} {\bibfnamefont {Christopher}\
  \bibnamefont {Dombrowski}}, \bibinfo {author} {\bibfnamefont {Luis}\
  \bibnamefont {Cisneros}}, \bibinfo {author} {\bibfnamefont {Sunita}\
  \bibnamefont {Chatkaew}}, \bibinfo {author} {\bibfnamefont {Raymond~E.}\
  \bibnamefont {Goldstein}}, \ and\ \bibinfo {author} {\bibfnamefont {John~O.}\
  \bibnamefont {Kessler}},\ }\bibfield  {title} {\enquote {\bibinfo {title}
  {Self-concentration and large-scale coherence in bacterial dynamics},}\
  }\href {\doibase 10.1103/PhysRevLett.93.098103} {\bibfield  {journal}
  {\bibinfo  {journal} {Phys. Rev. Lett.}\ }\textbf {\bibinfo {volume} {93}},\
  \bibinfo {pages} {098103} (\bibinfo {year} {2004})}\BibitemShut {NoStop}%
\bibitem [{\citenamefont {Wensink}\ \emph
  {et~al.}(2012{\natexlab{a}})\citenamefont {Wensink}, \citenamefont {Dunkel},
  \citenamefont {Heidenreich}, \citenamefont {Drescher}, \citenamefont
  {Goldstein}, \citenamefont {Lowen},\ and\ \citenamefont
  {Yeomans}}]{Wensink2012}%
  \BibitemOpen
  \bibfield  {author} {\bibinfo {author} {\bibfnamefont {H.~H.}\ \bibnamefont
  {Wensink}}, \bibinfo {author} {\bibfnamefont {J.}~\bibnamefont {Dunkel}},
  \bibinfo {author} {\bibfnamefont {S.}~\bibnamefont {Heidenreich}}, \bibinfo
  {author} {\bibfnamefont {K.}~\bibnamefont {Drescher}}, \bibinfo {author}
  {\bibfnamefont {R.~E.}\ \bibnamefont {Goldstein}}, \bibinfo {author}
  {\bibfnamefont {H.}~\bibnamefont {Lowen}}, \ and\ \bibinfo {author}
  {\bibfnamefont {J.~M.}\ \bibnamefont {Yeomans}},\ }\bibfield  {title}
  {\enquote {\bibinfo {title} {Meso-scale turbulence in living fluids},}\
  }\href {\doibase 10.1073/pnas.1202032109} {\bibfield  {journal} {\bibinfo
  {journal} {PNAS}\ }\textbf {\bibinfo {volume} {109}},\ \bibinfo {pages}
  {14308--14313} (\bibinfo {year} {2012}{\natexlab{a}})}\BibitemShut {NoStop}%
\bibitem [{\citenamefont {Doostmohammadi}\ \emph {et~al.}(2018)\citenamefont
  {Doostmohammadi}, \citenamefont {Ign{\'e}s-Mullol}, \citenamefont {Yeomans},\
  and\ \citenamefont {Sagu{\'e}s}}]{doostmohammadi2018active}%
  \BibitemOpen
  \bibfield  {author} {\bibinfo {author} {\bibfnamefont {Amin}\ \bibnamefont
  {Doostmohammadi}}, \bibinfo {author} {\bibfnamefont {Jordi}\ \bibnamefont
  {Ign{\'e}s-Mullol}}, \bibinfo {author} {\bibfnamefont {Julia~M}\ \bibnamefont
  {Yeomans}}, \ and\ \bibinfo {author} {\bibfnamefont {Francesc}\ \bibnamefont
  {Sagu{\'e}s}},\ }\bibfield  {title} {\enquote {\bibinfo {title} {Active
  nematics},}\ }\href@noop {} {\bibfield  {journal} {\bibinfo  {journal} {Nat.
  Commun.}\ }\textbf {\bibinfo {volume} {9}},\ \bibinfo {pages} {1--13}
  (\bibinfo {year} {2018})}\BibitemShut {NoStop}%
\bibitem [{\citenamefont {Blanch-Mercader}\ \emph {et~al.}(2018)\citenamefont
  {Blanch-Mercader}, \citenamefont {Yashunsky}, \citenamefont {Garcia},
  \citenamefont {Duclos}, \citenamefont {Giomi},\ and\ \citenamefont
  {Silberzan}}]{BlanchMercader2018}%
  \BibitemOpen
  \bibfield  {author} {\bibinfo {author} {\bibfnamefont {C.}~\bibnamefont
  {Blanch-Mercader}}, \bibinfo {author} {\bibfnamefont {V.}~\bibnamefont
  {Yashunsky}}, \bibinfo {author} {\bibfnamefont {S.}~\bibnamefont {Garcia}},
  \bibinfo {author} {\bibfnamefont {G.}~\bibnamefont {Duclos}}, \bibinfo
  {author} {\bibfnamefont {L.}~\bibnamefont {Giomi}}, \ and\ \bibinfo {author}
  {\bibfnamefont {P.}~\bibnamefont {Silberzan}},\ }\bibfield  {title} {\enquote
  {\bibinfo {title} {Turbulent dynamics of epithelial cell cultures},}\ }\href
  {\doibase 10.1103/physrevlett.120.208101} {\bibfield  {journal} {\bibinfo
  {journal} {Phys. Rev. Lett.}\ }\textbf {\bibinfo {volume} {120}} (\bibinfo
  {year} {2018}),\ 10.1103/physrevlett.120.208101}\BibitemShut {NoStop}%
\bibitem [{\citenamefont {Alert}\ \emph {et~al.}(2021)\citenamefont {Alert},
  \citenamefont {Casademunt},\ and\ \citenamefont {Joanny}}]{alert2021active}%
  \BibitemOpen
  \bibfield  {author} {\bibinfo {author} {\bibfnamefont {Ricard}\ \bibnamefont
  {Alert}}, \bibinfo {author} {\bibfnamefont {Jaume}\ \bibnamefont
  {Casademunt}}, \ and\ \bibinfo {author} {\bibfnamefont {Jean-Fran{\c{c}}ois}\
  \bibnamefont {Joanny}},\ }\bibfield  {title} {\enquote {\bibinfo {title}
  {Active turbulence},}\ }\href@noop {} {\bibfield  {journal} {\bibinfo
  {journal} {arXiv preprint arXiv:2104.02122}\ } (\bibinfo {year}
  {2021})}\BibitemShut {NoStop}%
\bibitem [{\citenamefont {Marchetti}\ \emph {et~al.}(2013)\citenamefont
  {Marchetti}, \citenamefont {Joanny}, \citenamefont {Ramaswamy}, \citenamefont
  {Liverpool}, \citenamefont {Prost}, \citenamefont {Rao},\ and\ \citenamefont
  {Simha}}]{marchetti2013hydrodynamics}%
  \BibitemOpen
  \bibfield  {author} {\bibinfo {author} {\bibfnamefont {M~Cristina}\
  \bibnamefont {Marchetti}}, \bibinfo {author} {\bibfnamefont
  {Jean-Fran{\c{c}}ois}\ \bibnamefont {Joanny}}, \bibinfo {author}
  {\bibfnamefont {Sriram}\ \bibnamefont {Ramaswamy}}, \bibinfo {author}
  {\bibfnamefont {Tanniemola~B}\ \bibnamefont {Liverpool}}, \bibinfo {author}
  {\bibfnamefont {Jacques}\ \bibnamefont {Prost}}, \bibinfo {author}
  {\bibfnamefont {Madan}\ \bibnamefont {Rao}}, \ and\ \bibinfo {author}
  {\bibfnamefont {R~Aditi}\ \bibnamefont {Simha}},\ }\bibfield  {title}
  {\enquote {\bibinfo {title} {Hydrodynamics of soft active matter},}\
  }\href@noop {} {\bibfield  {journal} {\bibinfo  {journal} {Rev. Mod. Phys.,}\
  }\textbf {\bibinfo {volume} {85}},\ \bibinfo {pages} {1143} (\bibinfo {year}
  {2013})}\BibitemShut {NoStop}%
\bibitem [{\citenamefont {Yaman}\ \emph {et~al.}(2019)\citenamefont {Yaman},
  \citenamefont {Demir}, \citenamefont {Vetter},\ and\ \citenamefont
  {Kocabas}}]{Yaman2019}%
  \BibitemOpen
  \bibfield  {author} {\bibinfo {author} {\bibfnamefont {Yusuf~Ilker}\
  \bibnamefont {Yaman}}, \bibinfo {author} {\bibfnamefont {Esin}\ \bibnamefont
  {Demir}}, \bibinfo {author} {\bibfnamefont {Roman}\ \bibnamefont {Vetter}}, \
  and\ \bibinfo {author} {\bibfnamefont {Askin}\ \bibnamefont {Kocabas}},\
  }\bibfield  {title} {\enquote {\bibinfo {title} {Emergence of active nematics
  in chaining bacterial biofilms},}\ }\href {\doibase
  10.1038/s41467-019-10311-z} {\bibfield  {journal} {\bibinfo  {journal} {Nat.
  Commun.}\ }\textbf {\bibinfo {volume} {10}} (\bibinfo {year} {2019}),\
  10.1038/s41467-019-10311-z}\BibitemShut {NoStop}%
\bibitem [{\citenamefont {Dell'Arciprete}\ \emph {et~al.}(2018)\citenamefont
  {Dell'Arciprete}, \citenamefont {Blow}, \citenamefont {Brown}, \citenamefont
  {Farrell}, \citenamefont {Lintuvuori}, \citenamefont {McVey}, \citenamefont
  {Marenduzzo},\ and\ \citenamefont {Poon}}]{DellArciprete2018}%
  \BibitemOpen
  \bibfield  {author} {\bibinfo {author} {\bibfnamefont {D.}~\bibnamefont
  {Dell'Arciprete}}, \bibinfo {author} {\bibfnamefont {M.~L.}\ \bibnamefont
  {Blow}}, \bibinfo {author} {\bibfnamefont {A.~T.}\ \bibnamefont {Brown}},
  \bibinfo {author} {\bibfnamefont {F.~D.~C.}\ \bibnamefont {Farrell}},
  \bibinfo {author} {\bibfnamefont {J.~S.}\ \bibnamefont {Lintuvuori}},
  \bibinfo {author} {\bibfnamefont {A.~F.}\ \bibnamefont {McVey}}, \bibinfo
  {author} {\bibfnamefont {D.}~\bibnamefont {Marenduzzo}}, \ and\ \bibinfo
  {author} {\bibfnamefont {W.~C.~K.}\ \bibnamefont {Poon}},\ }\bibfield
  {title} {\enquote {\bibinfo {title} {A growing bacterial colony in two
  dimensions as an active nematic},}\ }\href {\doibase
  10.1038/s41467-018-06370-3} {\bibfield  {journal} {\bibinfo  {journal} {Nat.
  Commun.}\ }\textbf {\bibinfo {volume} {9}} (\bibinfo {year} {2018}),\
  10.1038/s41467-018-06370-3}\BibitemShut {NoStop}%
\bibitem [{\citenamefont {Keber}\ \emph {et~al.}(2014)\citenamefont {Keber},
  \citenamefont {Loiseau}, \citenamefont {Sanchez}, \citenamefont {DeCamp},
  \citenamefont {Giomi}, \citenamefont {Bowick}, \citenamefont {Marchetti},
  \citenamefont {Dogic},\ and\ \citenamefont {Bausch}}]{Keber2014}%
  \BibitemOpen
  \bibfield  {author} {\bibinfo {author} {\bibfnamefont {F.~C.}\ \bibnamefont
  {Keber}}, \bibinfo {author} {\bibfnamefont {E.}~\bibnamefont {Loiseau}},
  \bibinfo {author} {\bibfnamefont {T.}~\bibnamefont {Sanchez}}, \bibinfo
  {author} {\bibfnamefont {S.~J.}\ \bibnamefont {DeCamp}}, \bibinfo {author}
  {\bibfnamefont {L.}~\bibnamefont {Giomi}}, \bibinfo {author} {\bibfnamefont
  {M.~J.}\ \bibnamefont {Bowick}}, \bibinfo {author} {\bibfnamefont {M.~C.}\
  \bibnamefont {Marchetti}}, \bibinfo {author} {\bibfnamefont {Z.}~\bibnamefont
  {Dogic}}, \ and\ \bibinfo {author} {\bibfnamefont {A.~R.}\ \bibnamefont
  {Bausch}},\ }\bibfield  {title} {\enquote {\bibinfo {title} {Topology and
  dynamics of active nematic vesicles},}\ }\href {\doibase
  10.1126/science.1254784} {\bibfield  {journal} {\bibinfo  {journal}
  {Science}\ }\textbf {\bibinfo {volume} {345}},\ \bibinfo {pages} {1135--1139}
  (\bibinfo {year} {2014})}\BibitemShut {NoStop}%
\bibitem [{\citenamefont {Opathalage}\ \emph {et~al.}(2019)\citenamefont
  {Opathalage}, \citenamefont {Norton}, \citenamefont {Juniper}, \citenamefont
  {Langeslay}, \citenamefont {Aghvami}, \citenamefont {Fraden},\ and\
  \citenamefont {Dogic}}]{opathalage2019self}%
  \BibitemOpen
  \bibfield  {author} {\bibinfo {author} {\bibfnamefont {Achini}\ \bibnamefont
  {Opathalage}}, \bibinfo {author} {\bibfnamefont {Michael~M}\ \bibnamefont
  {Norton}}, \bibinfo {author} {\bibfnamefont {Michael~PN}\ \bibnamefont
  {Juniper}}, \bibinfo {author} {\bibfnamefont {Blake}\ \bibnamefont
  {Langeslay}}, \bibinfo {author} {\bibfnamefont {S~Ali}\ \bibnamefont
  {Aghvami}}, \bibinfo {author} {\bibfnamefont {Seth}\ \bibnamefont {Fraden}},
  \ and\ \bibinfo {author} {\bibfnamefont {Zvonimir}\ \bibnamefont {Dogic}},\
  }\bibfield  {title} {\enquote {\bibinfo {title} {Self-organized dynamics and
  the transition to turbulence of confined active nematics},}\ }\href@noop {}
  {\bibfield  {journal} {\bibinfo  {journal} {PNAS}\ }\textbf {\bibinfo
  {volume} {116}},\ \bibinfo {pages} {4788--4797} (\bibinfo {year}
  {2019})}\BibitemShut {NoStop}%
\bibitem [{\citenamefont {Wu}\ \emph {et~al.}(2017)\citenamefont {Wu},
  \citenamefont {Hishamunda}, \citenamefont {Chen}, \citenamefont {DeCamp},
  \citenamefont {Chang}, \citenamefont {Fern{\'a}ndez-Nieves}, \citenamefont
  {Fraden},\ and\ \citenamefont {Dogic}}]{Wu2017transition}%
  \BibitemOpen
  \bibfield  {author} {\bibinfo {author} {\bibfnamefont {Kun-Ta}\ \bibnamefont
  {Wu}}, \bibinfo {author} {\bibfnamefont {Jean~Bernard}\ \bibnamefont
  {Hishamunda}}, \bibinfo {author} {\bibfnamefont {Daniel~TN}\ \bibnamefont
  {Chen}}, \bibinfo {author} {\bibfnamefont {Stephen~J}\ \bibnamefont
  {DeCamp}}, \bibinfo {author} {\bibfnamefont {Ya-Wen}\ \bibnamefont {Chang}},
  \bibinfo {author} {\bibfnamefont {Alberto}\ \bibnamefont
  {Fern{\'a}ndez-Nieves}}, \bibinfo {author} {\bibfnamefont {Seth}\
  \bibnamefont {Fraden}}, \ and\ \bibinfo {author} {\bibfnamefont {Zvonimir}\
  \bibnamefont {Dogic}},\ }\bibfield  {title} {\enquote {\bibinfo {title}
  {Transition from turbulent to coherent flows in confined three-dimensional
  active fluids},}\ }\href@noop {} {\bibfield  {journal} {\bibinfo  {journal}
  {Science}\ }\textbf {\bibinfo {volume} {355}},\ \bibinfo {pages} {eaal1979}
  (\bibinfo {year} {2017})}\BibitemShut {NoStop}%
\bibitem [{\citenamefont {Shendruk}\ \emph {et~al.}(2017)\citenamefont
  {Shendruk}, \citenamefont {Doostmohammadi}, \citenamefont {Thijssen},\ and\
  \citenamefont {Yeomans}}]{shendruk2017dancing}%
  \BibitemOpen
  \bibfield  {author} {\bibinfo {author} {\bibfnamefont {Tyler~N}\ \bibnamefont
  {Shendruk}}, \bibinfo {author} {\bibfnamefont {Amin}\ \bibnamefont
  {Doostmohammadi}}, \bibinfo {author} {\bibfnamefont {Kristian}\ \bibnamefont
  {Thijssen}}, \ and\ \bibinfo {author} {\bibfnamefont {Julia~M}\ \bibnamefont
  {Yeomans}},\ }\bibfield  {title} {\enquote {\bibinfo {title} {Dancing
  disclinations in confined active nematics},}\ }\href@noop {} {\bibfield
  {journal} {\bibinfo  {journal} {Soft Matter}\ }\textbf {\bibinfo {volume}
  {13}},\ \bibinfo {pages} {3853--3862} (\bibinfo {year} {2017})}\BibitemShut
  {NoStop}%
\bibitem [{\citenamefont {Duclos}\ \emph {et~al.}(2020)\citenamefont {Duclos},
  \citenamefont {Adkins}, \citenamefont {Banerjee}, \citenamefont {Peterson},
  \citenamefont {Varghese}, \citenamefont {Kolvin}, \citenamefont {Baskaran},
  \citenamefont {Pelcovits}, \citenamefont {Powers}, \citenamefont {Baskaran},
  \citenamefont {Toschi}, \citenamefont {Hagan}, \citenamefont {Streichan},
  \citenamefont {Vitelli}, \citenamefont {Beller},\ and\ \citenamefont
  {Dogic}}]{duclos2020topological}%
  \BibitemOpen
  \bibfield  {author} {\bibinfo {author} {\bibfnamefont {Guillaume}\
  \bibnamefont {Duclos}}, \bibinfo {author} {\bibfnamefont {Raymond}\
  \bibnamefont {Adkins}}, \bibinfo {author} {\bibfnamefont {Debarghya}\
  \bibnamefont {Banerjee}}, \bibinfo {author} {\bibfnamefont {Matthew~SE}\
  \bibnamefont {Peterson}}, \bibinfo {author} {\bibfnamefont {Minu}\
  \bibnamefont {Varghese}}, \bibinfo {author} {\bibfnamefont {Itamar}\
  \bibnamefont {Kolvin}}, \bibinfo {author} {\bibfnamefont {Arvind}\
  \bibnamefont {Baskaran}}, \bibinfo {author} {\bibfnamefont {Robert~A}\
  \bibnamefont {Pelcovits}}, \bibinfo {author} {\bibfnamefont {Thomas~R}\
  \bibnamefont {Powers}}, \bibinfo {author} {\bibfnamefont {Aparna}\
  \bibnamefont {Baskaran}}, \bibinfo {author} {\bibfnamefont {Federico}\
  \bibnamefont {Toschi}}, \bibinfo {author} {\bibfnamefont {Michael~F}\
  \bibnamefont {Hagan}}, \bibinfo {author} {\bibfnamefont {Sebastian~J}\
  \bibnamefont {Streichan}}, \bibinfo {author} {\bibfnamefont {Vincenzo}\
  \bibnamefont {Vitelli}}, \bibinfo {author} {\bibfnamefont {Daniel~A}\
  \bibnamefont {Beller}}, \ and\ \bibinfo {author} {\bibfnamefont {Zvonimir}\
  \bibnamefont {Dogic}},\ }\bibfield  {title} {\enquote {\bibinfo {title}
  {Topological structure and dynamics of three-dimensional active nematics},}\
  }\href@noop {} {\bibfield  {journal} {\bibinfo  {journal} {Science}\ }\textbf
  {\bibinfo {volume} {367}},\ \bibinfo {pages} {1120--1124} (\bibinfo {year}
  {2020})}\BibitemShut {NoStop}%
\bibitem [{\citenamefont {Ross}\ \emph {et~al.}(2019)\citenamefont {Ross},
  \citenamefont {Lee}, \citenamefont {Qu}, \citenamefont {Banks}, \citenamefont
  {Phillips},\ and\ \citenamefont {Thomson}}]{ross2019controlling}%
  \BibitemOpen
  \bibfield  {author} {\bibinfo {author} {\bibfnamefont {Tyler~D}\ \bibnamefont
  {Ross}}, \bibinfo {author} {\bibfnamefont {Heun~Jin}\ \bibnamefont {Lee}},
  \bibinfo {author} {\bibfnamefont {Zijie}\ \bibnamefont {Qu}}, \bibinfo
  {author} {\bibfnamefont {Rachel~A}\ \bibnamefont {Banks}}, \bibinfo {author}
  {\bibfnamefont {Rob}\ \bibnamefont {Phillips}}, \ and\ \bibinfo {author}
  {\bibfnamefont {Matt}\ \bibnamefont {Thomson}},\ }\bibfield  {title}
  {\enquote {\bibinfo {title} {Controlling organization and forces in active
  matter through optically defined boundaries},}\ }\href@noop {} {\bibfield
  {journal} {\bibinfo  {journal} {Nature}\ }\textbf {\bibinfo {volume} {572}},\
  \bibinfo {pages} {224--229} (\bibinfo {year} {2019})}\BibitemShut {NoStop}%
\bibitem [{\citenamefont {Zhang}\ \emph {et~al.}(2019)\citenamefont {Zhang},
  \citenamefont {Redford}, \citenamefont {Ruijgrok}, \citenamefont {Kumar},
  \citenamefont {Mozaffari}, \citenamefont {Zemsky}, \citenamefont {Dinner},
  \citenamefont {Vitelli}, \citenamefont {Bryant}, \citenamefont {Gardel},\
  and\ \citenamefont {de~Pablo}}]{zhang2019structuring}%
  \BibitemOpen
  \bibfield  {author} {\bibinfo {author} {\bibfnamefont {Rui}\ \bibnamefont
  {Zhang}}, \bibinfo {author} {\bibfnamefont {Steven~A}\ \bibnamefont
  {Redford}}, \bibinfo {author} {\bibfnamefont {Paul~V}\ \bibnamefont
  {Ruijgrok}}, \bibinfo {author} {\bibfnamefont {Nitin}\ \bibnamefont {Kumar}},
  \bibinfo {author} {\bibfnamefont {Ali}\ \bibnamefont {Mozaffari}}, \bibinfo
  {author} {\bibfnamefont {Sasha}\ \bibnamefont {Zemsky}}, \bibinfo {author}
  {\bibfnamefont {Aaron~R}\ \bibnamefont {Dinner}}, \bibinfo {author}
  {\bibfnamefont {Vincenzo}\ \bibnamefont {Vitelli}}, \bibinfo {author}
  {\bibfnamefont {Zev}\ \bibnamefont {Bryant}}, \bibinfo {author}
  {\bibfnamefont {Margaret~L}\ \bibnamefont {Gardel}}, \ and\ \bibinfo {author}
  {\bibfnamefont {Juan~J}\ \bibnamefont {de~Pablo}},\ }\bibfield  {title}
  {\enquote {\bibinfo {title} {Structuring stress for active materials
  control},}\ }\href@noop {} {\bibfield  {journal} {\bibinfo  {journal} {arXiv
  preprint arXiv:1912.01630}\ } (\bibinfo {year} {2019})}\BibitemShut {NoStop}%
\bibitem [{\citenamefont {Norton}\ \emph {et~al.}(2020)\citenamefont {Norton},
  \citenamefont {Grover}, \citenamefont {Hagan},\ and\ \citenamefont
  {Fraden}}]{norton2020optimal}%
  \BibitemOpen
  \bibfield  {author} {\bibinfo {author} {\bibfnamefont {Michael~M}\
  \bibnamefont {Norton}}, \bibinfo {author} {\bibfnamefont {Piyush}\
  \bibnamefont {Grover}}, \bibinfo {author} {\bibfnamefont {Michael~F}\
  \bibnamefont {Hagan}}, \ and\ \bibinfo {author} {\bibfnamefont {Seth}\
  \bibnamefont {Fraden}},\ }\bibfield  {title} {\enquote {\bibinfo {title}
  {Optimal control of active nematics},}\ }\href@noop {} {\bibfield  {journal}
  {\bibinfo  {journal} {Phys. Rev. Lett.}\ }\textbf {\bibinfo {volume} {125}},\
  \bibinfo {pages} {178005} (\bibinfo {year} {2020})}\BibitemShut {NoStop}%
\bibitem [{\citenamefont {Bowick}\ \emph {et~al.}(2021)\citenamefont {Bowick},
  \citenamefont {Fakhri}, \citenamefont {Marchetti},\ and\ \citenamefont
  {Ramaswamy}}]{bowick2021symmetry}%
  \BibitemOpen
  \bibfield  {author} {\bibinfo {author} {\bibfnamefont {Mark~J}\ \bibnamefont
  {Bowick}}, \bibinfo {author} {\bibfnamefont {Nikta}\ \bibnamefont {Fakhri}},
  \bibinfo {author} {\bibfnamefont {M~Cristina}\ \bibnamefont {Marchetti}}, \
  and\ \bibinfo {author} {\bibfnamefont {Sriram}\ \bibnamefont {Ramaswamy}},\
  }\bibfield  {title} {\enquote {\bibinfo {title} {Symmetry, thermodynamics and
  topology in active matter},}\ }\href@noop {} {\bibfield  {journal} {\bibinfo
  {journal} {arXiv preprint arXiv:2107.00724}\ } (\bibinfo {year}
  {2021})}\BibitemShut {NoStop}%
\bibitem [{\citenamefont {Graham}\ and\ \citenamefont
  {Floryan}(2020)}]{graham2020exact}%
  \BibitemOpen
  \bibfield  {author} {\bibinfo {author} {\bibfnamefont {Michael~D}\
  \bibnamefont {Graham}}\ and\ \bibinfo {author} {\bibfnamefont {Daniel}\
  \bibnamefont {Floryan}},\ }\bibfield  {title} {\enquote {\bibinfo {title}
  {Exact coherent states and the nonlinear dynamics of wall-bounded turbulent
  flows},}\ }\href@noop {} {\bibfield  {journal} {\bibinfo  {journal} {Annu.
  Rev. Fluid Mech.}\ }\textbf {\bibinfo {volume} {53}} (\bibinfo {year}
  {2020})}\BibitemShut {NoStop}%
\bibitem [{\citenamefont {Hopf}(1948)}]{hopf1948mathematical}%
  \BibitemOpen
  \bibfield  {author} {\bibinfo {author} {\bibfnamefont {Eberhard}\
  \bibnamefont {Hopf}},\ }\bibfield  {title} {\enquote {\bibinfo {title} {A
  mathematical example displaying features of turbulence},}\ }\href@noop {}
  {\bibfield  {journal} {\bibinfo  {journal} {Commun. Pure Appl. Math.}\
  }\textbf {\bibinfo {volume} {1}},\ \bibinfo {pages} {303--322} (\bibinfo
  {year} {1948})}\BibitemShut {NoStop}%
\bibitem [{\citenamefont {Ruelle}\ and\ \citenamefont
  {Takens}(1971)}]{ruelle1971nature}%
  \BibitemOpen
  \bibfield  {author} {\bibinfo {author} {\bibfnamefont {David}\ \bibnamefont
  {Ruelle}}\ and\ \bibinfo {author} {\bibfnamefont {Floris}\ \bibnamefont
  {Takens}},\ }\bibfield  {title} {\enquote {\bibinfo {title} {On the nature of
  turbulence},}\ }\href@noop {} {\bibfield  {journal} {\bibinfo  {journal} {Les
  rencontres physiciens-math{\'e}maticiens de Strasbourg-RCP25}\ }\textbf
  {\bibinfo {volume} {12}},\ \bibinfo {pages} {1--44} (\bibinfo {year}
  {1971})}\BibitemShut {NoStop}%
\bibitem [{\citenamefont {Cvitanovi{\'c}}(2013)}]{cvitanovic2013recurrent}%
  \BibitemOpen
  \bibfield  {author} {\bibinfo {author} {\bibfnamefont {Predrag}\ \bibnamefont
  {Cvitanovi{\'c}}},\ }\bibfield  {title} {\enquote {\bibinfo {title}
  {Recurrent flows: the clockwork behind turbulence},}\ }\href@noop {}
  {\bibfield  {journal} {\bibinfo  {journal} {J. Fluid. Mech.}\ }\textbf
  {\bibinfo {volume} {726}},\ \bibinfo {pages} {1--4} (\bibinfo {year}
  {2013})}\BibitemShut {NoStop}%
\bibitem [{\citenamefont {Park}\ and\ \citenamefont
  {Graham}(2015)}]{park2015exact}%
  \BibitemOpen
  \bibfield  {author} {\bibinfo {author} {\bibfnamefont {Jae~Sung}\
  \bibnamefont {Park}}\ and\ \bibinfo {author} {\bibfnamefont {Michael~D}\
  \bibnamefont {Graham}},\ }\bibfield  {title} {\enquote {\bibinfo {title}
  {Exact coherent states and connections to turbulent dynamics in minimal
  channel flow},}\ }\href@noop {} {\bibfield  {journal} {\bibinfo  {journal}
  {J. Fluid. Mech.}\ }\textbf {\bibinfo {volume} {782}},\ \bibinfo {pages}
  {430--454} (\bibinfo {year} {2015})}\BibitemShut {NoStop}%
\bibitem [{\citenamefont {Budanur}\ \emph {et~al.}(2019)\citenamefont
  {Budanur}, \citenamefont {Dogra},\ and\ \citenamefont
  {Hof}}]{budanur2019geometry}%
  \BibitemOpen
  \bibfield  {author} {\bibinfo {author} {\bibfnamefont {Nazmi~Burak}\
  \bibnamefont {Budanur}}, \bibinfo {author} {\bibfnamefont {Akshunna~Shaurya}\
  \bibnamefont {Dogra}}, \ and\ \bibinfo {author} {\bibfnamefont {Bj{\"o}rn}\
  \bibnamefont {Hof}},\ }\bibfield  {title} {\enquote {\bibinfo {title}
  {Geometry of transient chaos in streamwise-localized pipe flow turbulence},}\
  }\href@noop {} {\bibfield  {journal} {\bibinfo  {journal} {Phys. Rev.
  Fluids}\ }\textbf {\bibinfo {volume} {4}},\ \bibinfo {pages} {102401}
  (\bibinfo {year} {2019})}\BibitemShut {NoStop}%
\bibitem [{\citenamefont {Suri}\ \emph {et~al.}(2020)\citenamefont {Suri},
  \citenamefont {Kageorge}, \citenamefont {Grigoriev},\ and\ \citenamefont
  {Schatz}}]{suri2020capturing}%
  \BibitemOpen
  \bibfield  {author} {\bibinfo {author} {\bibfnamefont {Balachandra}\
  \bibnamefont {Suri}}, \bibinfo {author} {\bibfnamefont {Logan}\ \bibnamefont
  {Kageorge}}, \bibinfo {author} {\bibfnamefont {Roman~O}\ \bibnamefont
  {Grigoriev}}, \ and\ \bibinfo {author} {\bibfnamefont {Michael~F}\
  \bibnamefont {Schatz}},\ }\bibfield  {title} {\enquote {\bibinfo {title}
  {Capturing turbulent dynamics and statistics in experiments with unstable
  periodic orbits},}\ }\href@noop {} {\bibfield  {journal} {\bibinfo  {journal}
  {Phys. Rev. Lett.}\ }\textbf {\bibinfo {volume} {125}},\ \bibinfo {pages}
  {064501} (\bibinfo {year} {2020})}\BibitemShut {NoStop}%
\bibitem [{\citenamefont {Davis}\ and\ \citenamefont
  {Park}(2020)}]{davis2020dynamics}%
  \BibitemOpen
  \bibfield  {author} {\bibinfo {author} {\bibfnamefont {Ethan~A}\ \bibnamefont
  {Davis}}\ and\ \bibinfo {author} {\bibfnamefont {Jae~Sung}\ \bibnamefont
  {Park}},\ }\bibfield  {title} {\enquote {\bibinfo {title} {Dynamics of
  laminar and transitional flows over slip surfaces: effects on the
  laminar--turbulent separatrix},}\ }\href@noop {} {\bibfield  {journal}
  {\bibinfo  {journal} {J. Fluid. Mech.}\ }\textbf {\bibinfo {volume} {894}}
  (\bibinfo {year} {2020})}\BibitemShut {NoStop}%
\bibitem [{\citenamefont {Linkmann}\ \emph
  {et~al.}(2020{\natexlab{a}})\citenamefont {Linkmann}, \citenamefont
  {Knierim}, \citenamefont {Zammert},\ and\ \citenamefont
  {Eckhardt}}]{linkmann2020linear}%
  \BibitemOpen
  \bibfield  {author} {\bibinfo {author} {\bibfnamefont {Moritz}\ \bibnamefont
  {Linkmann}}, \bibinfo {author} {\bibfnamefont {Florian}\ \bibnamefont
  {Knierim}}, \bibinfo {author} {\bibfnamefont {Stefan}\ \bibnamefont
  {Zammert}}, \ and\ \bibinfo {author} {\bibfnamefont {Bruno}\ \bibnamefont
  {Eckhardt}},\ }\bibfield  {title} {\enquote {\bibinfo {title} {Linear
  feedback control of invariant solutions in channel flow},}\ }\href@noop {}
  {\bibfield  {journal} {\bibinfo  {journal} {J. Fluid. Mech.}\ }\textbf
  {\bibinfo {volume} {900}} (\bibinfo {year} {2020}{\natexlab{a}})}\BibitemShut
  {NoStop}%
\bibitem [{\citenamefont {Lucas}(2020)}]{lucas2020stabilisation}%
  \BibitemOpen
  \bibfield  {author} {\bibinfo {author} {\bibfnamefont {Dan}\ \bibnamefont
  {Lucas}},\ }\bibfield  {title} {\enquote {\bibinfo {title} {Stabilisation of
  exact coherent structures in two-dimensional turbulence using time-delayed
  feedback},}\ }\href@noop {} {\bibfield  {journal} {\bibinfo  {journal} {arXiv
  preprint arXiv:2008.08388}\ } (\bibinfo {year} {2020})}\BibitemShut {NoStop}%
\bibitem [{\citenamefont {Dubief}\ \emph {et~al.}(2020)\citenamefont {Dubief},
  \citenamefont {Page}, \citenamefont {Kerswell}, \citenamefont {Terrapon},\
  and\ \citenamefont {Steinberg}}]{dubief2020first}%
  \BibitemOpen
  \bibfield  {author} {\bibinfo {author} {\bibfnamefont {Yves}\ \bibnamefont
  {Dubief}}, \bibinfo {author} {\bibfnamefont {Jacob}\ \bibnamefont {Page}},
  \bibinfo {author} {\bibfnamefont {Rich~R}\ \bibnamefont {Kerswell}}, \bibinfo
  {author} {\bibfnamefont {Vincent~E}\ \bibnamefont {Terrapon}}, \ and\
  \bibinfo {author} {\bibfnamefont {Victor}\ \bibnamefont {Steinberg}},\
  }\bibfield  {title} {\enquote {\bibinfo {title} {A first coherent structure
  in elasto-inertial turbulence},}\ }\href@noop {} {\bibfield  {journal}
  {\bibinfo  {journal} {arXiv preprint arXiv:2006.06770}\ } (\bibinfo {year}
  {2020})}\BibitemShut {NoStop}%
\bibitem [{\citenamefont {Page}\ \emph {et~al.}(2020)\citenamefont {Page},
  \citenamefont {Dubief},\ and\ \citenamefont {Kerswell}}]{page2020exact}%
  \BibitemOpen
  \bibfield  {author} {\bibinfo {author} {\bibfnamefont {Jacob}\ \bibnamefont
  {Page}}, \bibinfo {author} {\bibfnamefont {Yves}\ \bibnamefont {Dubief}}, \
  and\ \bibinfo {author} {\bibfnamefont {Rich~R}\ \bibnamefont {Kerswell}},\
  }\bibfield  {title} {\enquote {\bibinfo {title} {Exact traveling wave
  solutions in viscoelastic channel flow},}\ }\href@noop {} {\bibfield
  {journal} {\bibinfo  {journal} {Phys. Rev. Lett.}\ }\textbf {\bibinfo
  {volume} {125}},\ \bibinfo {pages} {154501} (\bibinfo {year}
  {2020})}\BibitemShut {NoStop}%
\bibitem [{\citenamefont {Norton}\ \emph {et~al.}(2018)\citenamefont {Norton},
  \citenamefont {Baskaran}, \citenamefont {Opathalage}, \citenamefont
  {Langeslay}, \citenamefont {Fraden}, \citenamefont {Baskaran},\ and\
  \citenamefont {Hagan}}]{norton2018insensitivity}%
  \BibitemOpen
  \bibfield  {author} {\bibinfo {author} {\bibfnamefont {Michael~M}\
  \bibnamefont {Norton}}, \bibinfo {author} {\bibfnamefont {Arvind}\
  \bibnamefont {Baskaran}}, \bibinfo {author} {\bibfnamefont {Achini}\
  \bibnamefont {Opathalage}}, \bibinfo {author} {\bibfnamefont {Blake}\
  \bibnamefont {Langeslay}}, \bibinfo {author} {\bibfnamefont {Seth}\
  \bibnamefont {Fraden}}, \bibinfo {author} {\bibfnamefont {Aparna}\
  \bibnamefont {Baskaran}}, \ and\ \bibinfo {author} {\bibfnamefont
  {Michael~F}\ \bibnamefont {Hagan}},\ }\bibfield  {title} {\enquote {\bibinfo
  {title} {Insensitivity of active nematic liquid crystal dynamics to
  topological constraints},}\ }\href@noop {} {\bibfield  {journal} {\bibinfo
  {journal} {Phys. Rev. E}\ }\textbf {\bibinfo {volume} {97}},\ \bibinfo
  {pages} {012702} (\bibinfo {year} {2018})}\BibitemShut {NoStop}%
\bibitem [{\citenamefont {Walton}\ \emph {et~al.}(2020)\citenamefont {Walton},
  \citenamefont {McKay}, \citenamefont {Grinfeld},\ and\ \citenamefont
  {Mottram}}]{walton2020pressure}%
  \BibitemOpen
  \bibfield  {author} {\bibinfo {author} {\bibfnamefont {Joshua}\ \bibnamefont
  {Walton}}, \bibinfo {author} {\bibfnamefont {Geoffrey}\ \bibnamefont
  {McKay}}, \bibinfo {author} {\bibfnamefont {Michael}\ \bibnamefont
  {Grinfeld}}, \ and\ \bibinfo {author} {\bibfnamefont {Nigel~J}\ \bibnamefont
  {Mottram}},\ }\bibfield  {title} {\enquote {\bibinfo {title} {Pressure-driven
  changes to spontaneous flow in active nematic liquid crystals},}\ }\href@noop
  {} {\bibfield  {journal} {\bibinfo  {journal} {Eur. Phys. J. E}\ }\textbf
  {\bibinfo {volume} {43}},\ \bibinfo {pages} {1--14} (\bibinfo {year}
  {2020})}\BibitemShut {NoStop}%
\bibitem [{\citenamefont {Wensink}\ \emph
  {et~al.}(2012{\natexlab{b}})\citenamefont {Wensink}, \citenamefont {Dunkel},
  \citenamefont {Heidenreich}, \citenamefont {Drescher}, \citenamefont
  {Goldstein}, \citenamefont {L{\"o}wen},\ and\ \citenamefont
  {Yeomans}}]{wensink2012meso}%
  \BibitemOpen
  \bibfield  {author} {\bibinfo {author} {\bibfnamefont {Henricus~H}\
  \bibnamefont {Wensink}}, \bibinfo {author} {\bibfnamefont {J{\"o}rn}\
  \bibnamefont {Dunkel}}, \bibinfo {author} {\bibfnamefont {Sebastian}\
  \bibnamefont {Heidenreich}}, \bibinfo {author} {\bibfnamefont {Knut}\
  \bibnamefont {Drescher}}, \bibinfo {author} {\bibfnamefont {Raymond~E}\
  \bibnamefont {Goldstein}}, \bibinfo {author} {\bibfnamefont {Hartmut}\
  \bibnamefont {L{\"o}wen}}, \ and\ \bibinfo {author} {\bibfnamefont {Julia~M}\
  \bibnamefont {Yeomans}},\ }\bibfield  {title} {\enquote {\bibinfo {title}
  {Meso-scale turbulence in living fluids},}\ }\href@noop {} {\bibfield
  {journal} {\bibinfo  {journal} {PNAS}\ }\textbf {\bibinfo {volume} {109}},\
  \bibinfo {pages} {14308--14313} (\bibinfo {year}
  {2012}{\natexlab{b}})}\BibitemShut {NoStop}%
\bibitem [{\citenamefont {Linkmann}\ \emph {et~al.}(2019)\citenamefont
  {Linkmann}, \citenamefont {Boffetta}, \citenamefont {Marchetti},\ and\
  \citenamefont {Eckhardt}}]{linkmann2019phase}%
  \BibitemOpen
  \bibfield  {author} {\bibinfo {author} {\bibfnamefont {Moritz}\ \bibnamefont
  {Linkmann}}, \bibinfo {author} {\bibfnamefont {Guido}\ \bibnamefont
  {Boffetta}}, \bibinfo {author} {\bibfnamefont {M~Cristina}\ \bibnamefont
  {Marchetti}}, \ and\ \bibinfo {author} {\bibfnamefont {Bruno}\ \bibnamefont
  {Eckhardt}},\ }\bibfield  {title} {\enquote {\bibinfo {title} {Phase
  transition to large scale coherent structures in two-dimensional active
  matter turbulence},}\ }\href@noop {} {\bibfield  {journal} {\bibinfo
  {journal} {Phys. Rev. Lett.}\ }\textbf {\bibinfo {volume} {122}},\ \bibinfo
  {pages} {214503} (\bibinfo {year} {2019})}\BibitemShut {NoStop}%
\bibitem [{\citenamefont {Linkmann}\ \emph
  {et~al.}(2020{\natexlab{b}})\citenamefont {Linkmann}, \citenamefont
  {Marchetti}, \citenamefont {Boffetta},\ and\ \citenamefont
  {Eckhardt}}]{linkmann2020condensate}%
  \BibitemOpen
  \bibfield  {author} {\bibinfo {author} {\bibfnamefont {Moritz}\ \bibnamefont
  {Linkmann}}, \bibinfo {author} {\bibfnamefont {M~Cristina}\ \bibnamefont
  {Marchetti}}, \bibinfo {author} {\bibfnamefont {Guido}\ \bibnamefont
  {Boffetta}}, \ and\ \bibinfo {author} {\bibfnamefont {Bruno}\ \bibnamefont
  {Eckhardt}},\ }\bibfield  {title} {\enquote {\bibinfo {title} {Condensate
  formation and multiscale dynamics in two-dimensional active suspensions},}\
  }\href@noop {} {\bibfield  {journal} {\bibinfo  {journal} {Phys. Rev. E}\
  }\textbf {\bibinfo {volume} {101}},\ \bibinfo {pages} {022609} (\bibinfo
  {year} {2020}{\natexlab{b}})}\BibitemShut {NoStop}%
\bibitem [{\citenamefont {Mukherjee}\ \emph {et~al.}(2021)\citenamefont
  {Mukherjee}, \citenamefont {Singh}, \citenamefont {James},\ and\
  \citenamefont {Ray}}]{mukherjee2021anomalous}%
  \BibitemOpen
  \bibfield  {author} {\bibinfo {author} {\bibfnamefont {Siddhartha}\
  \bibnamefont {Mukherjee}}, \bibinfo {author} {\bibfnamefont {Rahul~K}\
  \bibnamefont {Singh}}, \bibinfo {author} {\bibfnamefont {Martin}\
  \bibnamefont {James}}, \ and\ \bibinfo {author} {\bibfnamefont
  {Samriddhi~Sankar}\ \bibnamefont {Ray}},\ }\bibfield  {title} {\enquote
  {\bibinfo {title} {Anomalous diffusion and l$\backslash$'evy walks
  distinguish active turbulence},}\ }\href@noop {} {\bibfield  {journal}
  {\bibinfo  {journal} {arXiv preprint arXiv:2105.07872}\ } (\bibinfo {year}
  {2021})}\BibitemShut {NoStop}%
\bibitem [{\citenamefont {Cvitanovic}\ \emph {et~al.}(2005)\citenamefont
  {Cvitanovic}, \citenamefont {Artuso}, \citenamefont {Mainieri}, \citenamefont
  {Tanner}, \citenamefont {Vattay}, \citenamefont {Whelan},\ and\ \citenamefont
  {Wirzba}}]{cvitanovic2005chaos}%
  \BibitemOpen
  \bibfield  {author} {\bibinfo {author} {\bibfnamefont {Predrag}\ \bibnamefont
  {Cvitanovic}}, \bibinfo {author} {\bibfnamefont {Roberto}\ \bibnamefont
  {Artuso}}, \bibinfo {author} {\bibfnamefont {Ronnie}\ \bibnamefont
  {Mainieri}}, \bibinfo {author} {\bibfnamefont {Gregor}\ \bibnamefont
  {Tanner}}, \bibinfo {author} {\bibfnamefont {G{\'a}bor}\ \bibnamefont
  {Vattay}}, \bibinfo {author} {\bibfnamefont {Niall}\ \bibnamefont {Whelan}},
  \ and\ \bibinfo {author} {\bibfnamefont {Andreas}\ \bibnamefont {Wirzba}},\
  }\bibfield  {title} {\enquote {\bibinfo {title} {Chaos: classical and
  quantum},}\ }\href@noop {} {\bibfield  {journal} {\bibinfo  {journal}
  {ChaosBook. org (Niels Bohr Institute, Copenhagen 2005)}\ }\textbf {\bibinfo
  {volume} {69}},\ \bibinfo {pages} {25} (\bibinfo {year} {2005})}\BibitemShut
  {NoStop}%
\bibitem [{\citenamefont {Koch}\ and\ \citenamefont
  {Wilczek}(2021)}]{koch2021role}%
  \BibitemOpen
  \bibfield  {author} {\bibinfo {author} {\bibfnamefont {Colin-Marius}\
  \bibnamefont {Koch}}\ and\ \bibinfo {author} {\bibfnamefont {Michael}\
  \bibnamefont {Wilczek}},\ }\bibfield  {title} {\enquote {\bibinfo {title}
  {The role of advective inertia in active nematic turbulence},}\ }\href@noop
  {} {\bibfield  {journal} {\bibinfo  {journal} {arXiv preprint
  arXiv:2107.14167}\ } (\bibinfo {year} {2021})}\BibitemShut {NoStop}%
\bibitem [{\citenamefont {Shankar}\ \emph {et~al.}(2018)\citenamefont
  {Shankar}, \citenamefont {Ramaswamy}, \citenamefont {Marchetti},\ and\
  \citenamefont {Bowick}}]{shankar2018defect}%
  \BibitemOpen
  \bibfield  {author} {\bibinfo {author} {\bibfnamefont {Suraj}\ \bibnamefont
  {Shankar}}, \bibinfo {author} {\bibfnamefont {Sriram}\ \bibnamefont
  {Ramaswamy}}, \bibinfo {author} {\bibfnamefont {M~Cristina}\ \bibnamefont
  {Marchetti}}, \ and\ \bibinfo {author} {\bibfnamefont {Mark~J}\ \bibnamefont
  {Bowick}},\ }\bibfield  {title} {\enquote {\bibinfo {title} {Defect unbinding
  in active nematics},}\ }\href@noop {} {\bibfield  {journal} {\bibinfo
  {journal} {Phys. Rev. Lett.}\ }\textbf {\bibinfo {volume} {121}},\ \bibinfo
  {pages} {108002} (\bibinfo {year} {2018})}\BibitemShut {NoStop}%
\bibitem [{\citenamefont {Blow}\ \emph {et~al.}(2017)\citenamefont {Blow},
  \citenamefont {Aqil}, \citenamefont {Liebchen},\ and\ \citenamefont
  {Marenduzzo}}]{blow2017motility}%
  \BibitemOpen
  \bibfield  {author} {\bibinfo {author} {\bibfnamefont {Matthew~L}\
  \bibnamefont {Blow}}, \bibinfo {author} {\bibfnamefont {Marco}\ \bibnamefont
  {Aqil}}, \bibinfo {author} {\bibfnamefont {Benno}\ \bibnamefont {Liebchen}},
  \ and\ \bibinfo {author} {\bibfnamefont {Davide}\ \bibnamefont
  {Marenduzzo}},\ }\bibfield  {title} {\enquote {\bibinfo {title} {Motility of
  active nematic films driven by “active anchoring”},}\ }\href@noop {}
  {\bibfield  {journal} {\bibinfo  {journal} {Soft matter}\ }\textbf {\bibinfo
  {volume} {13}},\ \bibinfo {pages} {6137--6144} (\bibinfo {year}
  {2017})}\BibitemShut {NoStop}%
\bibitem [{\citenamefont {Suri}\ \emph {et~al.}(2019)\citenamefont {Suri},
  \citenamefont {Pallantla}, \citenamefont {Schatz},\ and\ \citenamefont
  {Grigoriev}}]{suri2019heteroclinic}%
  \BibitemOpen
  \bibfield  {author} {\bibinfo {author} {\bibfnamefont {Balachandra}\
  \bibnamefont {Suri}}, \bibinfo {author} {\bibfnamefont {Ravi~Kumar}\
  \bibnamefont {Pallantla}}, \bibinfo {author} {\bibfnamefont {Michael~F}\
  \bibnamefont {Schatz}}, \ and\ \bibinfo {author} {\bibfnamefont {Roman~O}\
  \bibnamefont {Grigoriev}},\ }\bibfield  {title} {\enquote {\bibinfo {title}
  {Heteroclinic and homoclinic connections in a {K}olmogorov-like flow},}\
  }\href@noop {} {\bibfield  {journal} {\bibinfo  {journal} {Phys. Rev. E}\
  }\textbf {\bibinfo {volume} {100}},\ \bibinfo {pages} {013112} (\bibinfo
  {year} {2019})}\BibitemShut {NoStop}%
\bibitem [{\citenamefont {Burns}\ \emph {et~al.}(2020)\citenamefont {Burns},
  \citenamefont {Vasil}, \citenamefont {Oishi}, \citenamefont {Lecoanet},\ and\
  \citenamefont {Brown}}]{burns2020dedalus}%
  \BibitemOpen
  \bibfield  {author} {\bibinfo {author} {\bibfnamefont {Keaton~J}\
  \bibnamefont {Burns}}, \bibinfo {author} {\bibfnamefont {Geoffrey~M}\
  \bibnamefont {Vasil}}, \bibinfo {author} {\bibfnamefont {Jeffrey~S}\
  \bibnamefont {Oishi}}, \bibinfo {author} {\bibfnamefont {Daniel}\
  \bibnamefont {Lecoanet}}, \ and\ \bibinfo {author} {\bibfnamefont
  {Benjamin~P}\ \bibnamefont {Brown}},\ }\bibfield  {title} {\enquote {\bibinfo
  {title} {Dedalus: A flexible framework for numerical simulations with
  spectral methods},}\ }\href@noop {} {\bibfield  {journal} {\bibinfo
  {journal} {Phys. Rev. Res.}\ }\textbf {\bibinfo {volume} {2}},\ \bibinfo
  {pages} {023068} (\bibinfo {year} {2020})}\BibitemShut {NoStop}%
\bibitem [{\citenamefont {Viswanath}(2007)}]{viswanath2007recurrent}%
  \BibitemOpen
  \bibfield  {author} {\bibinfo {author} {\bibfnamefont {D}~\bibnamefont
  {Viswanath}},\ }\bibfield  {title} {\enquote {\bibinfo {title} {Recurrent
  motions within plane {C}ouette turbulence},}\ }\href@noop {} {\bibfield
  {journal} {\bibinfo  {journal} {J. Fluid. Mech.}\ }\textbf {\bibinfo {volume}
  {580}},\ \bibinfo {pages} {339--358} (\bibinfo {year} {2007})}\BibitemShut
  {NoStop}%
\bibitem [{\citenamefont {Dennis}\ and\ \citenamefont
  {Schnabel}(1996)}]{Dennis1996}%
  \BibitemOpen
  \bibfield  {author} {\bibinfo {author} {\bibfnamefont {J.~E.}\ \bibnamefont
  {Dennis}}\ and\ \bibinfo {author} {\bibfnamefont {Robert~B.}\ \bibnamefont
  {Schnabel}},\ }\href {\doibase 10.1137/1.9781611971200} {\emph {\bibinfo
  {title} {Numerical Methods for Unconstrained Optimization and Nonlinear
  Equations}}}\ (\bibinfo  {publisher} {Society for Industrial and Applied
  Mathematics},\ \bibinfo {year} {1996})\BibitemShut {NoStop}%
\bibitem [{\citenamefont {Saad}\ and\ \citenamefont
  {Schultz}(1986)}]{saad1986gmres}%
  \BibitemOpen
  \bibfield  {author} {\bibinfo {author} {\bibfnamefont {Youcef}\ \bibnamefont
  {Saad}}\ and\ \bibinfo {author} {\bibfnamefont {Martin~H}\ \bibnamefont
  {Schultz}},\ }\bibfield  {title} {\enquote {\bibinfo {title} {{GMRES}: A
  generalized minimal residual algorithm for solving nonsymmetric linear
  systems},}\ }\href@noop {} {\bibfield  {journal} {\bibinfo  {journal} {SIAM
  J. Sci. Comput.}\ }\textbf {\bibinfo {volume} {7}},\ \bibinfo {pages}
  {856--869} (\bibinfo {year} {1986})}\BibitemShut {NoStop}%
\bibitem [{\citenamefont {Chandler}\ and\ \citenamefont
  {Kerswell}(2013)}]{chandler2013invariant}%
  \BibitemOpen
  \bibfield  {author} {\bibinfo {author} {\bibfnamefont {Gary~J}\ \bibnamefont
  {Chandler}}\ and\ \bibinfo {author} {\bibfnamefont {Rich~R}\ \bibnamefont
  {Kerswell}},\ }\bibfield  {title} {\enquote {\bibinfo {title} {Invariant
  recurrent solutions embedded in a turbulent two-dimensional {K}olmogorov
  flow},}\ }\href@noop {} {\bibfield  {journal} {\bibinfo  {journal} {J. Fluid.
  Mech.}\ }\textbf {\bibinfo {volume} {722}},\ \bibinfo {pages} {554--595}
  (\bibinfo {year} {2013})}\BibitemShut {NoStop}%
\bibitem [{\citenamefont {Willis}\ \emph {et~al.}(2013)\citenamefont {Willis},
  \citenamefont {Cvitanovi{\'c}},\ and\ \citenamefont
  {Avila}}]{willis2013revealing}%
  \BibitemOpen
  \bibfield  {author} {\bibinfo {author} {\bibfnamefont {Ashley~P}\
  \bibnamefont {Willis}}, \bibinfo {author} {\bibfnamefont {P}~\bibnamefont
  {Cvitanovi{\'c}}}, \ and\ \bibinfo {author} {\bibfnamefont {Marc}\
  \bibnamefont {Avila}},\ }\bibfield  {title} {\enquote {\bibinfo {title}
  {Revealing the state space of turbulent pipe flow by symmetry reduction},}\
  }\href@noop {} {\bibfield  {journal} {\bibinfo  {journal} {J. Fluid. Mech.}\
  }\textbf {\bibinfo {volume} {721}},\ \bibinfo {pages} {514--540} (\bibinfo
  {year} {2013})}\BibitemShut {NoStop}%
\bibitem [{\citenamefont {Tan}\ \emph {et~al.}(2019)\citenamefont {Tan},
  \citenamefont {Roberts}, \citenamefont {Smith}, \citenamefont {Olvera},
  \citenamefont {Arteaga}, \citenamefont {Fortini}, \citenamefont {Mitchell},\
  and\ \citenamefont {Hirst}}]{tan2019topological}%
  \BibitemOpen
  \bibfield  {author} {\bibinfo {author} {\bibfnamefont {Amanda~J}\
  \bibnamefont {Tan}}, \bibinfo {author} {\bibfnamefont {Eric}\ \bibnamefont
  {Roberts}}, \bibinfo {author} {\bibfnamefont {Spencer~A}\ \bibnamefont
  {Smith}}, \bibinfo {author} {\bibfnamefont {Ulyses~Alvarado}\ \bibnamefont
  {Olvera}}, \bibinfo {author} {\bibfnamefont {Jorge}\ \bibnamefont {Arteaga}},
  \bibinfo {author} {\bibfnamefont {Sam}\ \bibnamefont {Fortini}}, \bibinfo
  {author} {\bibfnamefont {Kevin~A}\ \bibnamefont {Mitchell}}, \ and\ \bibinfo
  {author} {\bibfnamefont {Linda~S}\ \bibnamefont {Hirst}},\ }\bibfield
  {title} {\enquote {\bibinfo {title} {Topological chaos in active nematics},}\
  }\href@noop {} {\bibfield  {journal} {\bibinfo  {journal} {Nature Physics}\
  }\textbf {\bibinfo {volume} {15}},\ \bibinfo {pages} {1033--1039} (\bibinfo
  {year} {2019})}\BibitemShut {NoStop}%
\bibitem [{\citenamefont {Farano}\ \emph {et~al.}(2019)\citenamefont {Farano},
  \citenamefont {Cherubini}, \citenamefont {Robinet}, \citenamefont
  {De~Palma},\ and\ \citenamefont {Schneider}}]{farano2019computing}%
  \BibitemOpen
  \bibfield  {author} {\bibinfo {author} {\bibfnamefont {Mirko}\ \bibnamefont
  {Farano}}, \bibinfo {author} {\bibfnamefont {Stefania}\ \bibnamefont
  {Cherubini}}, \bibinfo {author} {\bibfnamefont {J-C}\ \bibnamefont
  {Robinet}}, \bibinfo {author} {\bibfnamefont {Pietro}\ \bibnamefont
  {De~Palma}}, \ and\ \bibinfo {author} {\bibfnamefont {TM}~\bibnamefont
  {Schneider}},\ }\bibfield  {title} {\enquote {\bibinfo {title} {Computing
  heteroclinic orbits using adjoint-based methods},}\ }\href@noop {} {\bibfield
   {journal} {\bibinfo  {journal} {J. Fluid. Mech.}\ }\textbf {\bibinfo
  {volume} {858}} (\bibinfo {year} {2019})}\BibitemShut {NoStop}%
\bibitem [{\citenamefont {Koon}\ \emph {et~al.}(2000)\citenamefont {Koon},
  \citenamefont {Lo}, \citenamefont {Marsden},\ and\ \citenamefont
  {Ross}}]{koon2000heteroclinic}%
  \BibitemOpen
  \bibfield  {author} {\bibinfo {author} {\bibfnamefont {Wang~Sang}\
  \bibnamefont {Koon}}, \bibinfo {author} {\bibfnamefont {Martin~W}\
  \bibnamefont {Lo}}, \bibinfo {author} {\bibfnamefont {Jerrold~E}\
  \bibnamefont {Marsden}}, \ and\ \bibinfo {author} {\bibfnamefont {Shane~D}\
  \bibnamefont {Ross}},\ }\bibfield  {title} {\enquote {\bibinfo {title}
  {Heteroclinic connections between periodic orbits and resonance transitions
  in celestial mechanics},}\ }\href@noop {} {\bibfield  {journal} {\bibinfo
  {journal} {Chaos}\ }\textbf {\bibinfo {volume} {10}},\ \bibinfo {pages}
  {427--469} (\bibinfo {year} {2000})}\BibitemShut {NoStop}%
\bibitem [{\citenamefont {Rivas}\ \emph {et~al.}(2020)\citenamefont {Rivas},
  \citenamefont {Shendruk}, \citenamefont {Henry}, \citenamefont {Reich},\ and\
  \citenamefont {Leheny}}]{rivas2020driventransition}%
  \BibitemOpen
  \bibfield  {author} {\bibinfo {author} {\bibfnamefont {David~P.}\
  \bibnamefont {Rivas}}, \bibinfo {author} {\bibfnamefont {Tyler~N.}\
  \bibnamefont {Shendruk}}, \bibinfo {author} {\bibfnamefont {Robert~R.}\
  \bibnamefont {Henry}}, \bibinfo {author} {\bibfnamefont {Daniel~H.}\
  \bibnamefont {Reich}}, \ and\ \bibinfo {author} {\bibfnamefont {Robert~L.}\
  \bibnamefont {Leheny}},\ }\bibfield  {title} {\enquote {\bibinfo {title}
  {Driven topological transitions in active nematic films},}\ }\href {\doibase
  10.1039/D0SM00693A} {\bibfield  {journal} {\bibinfo  {journal} {Soft Matter}\
  }\textbf {\bibinfo {volume} {16}},\ \bibinfo {pages} {9331--9338} (\bibinfo
  {year} {2020})}\BibitemShut {NoStop}%
  \bibitem [{Note1()}]{Note1}%
  \BibitemOpen
  \bibinfo {note} {Videos can be accessed at \url{https://www.youtube.com/channel/UCeNcHrYW6yAAQ95bdDqQjUw}}\BibitemShut {NoStop}%
\end{thebibliography}

\begin{thebibliography}{1}
\makeatletter
\renewcommand\@bibitem[1]{\item\if@filesw \immediate\write\@auxout
    {\string\bibcite{#1}{S\the\value{\@listctr}}}\fi\ignorespaces}
\def\@biblabel#1{[S#1]}
\makeatother
\bibitem{Gottwald2004}
Georg~A. Gottwald and Ian Melbourne.
\newblock A new test for chaos in deterministic systems.
\newblock {\em Proceedings of the Royal Society of London. Series A:
  Mathematical, Physical and Engineering Sciences}, 460(2042):603--611,
  February 2004.

\bibitem{Gottwald2016}
Georg~A. Gottwald and Ian Melbourne.
\newblock {\em The 0-1 Test for Chaos: A Review}, pages 221--247.
\newblock Springer Berlin Heidelberg, Berlin, Heidelberg, 2016.

\bibitem{suri2019heteroclinic}
Balachandra Suri, Ravi~Kumar Pallantla, Michael~F Schatz, and Roman~O
  Grigoriev.
\newblock Heteroclinic and homoclinic connections in a {K}olmogorov-like flow.
\newblock {\em Phys. Rev. E}, 100(1):013112, 2019.

\bibitem{verhulst2006nonlinear}
Ferdinand Verhulst.
\newblock {\em Nonlinear differential equations and dynamical systems}.
\newblock Springer Science \& Business Media, 2006.

\bibitem{viswanath2007recurrent}
D~Viswanath.
\newblock Recurrent motions within plane {C}ouette turbulence.
\newblock {\em J. Fluid. Mech.}, 580:339--358, 2007.

\end{thebibliography}

\end{document}